\newif\ifwordcount
\wordcountfalse

\documentclass{aastex631}
\usepackage{multirow,bigdelim}
\usepackage{cprotect}
\usepackage{xcolor}
\usepackage[thinc]{esdiff}
\usepackage{amsmath}
\usepackage{tablefootnote}
\usepackage{booktabs}
\usepackage{hhline}
\usepackage{color}

\def\x{{\bf x}}

\newcommand\beq{\begin{equation}}
\newcommand\eeq{\end{equation}}
\newcommand\beqn{\begin{eqnarray}}
\newcommand\eeqn{\end{eqnarray}}

\newcommand{\ba}{\begin{eqnarray}}
\newcommand{\ea}{\end{eqnarray}}
\newcommand{\be}{\begin{equation}}
\newcommand{\ee}{\end{equation}}

\usepackage{cprotect}
\usepackage{graphicx}
\usepackage{dcolumn}
\usepackage{bm}
\usepackage{mathtools}
\usepackage{fancyvrb}
\usepackage{hyperref}
\usepackage{graphicx}
\usepackage{caption}
\usepackage{subcaption}
\usepackage{ragged2e}

\DeclareCaptionJustification{justified}{\justifying}
\begin{document}

\renewcommand{\arraystretch}{1.1}

\title{Projected-Field Kinetic Sunyaev-Zel'dovich Cross-Correlations: Halo Model and Forecasts}

\author{Boris Bolliet}
\affiliation{Department of Physics, Columbia University, New York, NY, USA 10027}
\email{bb3028@columbia.edu}

\author{J.~Colin Hill}
\affiliation{Department of Physics, Columbia University, New York, NY, USA 10027}
\affiliation{Center for Computational Astrophysics, Flatiron Institute, New York, NY, USA 10010}

\author{Simone Ferraro}
\affiliation{Lawrence Berkeley National Laboratory, One Cyclotron Road, Berkeley, CA, USA 94720}
\affiliation{Berkeley Center for Cosmological Physics, Department of Physics, University of California, Berkeley, CA, USA 94720}

\author{Aleksandra Kusiak}
\affiliation{Department of Physics, Columbia University, New York, NY, USA 10027}

\author{Alex Krolewski}
\affiliation{AMTD Fellow, Waterloo Centre for Astrophysics, University of Waterloo, Waterloo, ON, Canada N2L 3G1}
\affiliation{Perimeter Institute for Theoretical Physics, 31 Caroline St. North, Waterloo, ON, Canada NL2 2Y5}


\date{\vspace{-7mm}{Accepted 2019 --. Received 2019 --}}

\begin{abstract}
 The kinetic Sunyaev-Zel'dovich (kSZ) effect, i.e., the Doppler boost of cosmic microwave background (CMB) photons caused by their scattering off free electrons in galaxy clusters and groups with non-zero bulk velocity, is a powerful window on baryons in the universe. We present the first halo-model computation of the cross-power spectrum of the ``projected-field'' kSZ signal with large-scale structure (LSS) tracers. We compare and validate our calculations against previous studies, which relied on $N$-body-calibrated effective formulas rather than the halo model.  We forecast results for  CMB maps from the Atacama Cosmology Telescope (AdvACT), Simons Observatory (SO), and CMB-S4, and LSS survey data from the Dark Energy Survey, the Vera C.~Rubin Observatory (VRO), and \textit{Euclid}.  
 In cross-correlation with galaxy number density, for AdvACT $\times$ \textit{unWISE} we forecast an 18$\sigma$ projected-field kSZ detection using data already in hand.  Combining SO CMB maps and \textit{unWISE} galaxy catalogs, we expect a $62\sigma$ detection, yielding precise measurements of the gas density profile radial slopes.  Additionally, we forecast first detections of the kSZ -- galaxy weak lensing cross-correlation with AdvACT $\times$ VRO/\textit{Euclid} (at 6$\sigma$) and of the kSZ -- CMB weak lensing cross-correlation with SO (at 16$\sigma$). Finally, $\approx 10-20$\% precision measurements of the shape of the gas density profile should be possible with CMB-S4 kSZ -- CMB lensing cross-correlation without using any external datasets.

\end{abstract}


\tableofcontents



\section{Introduction}\label{sec:preintro}

The Sunyaev-Zel'dovich (SZ) effect is the Compton scattering  of cosmic microwave background (CMB) photons off electrons during cosmological expansion. 
 When the electron population is thermal, the effect is proportional to the electron density-weighted temperature, i.e., the electron pressure, projected onto the line of sight (LOS) and is called the \textit{thermal} SZ (tSZ) effect. It was first predicted by \cite{1969Ap&SS...4..301Z}, shortly after the discovery of the CMB \citep[][]{1965ApJ...142..419P}. In their seminal work, Zel'dovich and Sunyaev used the tSZ effect, and observational data available at the time, to estimate the age of the universe at recombination. In a subsequent work \citep{SZ_1972}, they showed that the tSZ effect could be used as a probe of the hot, ionized gas around galaxy clusters and groups: the intracluster medium (ICM) and the circumgalactic medium (CGM), filled with hot electrons at temperature $k_B T_e\approx 1-10\,\mathrm{keV}$. In \cite{SZ_1980,SZ_1980MNRAS.190..413S} they identified the \textit{kinetic} component of the effect as a probe of the peculiar velocity of galaxy clusters and groups. The kinetic SZ (kSZ) effect is the Doppler shift of the CMB spectrum with respect to the mean CMB temperature, proportional to the electron density-weighted bulk velocity of the gas projected onto the LOS. At lowest order in the LOS velocity, the kSZ effect preserves the form of the blackbody spectrum, simply producing an overall shift up or down in the photon temperature, depending on the direction of the bulk motion with respect to the LOS.

The tSZ effect was measured for the first time nearly four decades ago at the Owens Valley Radio Observatory \citep[][]{BGH_1984}. Today, it is routinely used in astrophysics to learn about the thermodynamics of the ICM and CGM \citep[see, e.g.,][for recent analyses]{Makiya_2018,Koukoufilippas:2019ilu,Chiang_2020,Pandey_21} and the distribution of matter in clusters \citep[e.g.,][]{Ruppin:2017bnt,DES:2018qrh,2022EPJWC.25700031M} and filaments \citep[e.g.,][]{Tanimura_2018,de_Graaff_2019,Hincks_2021,Lokken:2021omd}. In addition, owing to the success of recent CMB instruments in producing wide-area and high-resolution multi-frequency CMB maps --- such as \emph{Planck} \citep[][]{Plc_2014_overview,Plc_2016_overview,Plc_2020_overview}, the Atacama Cosmology Telescope \citep[ACT;][]{Aiola2020,Naess_2020} and the South Pole Telescope \citep[SPT;][]{Schaffer_2011,SPT:2018whf} --- the tSZ effect is becoming a competitive low-redshift cosmological probe, via the power spectrum and bispectrum of the Compton-$y$ field \citep[e.g.,][]{Wilson_2012, Crawford_2014, Planck_ymap_2014, Planck_ymap_2016, Horowitz_2017, Bolliet:2017lha,Ravenni_2021,Tanimura_2021}, tSZ cluster abundance \citep[e.g.,][]{Hasselfield_2013,Planck_cluster_counts_2014,Planck_cluster_counts_2016,Salvati_2018_cc,Bocquet_2019,Bolliet_2020,Salvati_21_cc}, and cross-correlations with galaxy and weak lensing surveys \citep[e.g.,][]{Hill_2014, Hojjati_2017, Yan_2021,Gatti_2021, Tr_ster_2022}.

\begin{figure}
\centering
    \includegraphics[width=0.7\columnwidth]{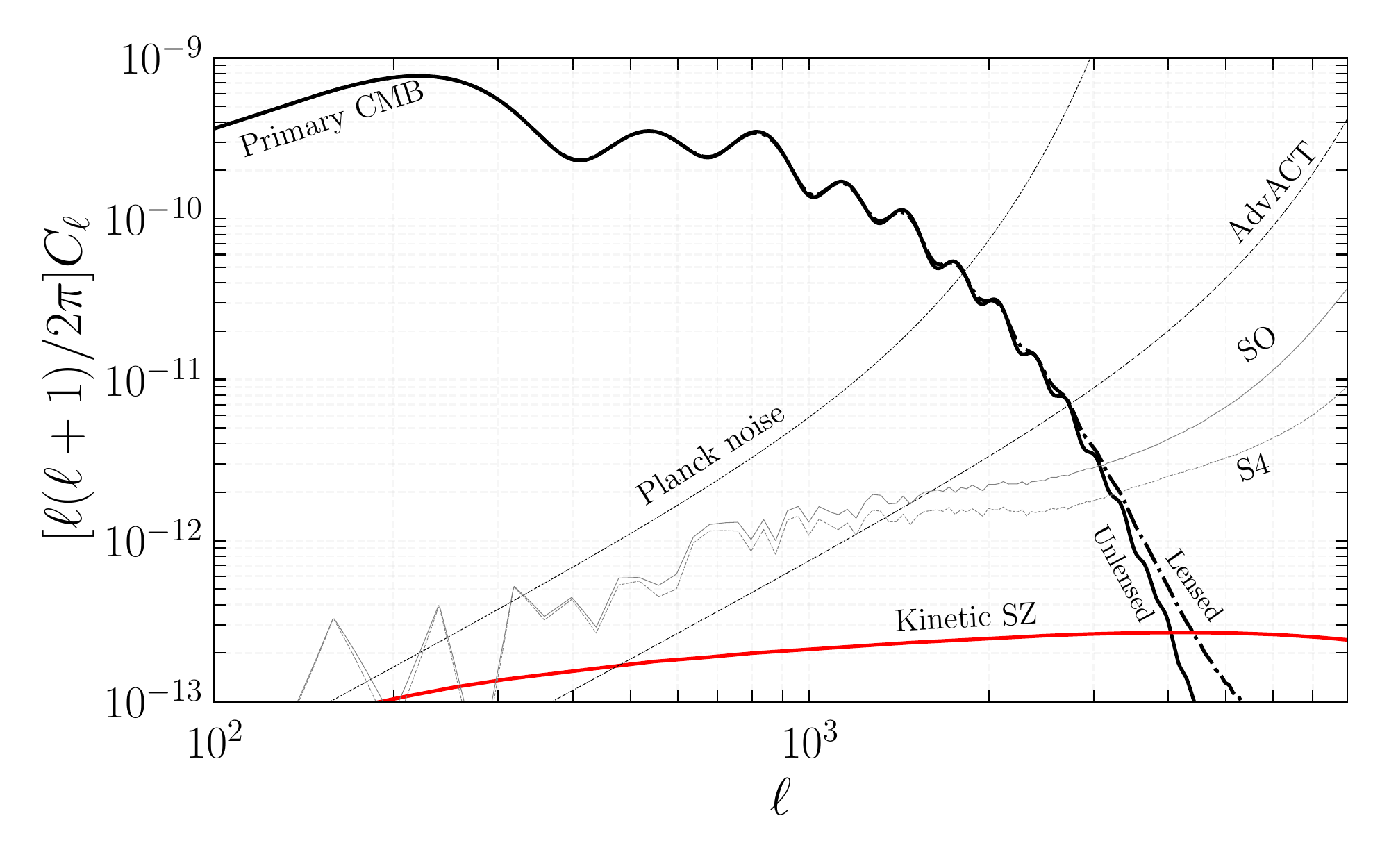}
    \vspace{-0.3cm}
  \caption{ The dimensionless CMB temperature angular anisotropy power spectrum, unlensed (solid line) and lensed (dash-dotted), with the four noise curves used in our forecast analysis (Section \ref{s:forecasts}) as labeled and the expected kSZ power spectrum computed in the halo model with Eq.~\eqref{eq:clksz2} (note that this approximation slightly underestimates the total kSZ power spectrum, particularly on large scales). The CMB and kSZ power spectra assume our fiducial model (Subsection~\ref{s:fid}) and the noise curves are computed according to specifications given in Table \ref{tab:exps}. 
  }  
    \label{fig:cls_ksz_tt}
\end{figure}

In contrast, the kSZ effect has been much more challenging to detect. Naturally, if a cluster has a large enough peculiar LOS velocity, the kSZ effect can be significant and detectable by targeted observations. This was achieved for the first time by \cite{Sayers_2013} with Bolocam data collected at the Caltech Submillimeter Observatory, more recently by \cite{Adam_2017} on the same system with NIKA observations  at the IRAM telescope, and lastly by \cite{Sayers_2019} for several galaxy clusters observed with Bolocam and AzTEC/ASTE. Nonetheless,  in order to understand  global properties of galaxy evolution or large-scale structure (LSS) formation, it is necessary to measure the kSZ effect averaged over a large number of objects, or over a large fraction of the sky. This can be done using large-area maps produced by CMB telescopes. The main challenges are that the resolution of CMB maps has only recently allowed us to probe the small angular scales where the kSZ effect is significant compared to the primary anisotropy of the CMB (see Figure \ref{fig:cls_ksz_tt}), and that since the kSZ effect preserves the blackbody distribution of CMB photons, it is impossible to separate clearly from the primary CMB anisotropy in multi-frequency power spectrum analyses. \emph{Planck} did not report a constraint on the amplitude of the kSZ power spectrum \citep[][]{planck_2020_ps_lkl}. With ACT, \cite{Choi_2020} reported an upper bound, and SPT obtained a tentative $3\,\sigma$ detection \citep{Reichardt_2021}. Subsequently, \cite{Gorce:2022cvb} reanalyzed the SPT data, incorporating information from \emph{Planck} for constraining the parameters controlling reionization, as well as cosmology-dependent kSZ and tSZ spectra.

Several alternative techniques have been proposed in order to overcome these challenges and obtain robust measurements of the kSZ effect. They fall into four main categories:
\begin{itemize}
\item The \textit{mean pairwise momentum} method \citep[][]{Ferreira_99} uses the fact that the averaged momentum of cluster pairs should be negative when clusters are separated by a distance smaller than $\approx 25-50 \, \mathrm{Mpc}$, as they tend to move towards each other due to gravity \citep[see][and references therein]{Calafut_2021}. This method requires high-resolution CMB maps and galaxy surveys with accurate redshifts in order to reconstruct the distances between clusters and groups.  \cite{Hand_2012} made the first detection of the kSZ effect, applying the pairwise momentum method to data from ACT and the Sloan Digital Sky Survey III (SDSS-III) DR9 Baryon Oscillation Spectroscopic Survey (BOSS) \citep{2011AJ....142...72E}, followed by \cite{plc_ksz_2016} using \emph{Planck} maps and SDSS DR7, \cite{Soergel_2016} with DES-Y1 and SPT at $4.2\,\sigma$ (using photometric rather than spectroscopic redshifts), \cite{Bernardis_2017} with ACT and SDSS DR11 at $4.1\,\sigma$,  and \cite{Calafut_2021} reaching $5.4\,\sigma$ with ACT DR5 maps and SDSS DR15 catalogs. This method can also be implemented in Fourier space --- see \cite{https://doi.org/10.48550/arxiv.1606.06367} for the formalism and \cite{Sugiyama:2017uvr} for a first measurement. 

\item The \textit{velocity-weighted stacking} method relies on reconstructing the velocity field from spectroscopic galaxy surveys before stacking cut-outs of CMB maps centered at the locations of galaxies, each weighted by their LOS velocity.  The velocity weights allow us to extract the correlations between the kSZ effect and the galaxies, while avoiding the velocity cancellation which would occur in a na\"{i}ve approach without the weights, since the LOS velocity is as likely to be positive or negative. See \cite{Ho2009,Shao_2011} for early developments of the idea and \cite{Li_2014} for a presentation of the method in its more modern formulation. \cite{plc_ksz_2016} and \cite{Schaan_2016} applied the method using \emph{Planck} maps with SDSS DR7, and ACT maps with CMASS galaxies, respectively, detecting the kSZ effect at $3$-$4\,\sigma$ significance.\footnote{In fact, \cite{plc_ksz_2016} does not use stacks, but rather they reconstruct a correlation function and use \textit{template fitting} for the kSZ detection, as proposed in \cite{Ho2009}. The idea remains the same as in stacking analyses: use the reconstructed velocity as a weight.}  Recently, \cite{schaan_2020_combined} reported a $6.5\sigma$ detection with ACT DR5 and \emph{Planck} in combination with CMASS galaxies ($z\approx 0.55$, $M_\mathrm{vir}\approx3\times 10^{13}M_\odot/h$), and \cite{tanimura2020direct} obtained a $3.5\,\sigma$ detection for \emph{Planck} maps and the more massive WHL clusters of the SDSS survey ($z\approx 0.25-0.55$, $M_\mathrm{500c}\approx10^{14}M_\odot/h$).

\item The \textit{projected-field} method measures the cross-correlation between the square of the kSZ temperature anisotropy and a LSS tracer field, projected onto the 2D sphere. Here, the velocity cancellation is avoided via the squaring operation.  In contrast to the methods described above, one challenge for the projected-field estimator is that the CMB maps must be thoroughly cleaned of foregrounds, as there is no external information about the LOS velocity field used to extract the kSZ signal from amongst the other signals in the small-scale mm-wave sky.  Although this method was introduced nearly two decades ago \citep[][]{Dore2004,DeDeo}, there are only two reported detections of the kSZ effect making use of it: \cite{Hill2016} cross-correlated CMB maps constructed from \emph{Planck} and \emph{WMAP} data with galaxies from the \emph{Wide-field Infrared Survey Explorer} (\emph{WISE}), and \cite{Kusiak:2021hai} used similar CMB maps and the \emph{unWISE} galaxy catalog \citep{Alex}, achieving a $5\,\sigma$ detection.

\item The \textit{velocity reconstruction} method measures the large-scale velocity modes via a quadratic estimator applied on a CMB map combined with a galaxy catalog \citep[e.g.,][]{Deutsch_2018,Munchmeyer:2018eey,Sato-Polito:2020cil,giri_20,Cayuso:2021ljq,Contreras:2022zdz}. (This is different from reconstructed velocities in the velocity-weighted stacking method. Here, velocities are reconstructed from the kSZ effect, rather than just the galaxy catalog.) There are no measurements with this method yet; however it is expected to become a competitive cosmological probe. For instance, \cite{Munchmeyer:2018eey} forecast primordial non-Gaussianity constraints with CMB-S4 and VRO that are three times more sensitive than with VRO alone.
\end{itemize}

\cite{Smith:2018bpn} proved that apart from the projected-field estimator, the other three methods are mathematically equivalent, involving different ways of estimating the bispectrum $\langle T \delta_g \delta_g \rangle$, where $T$ is the CMB temperature field and $\delta_g$ is the galaxy overdensity field.  In contrast, the projected-field estimator is a bispectrum of the form $\langle T T \delta_g \rangle$.  Measuring the former bispectrum generally requires precise redshifts for the LSS tracers, while the latter requires precise removal of foregrounds from the CMB map.

Each method has its specific areas of applications. For instance, the pairwise momentum method is expected to become a unique probe of dark energy and modified gravity models \citep[e.g.,][]{Kosowsky_2009,Bull_2012,Keisler_2013,Mueller_2015}. Furthermore, with assumptions about the velocity field and in combination with tSZ measurements, it allows us to constrain the relationship between clusters' optical depth  \citep{Battaglia_2016,Flender_2017} and Compton-$y$ parameters, probing the ICM/CGM thermodynamics, as was done for the first time using kSZ measurements in \cite{Vavagiakis_2021}.
With the velocity-weighted stacking method, one can use aperture photometry in order to measure the shape of the gas density profile \citep{schaan_2020_combined} and its thermodynamics. Remarkably, \cite{Amodeo_2021} constrained the parameters of the Ostriker-Bode-Babul (OBB) model \citep{Ostriker_2005} using the tSZ and kSZ measurements from \cite{schaan_2020_combined}, and found that cosmological simulations, such as Illustris TNG \citep{Springel_10.1093/mnras/stx3304,tng_18} and simulations by \cite{b2010}, underpredict the CGM density and pressure at large radii. 

In this paper we focus on the projected-field method for measuring the kSZ effect.  We extend the theoretical formalism developed in \cite{Dore2004, DeDeo, Hill2016, Ferraro2016, Kusiak:2021hai} so that we can use the projected-field power spectrum as a probe of the ionized gas density profile in and around massive halos.  We do so by developing a halo-model-based approach \citep[see, e.g.,][for reviews of the halo model]{Seljak:2000gq, Cooray:2002dia}.

For general reviews of the tSZ and kSZ effects and their applications to astrophysics and cosmology, we refer to \cite{Birkinshaw_1999,Carlstrom_2002,Mroczkowski_2019}. Nonetheless, these do not cover recent developments of applications of the tSZ and kSZ effects to cosmology, such as: growth reconstruction \citep[e.g.,][]{Alonso:2016jpy}; velocity reconstruction (see above); reionization probes \citep[e.g.,][]{Smith:2016lnt,Ferraro_smith_2018,2018MNRAS.476.4025M,LaPlante2020,LaPlante:2021ced,Chen:2022lhr}; or the cosmological applications of the projected-field estimator that we discuss here. 

The remainder of this paper is organized as follows. In the next subsections, we present  our fiducial model, e.g., fiducial cosmological and astrophysical parameter values, assumptions, and the notation used throughout the paper (Subsection \ref{s:fid}).  In Subsection \ref{sec:kszani}, we review key aspects of kSZ temperature fluctuations. In Subsection \ref{sec:pf_estimator}, we review the general formalism for the projected-field estimator. In Subsection \ref{ss:eff}, we review the \textit{effective approach} for modeling this signal \citep[as in][]{Dore2004,DeDeo,Ferraro2016} and its numerical implementation  \verb|class_sz|. Section \ref{s:hm} is dedicated to the halo model formulation and implementation. The halo model  code \verb|class_sz| is described in Subsection \ref{s:class_sz}.  Subsection \ref{ss:hybrid_bispec}-\ref{ss:hmeqs} contain the main material needed for the halo model calculations. The covariance matrix and lensing contribution  are discussed in Subsection \ref{ss:covmat} and \ref{ss:cmb_lens_cont}, respectively. We present forecasts for several experimental configurations and cross-correlations in Section \ref{s:forecasts}. Our main results and conclusions are summarized in Section \ref{s:conclusions}. Finally, we report a number of useful halo model tools and comparisons in the Appendix. In  Appendix \ref{s:veldisp} we discuss different assumptions for the velocity dispersion. In Appendix \ref{app:hm} we present various halo model quantities and their \verb|class_sz| implementation. In Appendix \ref{app:comp} we discuss in more detail some of the differences between previous studies and the new results presented here: choice of Wiener filter in Appendix \ref{app:filters} and comparison with forecasts from \cite{Ferraro2016} in Appendix \ref{ap:comp_ferraro}.

\section{Background}

\subsection{Fiducial Model,  Assumptions, and Notations}\label{s:fid}

We assume the homogeneous $\Lambda$ Cold Dark Matter (CDM) cosmology, on a spatially flat Friedmann-Lema\^{i}tre-Robertson-Walker geometry with scale factor $a$. Our fiducial model corresponds to the  \emph{Planck} 2018 cosmology \citep[last column of Table 1 in][]{Planck2018} with parameters:  $\Omega_ch^2=0.11933$ and $\Omega_bh^2=0.02242$, the CDM and baryon density, respectively, where $h$ is the reduced Hubble parameter; $\ln(10^{10}A_s)=3.047$ and $n_s=0.9665$, the amplitude and spectral index of the primordial scalar perturbation power spectrum, defined at pivot scale $k_\mathrm{pivot}=0.05\,\mathrm{Mpc}^{-1}$; and $\tau_\mathrm{reio}=0.0561$, the reionization optical depth. In our fiducial model, we consider one massive and two massless neutrino states, with total mass $\sum m_\nu=0.06\,\mathrm{eV}$ and an effective number of extra relativistic degrees of freedom  $N_\mathrm{eff}=3.046$ (in order to obtain these $N_\mathrm{eff}$ and $\sum m_\nu$ values, we set the parameters $N_\mathrm{ur}=2.0328$ and $N_\mathrm{ncdm}=1$ in \verb|class_sz|). As derived parameters, we have $\sigma_8=0.81$ for the standard deviation of the linear-theory matter density field smoothed over a sphere of $8\,\mathrm{Mpc}/h$ at $z=0$, a matter fraction $\Omega_\mathrm{m}=0.311$, and a baryon fraction $\Omega_\mathrm{e}=0.0490$, corresponding to a cosmological baryon fraction $f_\mathrm{b}\equiv \Omega_\mathrm{b}/\Omega_\mathrm{m}=0.156$.  Furthermore, our fiducial model assumes a fully ionized ICM, with the free electron fraction $f_\mathrm{free}=1$ (in reality, $f_\mathrm{free}$ is slightly smaller than unity since some of the baryons are in the form of stars and neutral gas), and the primordial Helium abundance $Y_\mathrm{He}=0.245$.

For the halo abundance, we use the \cite{Tinker_2008} halo mass function computed for overdensity masses $m_\mathrm{200c}$, i.e., the mass within the sphere enclosing 200 times the critical density at the halo redshift (see Appendix \ref{app:hmf} for details on the halo mass function). When needed, we convert between mass definitions and compute halo concentrations using the \cite{2013ApJ...766...32B} concentration-mass relation (see Appendix \ref{ss:mass_conv} for details on the mass conversion). This is also our choice for calculating the halo concentration entering the NFW profile. 

The goal of this paper is to apply the halo model formalism to the projected-field kSZ power spectrum. This involves computing cross-correlations between the kSZ effect and LSS tracers. In this work, we shall focus on cross-correlations with three different LSS tracers: the galaxy overdensity, $\delta_\mathrm{g}$; galaxy weak lensing convergence, $\kappa_\mathrm{g}$; and CMB weak lensing convergence, $\kappa_\mathrm{cmb}$. 

Each tracer depends on a specific physical property of halos. The galaxy overdensity is based on a galaxy Halo Occupation Distribution (HOD), galaxy weak lensing and CMB weak lensing are based on the halo mass profiles, which we parameterize with the Navarro-Frenk-White density profile \citep{Navarro_1997}, while the kSZ effect is based on the electron density profile (see Appendix \ref{sss:baryons}). In general, a physical property of a tracer $X$ can be written as a mass-independent redshift-dependent kernel, which we denote $W^X$, and a radial profile denoted $u^X$ that depends on the halo mass. For the Fourier transform of the profile, we use the notation $\hat{u}^X_k$.

The building blocks of halo model power spectra and bispectra are ensemble averages  of Fourier transforms of the tracers' physical properties. To simplify our expressions, we write the ensemble average over halo masses (at fixed redshift) of the Fourier transforms of the tracers' radial profiles with the equivalent notations

\begin{equation}
    \langle X \rangle_n = \langle \hat{u}^X_k  \rangle_n = \int \hat{u}^X_k \mathrm{d}n =\int_{m_\mathrm{min}}^{m_\mathrm{max}} \hat{u}^X_k \frac{\mathrm{d}n}{\mathrm{d}m} \mathrm{d}m = \int_{\ln m_\mathrm{min}}^{\ln m_\mathrm{max}} \hat{u}^X_k \frac{\mathrm{d}n}{\mathrm{d}\ln m}\mathrm{d}\ln m,\label{eq:massint}
\end{equation}
where $\mathrm{d}n/\mathrm{d}m$ represents the comoving differential number density of halos per unit mass. In \verb|class_sz|, we evaluate mass integrals on a logarithmic grid, as indicated in the right hand side (RHS) of the equation. Throughout the paper, we set the mass integral bounds to $m_\mathrm{min}=10^{10} M_\odot/h$ and $m_\mathrm{max}=3.5\times 10^{15} M_\odot/h$.

We write the differential number density of halos as $\mathrm{d}n/\mathrm{d}m=\mathrm{d}^2 N/(\mathrm{d}m \, \mathrm{dv})$ where $N$ is the number of halos of mass $m$ in comoving volume per steradian  $\mathrm{d}\mathrm{v}$.  The comoving volume per steradian is related to the comoving distance $\chi$ via $\mathrm{dv}=\chi^2\mathrm{d}\chi$, with $\mathrm{d}\chi=(c/H(z))\mathrm{d}z$ where $c$ is the speed of light, $H(z)$ is the Hubble parameter, and $z$ denotes redshift. 

The integrals over comoving volume, i.e., the redshift integrals, are also carried out on a logarithmic grid. In the paper, we shall use the following equivalent notations:
\begin{equation}
    \int [\cdot\cdot\cdot]\mathrm{dv}=\int [\cdot\cdot\cdot]\chi^2\mathrm{d}\chi=\int_{\ln(1+z_\mathrm{min})}^{\ln(1+z_\mathrm{max})}[\cdot\cdot\cdot]\frac{c\chi^2}{H}(1+z)\mathrm{d}\ln(1+z).\label{eq:zint}  
\end{equation}
For the redshift bounds we set $z_\mathrm{min}=0.005$ and $z_\mathrm{max}=3$. The lower bound is chosen to avoid spurious numerical divergence at $z=0$. We checked that as long as the value is small enough, our predictions are unchanged by different $z_\mathrm{min}$. The upper bound is chosen so that we do not need to extrapolate the redshift dependence of quantities calibrated on simulations (e.g., the halo mass function, gas density profiles, etc.).  For kSZ${}^2$ - CMB lensing cross-correlation, although the lensing kernel peaks at $z\sim 2$, we note that we are missing some contribution from $z>3$. We leave a detailed study of the high redshift contribution to future works, as it implies using a suited optical depth evolution and halo mass function which are outside the scope of this paper.

To go from 3D spectra (power spectra or bispectra) to 2D angular spectra, i.e., by projecting along the LOS, we work within the Limber approximation on the flat sky \citep{Limber}, such that 3D wavenumber $k$ is mapped to 2D angular multipole $\ell$ via
\begin{equation}
k=(\ell+1/2)/\chi.\label{eq:limber}
\end{equation}
Thus, in the paper, we often use $k$ and $\ell$ interchangeably. We refer the reader to Appendix A of \cite{hp2013} for a detailed discussion on the Limber/flat-sky approximation in this context.

\subsection{Kinetic SZ Anisotropy}\label{sec:kszani}

The CMB temperature fluctuation due to the kSZ effect,   $\Theta^\mathrm{kSZ}=\Delta T/\bar{T}$ where $\bar{T}$ is the mean CMB temperature, is sourced by the visibility-weighted electron velocity field projected along the LOS. In a direction ${\mathbf{n}}$, it can be written as
\begin{equation}
    \Theta^\mathrm{kSZ}({\mathbf{n}})=-\frac{1}{c}\int \mathrm{d}\chi \, g(\bm{x}){\mathbf{n}}\cdot\bm{v_e}(\bm{x}),\label{eq:theta_ksz}
\end{equation}
where $\bm{v_e}$ is the electron velocity at $\bm{x}=(\chi,{\mathbf{n}}$)  and $g$ is the visibility function, i.e., the probability of scattering within $\mathrm{d}\chi$, given by 
\begin{equation}
    g= \frac{\mathrm{d}\tau}{\mathrm{d}\chi}e^{-\tau}\quad\mathrm{with}\quad \frac{\mathrm{d}\tau}{\mathrm{d}\chi} = \frac{\dot{\tau}}{c} = a \sigma_\mathrm{T}n_\mathrm{e}\label{eq:g}
\end{equation}
where the overdot represents a derivative with respect to conformal time $\eta=\int\mathrm{d}t/a$, and where $\sigma_\mathrm{T}=6.985\times 10^{-74} \, \mathrm{Mpc}^2$ is the Thomson scattering cross-section.\footnote{In terms of coordinate (or cosmological) time $t$, the optical depth definition is simply $\mathrm{d}\tau=c\sigma_\mathrm{T}n_\mathrm{e}\mathrm{d}t$.}
We express the electron number density along the LOS as
\begin{equation}
    n_\mathrm{e} =\frac{\rho_\mathrm{gas,free}}{m_\mathrm{u}\mu_\mathrm{e}}\quad\mathrm{with}\quad\rho_\mathrm{gas,free}= f_\mathrm{free}\rho_\mathrm{gas}:=\rho_\mathrm{e}
\label{eq:ne}
\end{equation}
where $m_\mathrm{u}$ is the atomic mass unit and $\mu_\mathrm{e}= (1-Y_\mathrm{He}/2)^{-1}\simeq1.14$ is the mean molecular weight per electron.\footnote{This expression is valid at  $z\lesssim3$, when the gas is fully ionized, that is when Helium is doubly ionized \citep[see][and references therein]{2017ApJ...841...87L}. For a gas of Hydrogen and Helium, the general expression is $\mu_e^{-1}=(y_\mathrm{H}\frac{1}{1}X_\mathrm{H}+y_\mathrm{He}\frac{2}{4}Y_\mathrm{He})$, with $y_\mathrm{H}$ and $y_\mathrm{He}$ the ionization fraction of Hydrogen and Helium, and $X_\mathrm{H}=1-Y_\mathrm{He}$ the Hydrogen mass fraction. Assuming fully ionized gas, we then set $y_\mathrm{H}=y_\mathrm{He}=1$.}  We emphasize that in this expression, the free electron fraction $f_\mathrm{free}$ represents the fraction of the gas that is fully ionized --- for instance, this excludes stars. For galaxy clusters, we assume $\tau\ll1$ \citep[e.g.,][]{Battaglia_2016,Flender_2017} and $e^{-\tau}\approx1$ throughout the paper. In Eq.~\eqref{eq:ne} the homogeneous part of electron density is given by (assuming $\bar{\rho}_\mathrm{gas}=\bar{\rho}_\mathrm{b}$)
\begin{equation}
\bar{\rho}_\mathrm{e} = f_{\rm free} \bar{\rho}_\mathrm{b} = f_{\rm free} \Omega_\mathrm{b}\rho_\mathrm{crit,0}(1+z)^3=f_{\rm free}f_{\rm b}\rho_\mathrm{m,0}(1+z)^3
\end{equation}
where $f_\mathrm{b}=\Omega_\mathrm{b}/\Omega_\mathrm{m}$ is the baryon fraction.

Since the visibility function is proportional to the baryon density, inhomogeneities of the baryon density field are associated with inhomogeneities of the visibility function. At leading order, we have
\begin{equation}
 g=\bar{g}+\delta g \quad\mathrm{with}\quad \delta g=\frac{\bar{g}}{f_\mathrm{b}f_\mathrm{free}}\delta_\mathrm{e}\quad\mathrm{and}\quad\delta_\mathrm{e}=\frac{\delta\rho_\mathrm{e}}{\rho_\mathrm{m}},\label{eq:dg}
 \end{equation}
 where $\bar{g}$ is the homogeneous part of the visibility function computed with Eq.~\eqref{eq:g} using $ \bar{\rho}_\mathrm{e}$ in Eq.~\eqref{eq:ne}, and $\delta g$ is the fluctuation caused by inhomogeneities of the electron density. Note the here, we do not consider fluctuations caused by patchy reionization, as we focus on the LSS formation era at lower redshift.

In Eq.~\eqref{eq:dg} we wrote our definition of the electron density perturbation with respect to the mean matter density. With this convention, the power spectrum $P_{\delta_\mathrm{e}\delta_\mathrm{e}}$ vanishes for $\rho_\mathrm{e}\rightarrow 0$. Moreover, with $\rho_
\mathrm{e}=f_\mathrm{b}f_\mathrm{free}\rho_\mathrm{m}$, we have $P_{\delta_\mathrm{e}\delta_\mathrm{e}}\sim (f_\mathrm{b}f_\mathrm{free})^2P_L$ (on large scales). 

Plugging Eq.~\eqref{eq:dg} into Eq.~\eqref{eq:theta_ksz}, we can rewrite the temperature fluctuation as
 \begin{equation}
    \Theta^\mathrm{kSZ}({\mathbf{n}})=-\frac{1}{c}\int \mathrm{d}\chi \, \bar{g}(\chi){\mathbf{n}}\cdot\bm{\tilde{v}_\mathrm{e}}(\bm{x})\quad\mathrm{with}\quad\bm{\tilde{v}_\mathrm{e}}(\bm{x})=\bm{{v}_\mathrm{e}}(\bm{x})+\frac{1}{f_\mathrm{b}f_\mathrm{free}}\delta_\mathrm{e}(\bm{x})\bm{v_\mathrm{e}}(\bm{x}).\label{eq:theta_ksz_vg}
\end{equation}
In Fourier space, the velocity field $\bm{\tilde{v}_\mathrm{e}}$ is
\begin{equation}
\hat{\bm{\tilde{v}}}_{\bm{e}}(\bm{k}) = \hat{\bm{v}}_{\bm{\mathrm{e}}}(\bm{k})+\frac{1}{f_\mathrm{b}f_\mathrm{free}}\int\frac{\mathrm{d}^3\bm{k}^\prime}{(2\pi)^3} \hat{\delta}_\mathrm{e}(|\bm{k}-\bm{k}^\prime|)\hat{\bm{v}}_{\bm{\mathrm{e}}}(\bm{k}^\prime).\label{eq:total_v}
\end{equation}
The second term on the RHS is the velocity field modulated by the density perturbation. Note that since $\bm{v_\mathrm{e}}$ is a first-order quantity, this term is second order in perturbations. Let us introduce the longitudinal and transverse components of the velocity field in Fourier space, with respect to the mode $\bm{k}=k\bf{k}$:
\begin{equation}
\hat{\bm{\tilde{v}}}_{\bm{e}}=\hat{\bm{\tilde{v}}}_{\bm{e}}^{||}+\hat{\bm{\tilde{v}}}_{\bm{e}}^{\perp}\quad\mathrm{with}\quad\hat{\bm{\tilde{v}}}_{\bm{e}}^{||}=(\hat{\bm{\tilde{v}}}_{\bm{e}}\cdot{\bf{k}}){\bf{k}}\quad\mathrm{and}\quad\hat{\bm{\tilde{v}}}_{\bm{e}}^{\perp}=\hat{\bm{\tilde{v}}}_{\bm{e}}-(\hat{\bm{\tilde{v}}}_{\bm{e}}\cdot\bf{k})\bf{k}
\end{equation}
We note that in standard cosmological perturbation theory, the vortical components of the matter velocity field can be neglected, hence we can assume that $\bm{{v}_\mathrm{e}}$ is longitudinal, i.e., $\hat{\bm{{v}}}_{\bm{\mathrm{e}}}(\bm{k})=\hat{v}_{{\mathrm{e}}}(k)\bf{k}$ \citep[this is valid to all orders in perturbation, see, e.g.,][]{dodelson2020modern}. Then the longitudinal component is 
\begin{equation}
\hat{\tilde{v}}_{e}^{||}(k) = \hat{v}_{{\mathrm{e}}}(k)+\frac{1}{f_\mathrm{b}f_\mathrm{free}}\int\frac{\mathrm{d}^3\bm{k}^\prime}{(2\pi)^3} \hat{\delta}_\mathrm{e}(|\bm{k}-\bm{k}^\prime|) \hat{v}_{{\mathrm{e}}}(k^\prime)\mu, \quad\mathrm{with}\quad\mu=\bf{k}^\prime\cdot\bf{k}.\label{eq:vparallel}
\end{equation}
For the transverse component, we get
\begin{equation}
\hat{\bm{\tilde{v}}}_{\bm{e}}^{\perp} = \frac{1}{f_\mathrm{b}f_\mathrm{free}}\int \frac{\mathrm{d}^3\bm{k}^\prime}{(2\pi)^3} \hat{\delta}_\mathrm{e}(|\bm{k}-\bm{k}^\prime|) \hat{v}_{{\mathrm{e}}}(k^\prime)({\bf{k}}^\prime-\mu{\bf{k}}).
\end{equation}
Thus, at leading order, the total velocity field is purely longitudinal, with amplitude  (in linear theory):
\begin{equation}
\hat{v}_\mathrm{e}(k)=-iaHf\hat{\delta}_\mathrm{m}(k)/k \,\quad\mathrm{with}\quad f=\frac{\mathrm{d}\ln D}{\mathrm{d}\ln a} \label{eq:linvel}
\end{equation}
the growth rate and
where $D$ is the growth factor \citep[e.g.,][]{dodelson2020modern}.

Next, there is a second-order contribution of the form $\delta v$ (the last term in Eq.~\eqref{eq:vparallel}). The transverse component $\hat{\bm{\tilde{v}}}_{\bm{e}}^{\perp}$ is purely of the form $ \delta v$, i.e., a second-order quantity.
The longitudinal component $\hat{\bm{\tilde{v}}}_{\bm{e}}^{||}$ cannot contribute to the kSZ anisotropy at small scales, because of the cancellation of crests and troughs when projected along the line of sight \citep[e.g.,][]{Jaffe_1998}.\footnote{At large angular scales, the cancellation is not complete and both the linear Doppler term $\propto v_e$ (leading order) and the density-weighted term $\propto\delta_e v_e$  (next order) of $\hat{\bm{\tilde{v}}}_{\bm{e}}^{||}$ contribute to the temperature anisotropy. We refer to \cite{Alvarez_2016_ksz} for a derivation of the large-scale anisotropy from longitudinal modes during reionization.} Therefore, at leading order in perturbations, the kSZ effect vanishes for small inhomogenieties \citep{1984ApJ...282..374K}. Hence, the kSZ anisotropy, on small scales, is generated by the transverse, second-order term, $\hat{\bm{\tilde{v}}}_{\bm{e}}^{\perp}$. (Schematically, ${\bf{n}}\cdot\hat{\bm{\tilde{v}}}_{\bm{e}}\approx{\bf{n}}\cdot\hat{\bm{\tilde{v}}}_{\bm{e}}^{\perp}$ for small inhomogeneities.) Then, when $\delta_\mathrm{e}$ is in the linear regime, the kSZ anisotropy is called the Ostriker-Vishniac (OV) effect \citep[][]{1986ApJ...306L..51O,1987ApJ...322..597V}.  The large-scale kSZ anisotropy from the linear Doppler effect and the OV effect are important during reionization. They are not of direct interest here, since we focus on the anisotropy generated in collapsed regions of the density field (halos) where $\delta_\mathrm{e}$ is non-linear, i.e., \textit{the kSZ effect from large-scale structure}. Nonetheless, to understand how different terms and scales play out, even in the non-linear regime, it is instructive to review the main aspects of the OV power spectrum calculation --- see \cite{2000ApJ...529...12H,Ma:2001xr}, and \cite{Park_2013} or Appendix B of \cite{Alvarez_2016_ksz} for a thorough derivation. 

The key quantity to evaluate is the 3D power spectrum of the transverse component (i.e., the curl) of the density-modulated velocity field, $P_\perp$, defined via
\begin{equation}
\langle \hat{\bm{\tilde{v}}}_{\bm{e}}^{\perp}(\bm{k}) \hat{\bm{\tilde{v}}}_{\bm{e}}^{\perp} (\bm{k}^\prime)^*\rangle = (2\pi)^3 P_\perp(k,\chi)\delta_D(\bm{k}^\prime-\bm{k})\label{eq:p_perp_def}
\end{equation}
where $\delta_D$ is the Dirac delta function. Once this is known, the angular anisotropy power spectrum of the kSZ effect can be written as 
\begin{equation}
C_\ell^\mathrm{kSZ} = \frac{1}{2c^2}\int\mathrm{d}\chi\frac{\bar{g}^2}{\chi^2}P_\perp(k,\chi)\quad\mathrm{with}\quad k=(\ell+1/2)/\chi,
\end{equation}
where the pre-factor 1/2 comes from the projection along the LOS \citep[see, e.g.,][for the derivation of the pre-factor]{Jaffe_1998,Park_2013}. Schematically, $P_\perp$ has three terms since
\begin{equation}
\langle \delta v\delta v\rangle \sim \langle \delta \delta \rangle \langle v v\rangle + \langle \delta v\rangle\langle \delta v\rangle + \langle \delta v\delta v\rangle_c.
\end{equation}
The last term is the connected term,  expected to be subdominant \citep[see][]{Ma:2001xr,McQuinn_2005,Park__2016}, and is neglected hereafter. The first and second terms can be evaluated straightforwardly, yielding
\begin{equation}
P_\perp(k,\chi) = \frac{1}{(2\pi f_\mathrm{b}f_\mathrm{free})^2}\int \mathrm{d}k^\prime k^{\prime2}\int\mathrm{d}\mu(1-\mu^2)\left[P_{\delta_\mathrm{e}\delta_\mathrm{e}}(|\bm{k}-\bm{k}^\prime|)P_{v_\mathrm{e}v_\mathrm{e}}(k^\prime)-\frac{k^\prime}{|\bm{k}-\bm{k}^\prime|}P_{\delta_\mathrm{e}v_\mathrm{e}}(|\bm{k}-\bm{k}^\prime|)P_{\delta_\mathrm{e}v_\mathrm{e}}(k^\prime)\right]\label{eq:pperp}
\end{equation}
where we used the definitions
\begin{equation}
\langle\hat{\delta}_\mathrm{e}(\bm{k})\hat{\delta}_\mathrm{e}(\bm{k}^\prime)\rangle = (2\pi)^3P_{\delta_\mathrm{e}\delta_\mathrm{e}}\delta_D(\bm{k}+\bm{k}^\prime)\quad\mathrm{and}\quad\langle\hat{\delta}_\mathrm{e}(\bm{k})\hat{v}_\mathrm{e}(\bm{k}^\prime)\rangle = (2\pi)^3P_{\delta_\mathrm{e}v_\mathrm{e}}\delta_D(\bm{k}+\bm{k}^\prime).
\end{equation}
With the velocity from linear perturbation theory (Eq.~\eqref{eq:linvel}), we can further simplify Eq.~\eqref{eq:pperp}, since
\begin{equation}
P_{v_\mathrm{e}v_\mathrm{e}}(k) = (aHf/k)^2 P_{\delta_\mathrm{m}\delta_\mathrm{m}}(k)\quad\mathrm{and}\quad P_{\delta_\mathrm{e}v_\mathrm{e}} (k)= (aHf/k)P_{\delta_\mathrm{e}\delta_\mathrm{m}}(k) \,. \label{eq:pvv_lin}
\end{equation}
The final expression reads
\begin{equation}
P^{\mathrm{OV}}_\perp(k,\chi) =\frac{(aHf)^2}{(2\pi f_\mathrm{b}f_\mathrm{free})^2}\int \mathrm{d}k^\prime\int\mathrm{d}\mu(1-\mu^2)P_{\delta_\mathrm{e}\delta_\mathrm{e}}(|\bm{k}-\bm{k}^\prime|)P_{\delta_\mathrm{e}\delta_\mathrm{e}} (k^\prime)\frac{k(k-2k^\prime\mu)}{k^2+k^{\prime2}-2kk^\prime\mu} \,,
\end{equation}
which is the well-known OV approximation of Eq.~\eqref{eq:pperp}. 

Let us now turn to the non-linear regime. First, we define the volume-averaged velocity dispersion
\begin{equation}
         \sigma_v^2(\chi)\equiv \frac{v_\mathrm{rms}^2(\chi)}{3c^2}\quad\mathrm{with}\quad v_\mathrm{rms}^2(\chi) = \frac{1}{2\pi^{2}}\int\mathrm{d}k \, k^2P_{v_\mathrm{e}v_\mathrm{e}}\left(k,\chi\right).
         \label{eq:vrms2}
\end{equation}
  In the high-$k$ (non-linear) regime, we can drop terms of $\mathcal{O}(k^\prime/k)$ in Eq.~\eqref{eq:pperp}, including $P_{\delta_\mathrm{e}v_\mathrm{e}}$ \cite[e.g.,][]{2000ApJ...529...12H,Ma:2001xr,McQuinn_2005}. Hence, $P_{\delta_\mathrm{e}\delta_\mathrm{e}}(|\bm{k}-\bm{k}^\prime|)\approx P_{\delta_\mathrm{e}\delta_\mathrm{e}}(k)$ and the integral over $\mu$ can be carried out (yielding a factor $4/3$). One finds
\begin{equation}
P^{\mathrm{high-}k}_\perp(k,\chi) =\frac{2}{3}\frac{(v_\mathrm{rms}/c)^2}{(f_\mathrm{b}f_\mathrm{free})^2}P_{\delta_\mathrm{e}\delta_\mathrm{e}}(k,\chi)=\frac{2\sigma_v^2}{(f_\mathrm{b}f_\mathrm{free})^2}P_{\delta_\mathrm{e}\delta_\mathrm{e}}(k,\chi)\label{eq:p_perp_highk}
\end{equation}
so that in this regime, the angular anisotropy power spectrum reads
\begin{equation}
C_\ell^{\mathrm{kSZ,\,high-}k} =\int \mathrm{d}\mathrm{v}  W^\mathrm{kSZ}(\chi)^2P_{\delta_\mathrm{e}\delta_\mathrm{e}}\left(k,\chi\right)\quad\mathrm{with}\quad W^\mathrm{kSZ}(\chi)=\frac{a\sigma_\mathrm{T}\bar{\rho}_\mathrm{m}\sigma_v}{m_\mu \mu_e \chi^2}.\label{eq:clksz2}
\end{equation}
In the equation above, the $(f_\mathrm{b}f_\mathrm{free})^2$ factor of Eq.~\eqref{eq:p_perp_highk} is canceled by the same factor that appears in $\bar{g}$.
In order to evaluate the velocity dispersion $\sigma_v$, we can use the linear theory ansatz for $P_{v_\mathrm{e}v_\mathrm{e}}$ (see Eq.~\eqref{eq:pvv_lin}) and assume $P_{\delta_\mathrm{e}\delta_\mathrm{e}}\approx (f_b f_\mathrm{free})^2P_{\delta_\mathrm{m}\delta_\mathrm{m}}$, where $P_{\delta_\mathrm{m}\delta_\mathrm{m}}$ is the matter power spectrum. Here, the matter power spectrum can either be the linear matter power spectrum $P_L$ or the non-linear matter power spectrum $P_{NL}$ from \verb|halofit| \citep{Smith_2003,Takahashi_2012} or \verb|hmcode| \citep{mead2015,mead2021}. See Appendix \ref{s:veldisp} and Figure~\ref{fig:vrms2} for a comparison of these different choices. Meanwhile, the electron power spectrum  $P_{\delta_\mathrm{e}\delta_\mathrm{e}}$ that appears explicitly in Eq.~\eqref{eq:clksz2} is computed using the halo model (see Appendix \ref{sss:baryons}). 

\subsection{Projected-field kSZ Estimator}\label{sec:pf_estimator}
Consider a large-scale structure tracer $X$ that does not depend linearly on the velocity field. For instance, in this paper, we shall work with the galaxy number density $X=\delta_\mathrm{g}$, galaxy weak lensing convergence $X=\kappa_\mathrm{g}$, and CMB lensing convergence $X=\kappa_\mathrm{cmb}$. Since the kSZ anisotropy is proportional to the bulk velocity of the electrons, the cross-correlation $\langle X \Theta^\mathrm{kSZ}\rangle$ vanishes (at first order) simply due to sign cancellation because of the isotropy of the velocity field: collapsed regions are as likely to be in a ``positive'' or ``negative'' bulk motion with respect to the LOS.

To probe such correlations, we can resort to higher-order statistics \citep[e.g.,][]{Cooray:2001wa}. \cite{Dore2004} proposed to work with a three-point statistic: two kSZ points, \textit{so that the statistic is even in the velocity}, and one point in $X$. They introduced a collapsed three-point function, that is, the cross-correlation between the squared kSZ field and $X$. This condenses the information of the full bispectrum into an angular power spectrum that is arguably more easily measurable, the so-called \textit{projected-field kSZ power spectrum} denoted $C_{\ell}^{\mathrm{kSZ^2}X}$.

\begin{figure}
    \includegraphics[width=1.\columnwidth]{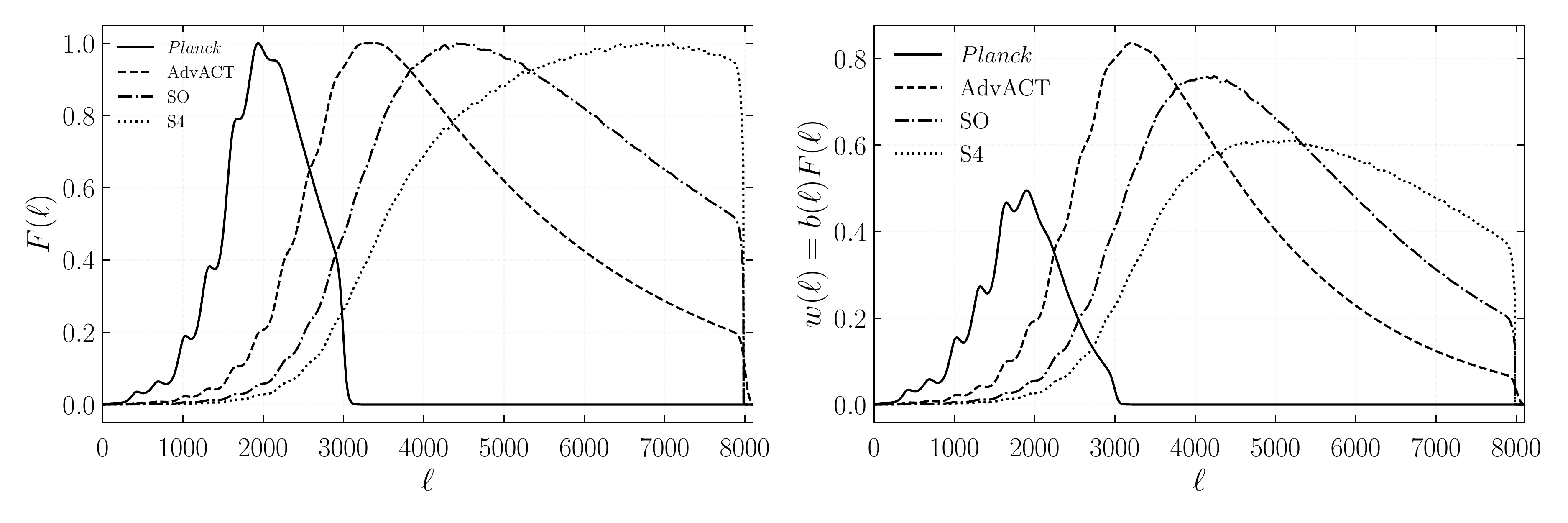}
    \vspace{-0.6cm}
     \caption{(\textit{Left:}) The Wiener filter of Eq.~\eqref{eq:filter}. (\textit{Right:}) The product of the filter and beam (see Eq.~\ref{eq:bwf}). We use the specifications of Table \ref{tab:exps} (see also Section~\ref{s:forecasts}) and our fiducial model to compute these filters. Except for \textit{Planck} that has $\ell_\mathrm{max}=3000$, we use $\ell_\mathrm{max}=8000$ for all other filters. (See footnote \ref{fn:taper} for the high-$\ell$ taper.)}
    \label{fig:filter_figure}
\end{figure}

In practice, the projected-field kSZ power spectrum is measured by squaring a CMB temperature map in real space before cross-correlating it with a map of tracer $X$ in harmonic space. Furthermore,  in order to maximize the contribution from arcminute scales where the kSZ effect is more significant, the temperature map is Wiener-filtered in harmonic space beforehand: 
\begin{equation}
    \Theta_f(\bm{\ell})\equiv  \Theta(\bm{\ell})\quad\mathrm{with}\quad w(\ell)=b(\ell) F(\ell)\quad\mathrm{where}\quad F(\ell) \equiv \frac{\sqrt{C_{\ell}^{\mathrm{kSZ}}}/C_{\ell}^\mathrm{tot}}{\mathrm{max}(\sqrt{C_{\ell}^{\mathrm{kSZ}}}/C_{\ell}^\mathrm{tot})}\label{eq:filter}
\end{equation}
is our ansatz for the  optimal filter.\footnote{Note that in practice we also use a taper in the filter definition, i.e., we multiply by  $\mathcal{T}(\ell,\ell_\mathrm{max},\Delta\ell)=\frac{1}{2}\left[1-\tanh\left(\frac{\ell-\ell_\mathrm{max}}{\Delta \ell}\right)\right]$
to regularize the transition to zero (we use $\Delta\ell=50$ throughout, see Table~\ref{tab:exps} for the $\ell_\mathrm{max}$ values used in the filters).\label{fn:taper}} This filter choice differs from previous works \citep[e.g.][]{Hill2016,Ferraro2016,Kusiak:2021hai} which did not use a square root --- we explain this difference in Appendix~\ref{app:filters}, justifying why it is more optimal. In Eq.~\eqref{eq:filter}, $\Theta_f$ is the temperature anisotropy of the filtered  CMB map, $\Theta$ is the temperature anisotropy of the unfiltered CMB map, $C_{\ell}^{\mathrm{kSZ}}$ is the power spectrum of the kSZ effect which we compute according to Eq.~\eqref{eq:clksz2} within our fiducial model, and
$C_{\ell}^\mathrm{tot}$ is the angular power spectrum of the map, which in principle contains contributions from the lensed primary CMB, ISW, the kSZ anisotropy (from both reionization and galaxy clusters), as well as all residual foregrounds (e.g., tSZ and CIB), and instrumental and atmospheric noise.  In this paper, we compute $C_{\ell}^\mathrm{tot}$ as a sum of the lensed CMB power spectrum from \verb|class|/\verb|class_sz| (with our fiducial parameters), the kSZ power spectrum, and the noise curves for a given CMB experiment, e.g., \textit{Planck}, AdvACT, SO, or CMB-S4, namely:
\begin{equation}
C_{\ell}^\mathrm{tot} = C_\ell^{\Theta\Theta,\mathrm{lensed}}+C_\ell^{\mathrm{kSZ}}+N_\ell^{\Theta\Theta}.
\end{equation}

In filtering operation, to mitigate divergences due to large noise at small scales,  we also include a Gaussian beam window function:
\begin{equation}
    b(\ell)=\mathrm{exp}\left\{-\tfrac{1}{2}\ell(\ell+1)\frac{\theta_\mathrm{FWHM}^2}{8\ln 2}\right\}\label{eq:bwf}
\end{equation}
where $\theta_\mathrm{FWHM}$ is the full width at half maximum (FWHM) of the telescope's beam in radians (in terms of the analysis, it is equivalent to leaving the beam in the observed CMB temperature map when computing the power spectra). The experimental configurations considered in this analysis are specified in Table~\ref{tab:exps}.

Then, in harmonic space (where the real space product becomes a convolution) the projected-field power spectrum estimator dubbed  $\bar{C}_{\ell}^{\mathrm{kSZ^2}X}$ is defined via
\begin{equation}
    \langle  \Theta_f^2(\bm{\ell}^\prime)  X(\bm{\ell})\rangle=\int\frac{\mathrm{d}^2\bm{\ell}^{\prime\prime}}{(2\pi)^2}\langle\Theta_f(\bm{\ell}^{\prime\prime})\Theta_f(\bm{\ell}^{\prime}-\bm{\ell}^{\prime\prime}) X(\bm{\ell})\rangle= (2\pi)^2\delta_D(\bm{\ell}+\bm{\ell}^\prime)\bar{C}_{\ell}^{\mathrm{kSZ^2}X}.\label{eq:cldef}
\end{equation}

Let us now make the connection between the projected-field kSZ power spectrum and the kSZ anisotropy. As we saw in the previous subsection,  the kSZ anisotropy is sourced by the velocity field $\bm{v_e}$ projected along the LOS. Moreover, we saw that only transverse modes contribute to the kSZ effect in the high-$k$ regime. 
Thus, it is natural to introduce the three-point function
\begin{equation}
    \langle \hat{\bm{\tilde{v}}}_{\bm{e}}^{\perp}(\bm{k}) \hat{\bm{\tilde{v}}}_{\bm{e}}^{\perp} (\bm{k}^\prime) \hat{X}(\bm{k}^{\prime\prime})\rangle=(2\pi)^3\delta_D(\bm{k}+\bm{k}^\prime+\bm{k}^{\prime\prime})\mathcal{C}_{v_e^2X}(\bm{k},\bm{k}^\prime,\bm{k}^{\prime\prime}),\label{eq:c_def}
\end{equation}
where $\mathcal{C}_{v_e^2X}$ is the analogue of $P_\perp$ (see Eq.~\eqref{eq:p_perp_def}). Since $\hat{\bm{\tilde{v}}}_{\bm{e}}^{\perp}\sim\delta v$, it is a contraction of the five-point function that can be schematically written as 
\begin{equation}
\langle \delta v\delta v \delta_X\rangle \sim \langle \delta \delta \rangle \langle v v \delta_X\rangle + \langle \delta v\rangle\langle \delta v\delta_X\rangle + \langle  vv\rangle\langle\delta \delta \delta_X\rangle
+\dotsb+
\langle \delta v\delta v\delta_X\rangle_c,
\label{eq:5ptfunction}
\end{equation}
where the last term is the connected term and where $\delta_X$ denotes the perturbation for the tracer $X$. \cite{Dore2004} only took into account the third term in this expansion. \cite{DeDeo} studied the relative importance of each of these terms (see their Figure 1), following a similar approach as \cite{Ma:2001xr} did for the power spectrum, and concluded that the \cite{Dore2004} approach was a good approximation.  Its validity was further confirmed by comparison to numerical simulations in \cite{Hill2016} and \cite{Ferraro2016}.  Here, we shall work with the same approximation, namely, $\mathcal{C}_{v_e^2X}\approx  \langle  vv\rangle\langle\delta \delta \delta_X\rangle$, so that (as in Eq.~\eqref{eq:p_perp_highk}):
\begin{equation}
\mathcal{C}_{v_e^2X}\approx\frac{2\sigma_v^2}{(f_\mathrm{b} f_\mathrm{free})^2}B_{\delta_\mathrm{e}\delta_\mathrm{e}X}\label{eq:c_approx}
\end{equation}
 where $\sigma_v^2$ is the velocity dispersion defined in Eq.~\eqref{eq:vrms2} and $B_{\delta_\mathrm{e}\delta_\mathrm{e}X}$ is the \textit{hybrid bispectrum} of baryon density perturbations and $X$ \citep{Dore2004}. Assuming temporarily  $\Theta=\Theta^\mathrm{kSZ}$ and combining Eqs.~\eqref{eq:c_approx}, \eqref{eq:c_def}, and \eqref{eq:cldef}, we can write the projected-field kSZ power spectrum as
\begin{equation}
    C_{\ell}^{\mathrm{kSZ}^2 X} = \int\mathrm{dv}W^\mathrm{kSZ}(\chi)^2W^X(\chi)T(\ell,\chi)\quad\mathrm{with}\quad T(\ell,\chi)=\int \frac{\mathrm{d}^2\bm{\ell}^{\prime}}{(2\pi)^2}w(\ell^\prime)w(\left|\bm{\ell}+\bm{\ell}^{\prime}\right|)B_{\delta_\mathrm{e}\delta_\mathrm{e}X}(\mathbf{k}_1,\mathbf{k}_2,\mathbf{k}_3)\label{eq:clpf}
\end{equation}
with wavenumbers $\mathbf{k}_1\chi = \bm{\ell}^{\prime}$, $\mathbf{k}_2\chi=-(\bm{\ell}+\bm{\ell}^{\prime})$, $\mathbf{k}_3\chi = \bm{\ell}$ such that $\mathbf{k}_1+\mathbf{k}_2+\mathbf{k}_3=\vec{0}$, where $W^X(\chi)$ is the projection kernel of the tracer $X$, and $W^\mathrm{kSZ}(\chi)$ is the projection kernel of the kSZ effect defined in Eq.~\eqref{eq:clksz2}. Here, $T(\ell,\chi)$ is the so-called \textit{triangle power spectrum}: it is a sum over all triangle configurations with sides $(\bm{\ell},\bm{\ell}^{\prime},-\bm{\ell}-\bm{\ell}^{\prime})$ in planes perpendicular to the LOS, at constant redshift \citep{Dore2004}. 

The remaining task is to find an expression for the hybrid bispectrum. With this aim, in the next Subsection we revisit the approach that was adopted in previous works \citep[][]{Dore2004,DeDeo,Hill2016,Ferraro2016,Kusiak:2021hai}, and in Section~\ref{s:hm} we present our newly developed halo model.

\begin{table}
\begin{centering}
\begin{tabular}{c|cccc}
 & $\Theta_\mathrm{FWHM}$ & $\Delta_T$ & \tabularnewline
 & (arcmin) & ($\mu$K-arcmin) & $\ell_\mathrm{max}$ & $f_\mathrm{sky}$
\tabularnewline
\hline 
\textit{Planck} & $5$ & $47$ & $3000$ & 0.6 \tabularnewline
AdvACT & $1.5$ & $20$ & $8000$ & 0.3 \tabularnewline
SO and CMB-S4 & $1.4$ & ILC & $8000$ 
& 0.4 \tabularnewline
\end{tabular}
\par\end{centering}
\caption{Specifications of the CMB experiments (\textit{Planck}, AdvACT, SO, and CMB-S4) used in the forecast analysis. $\Theta_\mathrm{FWHM}$ denotes the full width at half maximum (FWHM) of the telescope beam, $\ell_\mathrm{max}$ is the multipole that determines the transition to zero of the Wiener filter (see Eq.~\ref{eq:filter}), and $f_\mathrm{sky}$ denotes the sky fraction after relevant masks have been applied to the surveyed CMB sky area. See Section~\ref{s:forecasts} for further details.}
\label{tab:exps}
\end{table}

\subsection{Effective Approach}\label{ss:eff}
The task of finding an expression for the hybrid bispectrum $B_{\delta_\mathrm{e}\delta_\mathrm{e}X}$ can be simplified significantly under some assumptions. First, if the baryons are assumed to perfectly trace the dark matter density field, we can write $\delta_\mathrm{e}=f_\mathrm{b}f_\mathrm{free}\delta_\mathrm{m}$. Second, if the tracer $X$ is also a tracer of the density field, we have $\delta_X=b_X\delta_\mathrm{m}$, where $b_X$ is the bias of $X$. For instance, for the galaxy density field $b_X$ is the galaxy bias, while for CMB or galaxy weak lensing $b_X=1$. With these two assumptions, the hybrid bispectrum reduces to $B_{\delta_\mathrm{e}\delta_\mathrm{e}X}=b_X (f_\mathrm{b}f_\mathrm{free})^2B_\mathrm{m}$, where $B_\mathrm{m}$ is the matter bispectrum. 

In analogy with the calculation of the kSZ power spectrum in the high-$k$ regime, where the baryon density power spectrum can by approximated with the non-linear matter power spectrum calibrated on $N$-body simulations, \cite{Dore2004} and \cite{DeDeo} used a non-linear matter bispectrum fitting formula from \cite{Scoccimarro:2000ee}. Similarly, \cite{Hill2016,Ferraro2016,Kusiak:2021hai} used the fit from \cite{Gil_Mar_n_2012}, which improves upon \cite{Scoccimarro:2000ee} while keeping the same functional form. These bispectrum fitting functions are based on the expression of the tree-level matter bispectrum in Eulerian perturbation theory, for an Einstein de-Sitter Universe \citep{1984ApJ...279..499F}
\begin{equation}
    B_\mathrm{TL}(k_1,k_2,k_3) = 2F_2(k_1,k_2,k_3)P_{L}(k_1)P_{L}(k_2)+2\,\mathrm{cyc.}\label{eq:bt}
\end{equation}
where we did not write explicitly the permutations between modes and where the  $F_2$ kernel is given by \citep{1984ApJ...279..499F,Goroff:1986ep}\footnote{See Chapter 12 of \cite{dodelson2020modern} for a presentation of second-order cosmological perturbation theory and \cite{bernardeau2002} for further details.}
\begin{equation}
    F_2(k_1,k_2,k_3)=F_2(\mathbf{k}_1,\mathbf{k}_2)=\frac{5}{7}+\frac{1}{2}\cos\theta_{12}\left(\frac{k_1}{k_2}+\frac{k_2}{k_1}\right)+\frac{2}{7}(\cos\theta_{12})^2\quad\mathrm{with}\quad \cos\theta_{12}=\frac{\mathbf{k}_1\cdot\mathbf{k}_2}{k_1 k_2}=\frac{k_3^2-k_2^2-k_1^2}{2k_1 k_2}\label{eq:f2s}.
\end{equation}
(This is the expression as implemented in \verb|class_sz|, which takes the three wavenumber moduli as an input.) The \cite{Gil_Mar_n_2012} bispectrum fitting formula has the same form as Eq.~\eqref{eq:bt}, except that the linear matter power spectrum is replaced by its non-linear counterpart and that it includes extra scale- and redshift-dependent coefficients in front of the terms in the expression of the $F_2$ kernel. There are nine parameters that control the scale and redshift dependence of those coefficients, whose values are found by fitting the data from $N$-body simulations. Thus, the non-linear matter bispectrum is written as
 \begin{equation}
    B_{\mathrm{eff}}^{NL}(k_1,k_2,k_3) = 2F_2^\mathrm{eff}(k_1,k_2,k_3)P_{NL}(k_1)P_{NL}(k_2)+2\,\mathrm{cyc}.\label{eq:b_gilma}
\end{equation}
See Eqs.~2.6-2.12 of \cite{Gil_Mar_n_2012} for the expression of the effective kernel $F_2^\mathrm{eff}$, as well as the values of the fitting parameters.  

\begin{figure}
    \includegraphics[width=1.\columnwidth]{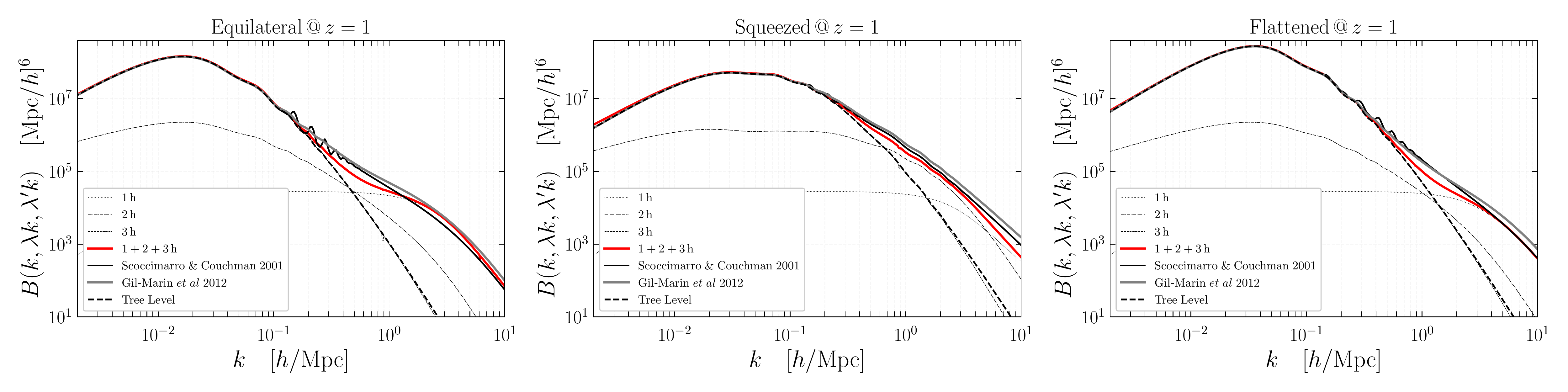}
    \includegraphics[width=1.\columnwidth]{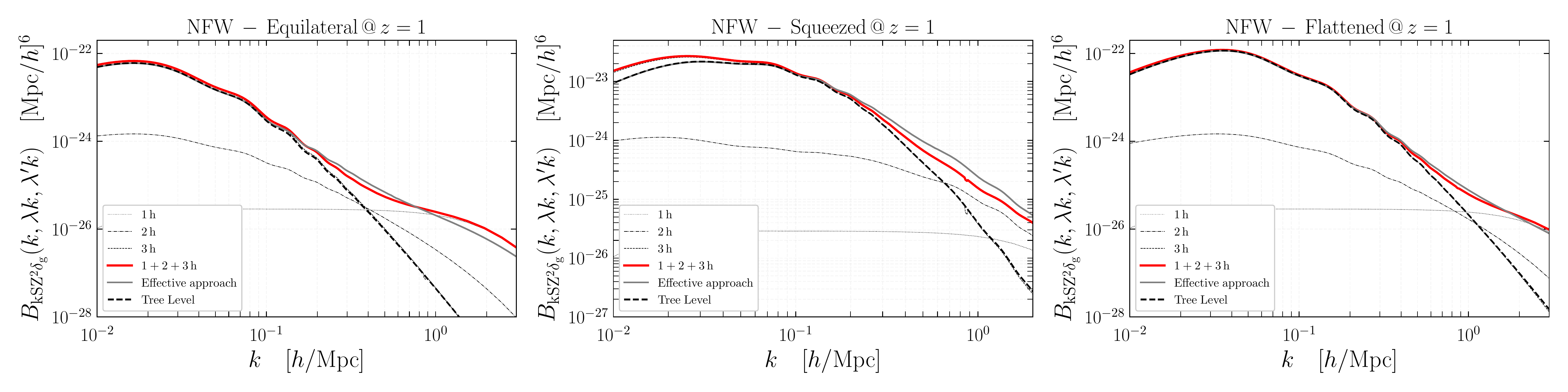}
    \includegraphics[width=1.\columnwidth]{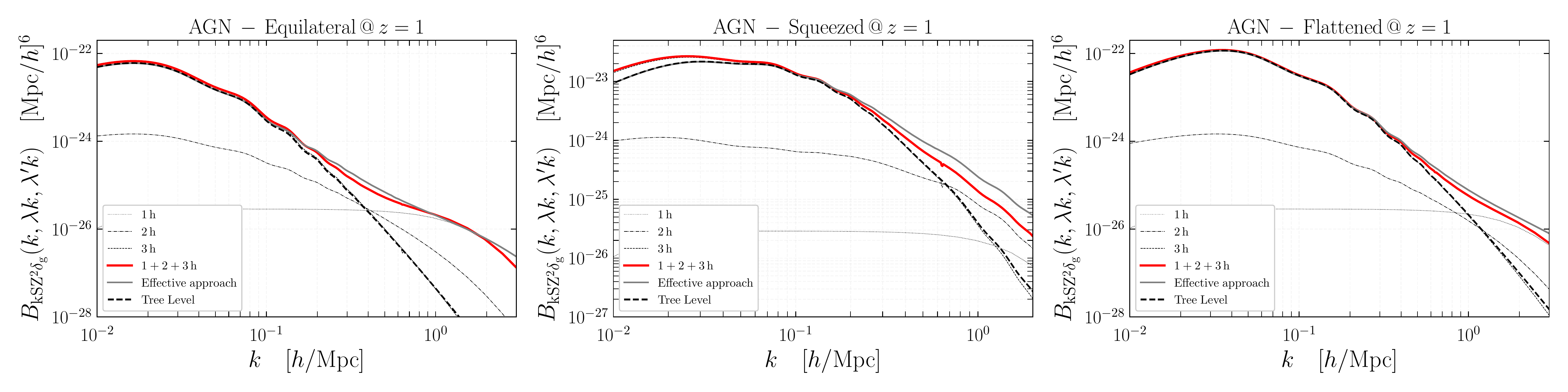}
    \includegraphics[width=1.\columnwidth]{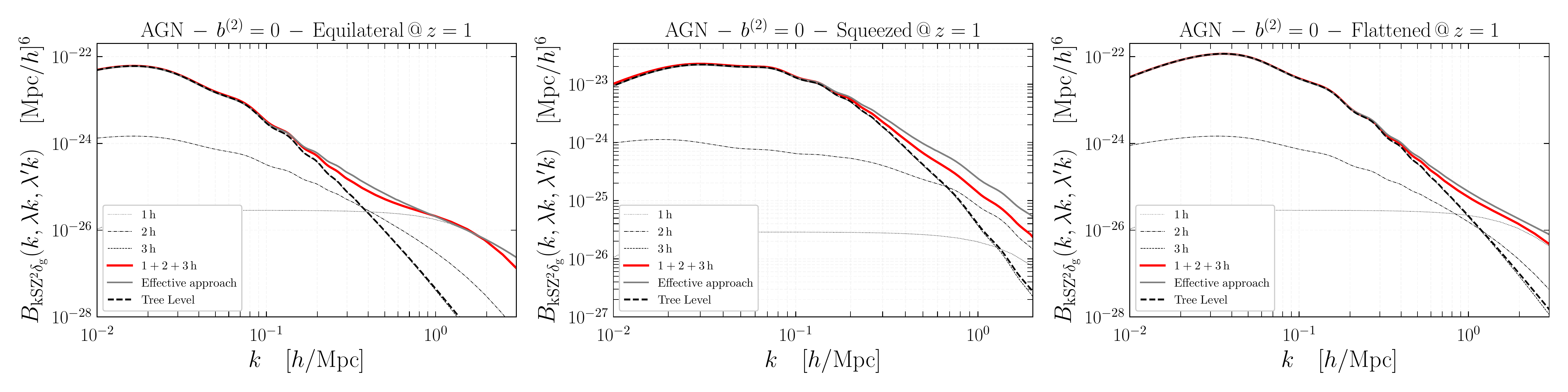}
    \caption{
    Matter bispectrum (top row) and hybrid bispectrum (cross-correlation with galaxy number density, bottom three rows) for \textit{Equilateral} ($\lambda=\lambda^\prime=1$), \textit{Squeezed} ($\lambda=1,\lambda^\prime=0.1$) and \textit{Flattened} ($\lambda=\lambda^\prime=0.5$) configurations, from left to right, at $z=1$. The halo model contributions are shown as the thin solid (1h), dashed (2h), and dotted (3h) lines, and their sum is shown as the red, thick line. We use our fiducial halo model and cosmology settings (see Subsection~\ref{s:fid}). The fitting formula predictions for the matter bispectrum \cite[][]{Gil_Mar_n_2012} and the corresponding  \textit{effective approach}  hybrid bispectrum predictions are the thick grey lines. The second and third rows show the hybrid bispectrum when the gas follows the \textit{NFW} or the \textit{AGN Feedback} model, respectively. The fourth row is with the \textit{AGN Feedback} model, where the second order bias is artificially set to zero. The galaxy HOD parameters are given below Eq.~\eqref{eq:nsat}.    }  
    \label{fig:bks}
\end{figure}

The non-linear matter bispectrum computed with the \cite{Gil_Mar_n_2012} and \cite{Scoccimarro:2000ee} fitting formulas are plotted in the top panels of Figure~\ref{fig:bks}. 
The main difference between these formulas is that the \cite{Gil_Mar_n_2012} formula corrects the unphysical oscillations associated with the BAOs in the power spectrum which are visible in the \cite{Scoccimarro:2000ee} prediction \citep[see, e.g.,][for details]{Gil_Mar_n_2012}. On large scales, both formulas match with the tree-level bispectrum of Eq.~\eqref{eq:bt}. Note that the \cite{Gil_Mar_n_2012} formula is calibrated on a fairly restricted $k$- and $z$-range compared to what may be needed for kSZ applications, namely:  $0.03\,h/\mathrm{Mpc}\leq k \leq 0.4 h/\mathrm{Mpc}$, and $0\leq z\leq 1.5$. Recent  matter bispectrum fitting formulas have been derived by \cite{Takahashi_2020} on a broader $k$- and $z$-range ($k\lesssim 3 h/\mathrm{Mpc}$ and $z<3$) but we do not discuss them here as they have not been used in the context of the kSZ effect.

In what we shall refer to as the \textit{effective approach projected-field kSZ power spectrum} hereafter, the hybrid bispectrum  in Eq.~\eqref{eq:clpf} is approximated by $b_X f_\mathrm{b}^2f_\mathrm{free}^2 B_{\mathrm{eff}}^{NL}$ of Eq.~\eqref{eq:b_gilma}. While the numerical evaluation of Eq.~\eqref{eq:b_gilma} is fast and straightforward, the computation of $C_{\ell}^{\mathrm{kSZ}^2 X}$ is more involved and takes several minutes on a laptop.  In \verb|class_sz|, the computation is parallelized with respect to multipoles $\ell$. Then, at each $\ell$ we tabulate the redshift integrals 
\begin{equation}
    I_\ell(\ell^\prime,\theta)=\int \mathrm{dv} \, W^\mathrm{kSZ}(\chi)^2W^X(\chi)b_X f_\mathrm{b}^2f_\mathrm{free}^2 B_{\mathrm{eff}}^{NL}(\ell^\prime/\chi,\left|\bm{\ell}+\bm{\ell}^\prime\right|/\chi,\ell/\chi)
\end{equation}
on a 2D grid spanned by $\ln \ell^\prime$ and the polar angle $\theta=(\hat{\bm{\ell},\bm{\ell}^\prime})$. Note that the dependence on $\theta$  arises from the modulus  $\left|\bm{\ell}+\bm{\ell}^\prime\right|$ that appears in the bispectrum (since $\left|\bm{\ell}+\bm{\ell}^\prime\right|^2=\ell^2+\ell^{\prime2}+2\ell\ell^\prime \cos\theta$). Then we integrate over $\ln \ell^\prime$ at fixed $\theta$, and eventually over $\theta$, i.e.,
\begin{equation}
    C_{\ell}^{\mathrm{kSZ}^2 X} =\frac{1}{2\pi^2}\int _0^{\pi} \mathrm{d}\theta  \int \mathrm{d}\ln \ell^\prime \ell^{\prime2} w(\ell^\prime)w(\left|\bm{\ell}+\bm{\ell}^{\prime}\right|)I_\ell(\ell^\prime,\theta),\label{eq:pfexpl}
\end{equation}
where we used the fact that $I_\ell(\ell^\prime,\theta)=I_\ell(\ell^\prime,-\theta)$ to integrate only over half the polar plane. Note that in principle, the computation could be made faster  using Fast Fourier Transform methods as in Section~\ref{s:hm}. See also Subsection 3.3 of \cite{Dore2004} for an approximation that requires only the radial integration, similar to the high-$k$ approximation of the kSZ auto power spectrum (see below Eq.~\eqref{eq:vrms2}).

We validate the \verb|class_sz| implementation of the effective approach against the code used in \cite{Hill2016,Ferraro2016,Kusiak:2021hai} by comparing the predictions of both codes, in two cases. For both cases  (\textit{Planck} $\times$ \textit{WISE} and AdvACT $\times$ \textit{WISE}), the codes agree within $\approx 10\%$ (see Appendix \ref{ap:comp_ferraro} and Figure \ref{fig:cls_comp_snr}). For the experimental configurations and cross-correlations of interest here,  we show the effective approach projected-field kSZ power spectrum in Figures~\ref{fig:fls_cls_ttg}, \ref{fig:fls_cls_ttkg}, and \ref{fig:fls_cls_ttkc} as the thick dashed lines.

The main limitations of the effective approach are its range of validity and its range of applications. It is well established that the gas in the ICM and CGM does not follow the underlying dark matter density profile \citep[e.g.,][]{schaan_2020_combined}, because of energetic feedback mechanisms or clusters' formation history. Therefore, the assumption $\delta_\mathrm{e}=f_\mathrm{b}f_\mathrm{free}\delta_\mathrm{m}$ is not a valid assumption in the high-$k$ regime, within halos. In addition, due to that same assumption, the effective approach does not enable us to probe the scale dependence of the gas density profile. By relaxing this assumption and using physical models for the gas distribution, the halo model allows us to overcome these limitations. 

\begin{figure}
    \includegraphics[width=0.95\columnwidth]{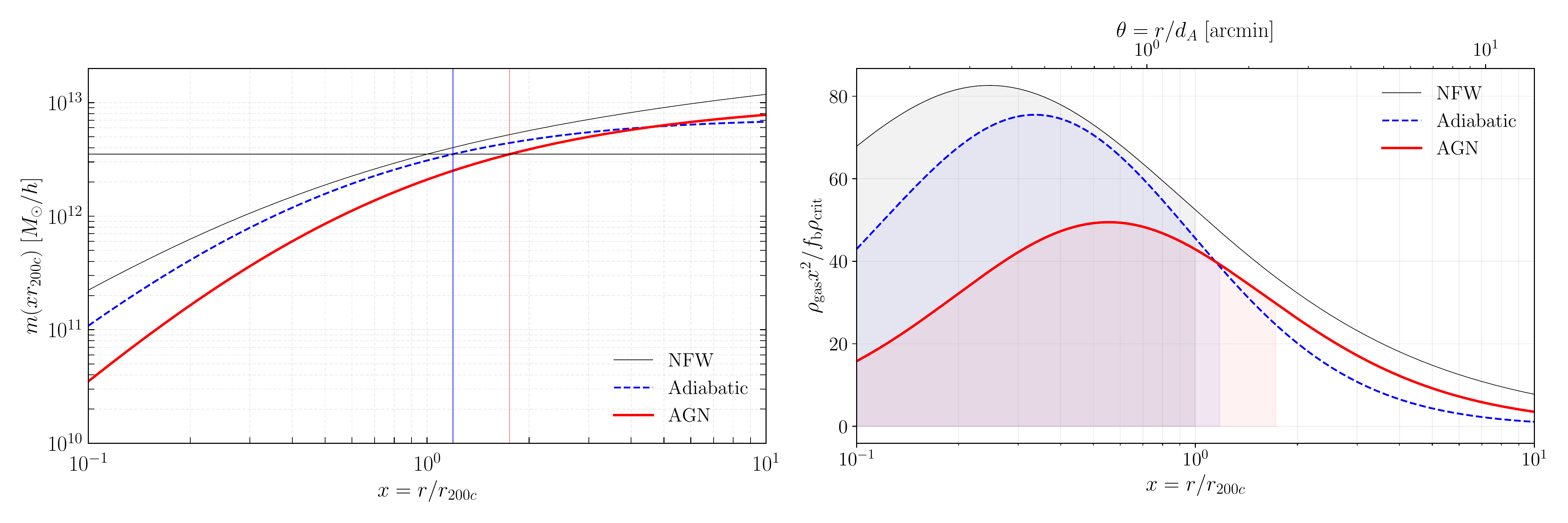}
    \vspace{-0.4cm}
   \caption{\textit{Left:} The mass of the gas (baryons) enclosed within a radius $r=x r_\mathrm{200c}$ for three different profile shapes. \textit{NFW} is the standard NFW formula, while \textit{Adiabatic} and \textit{AGN} are the models of Table \ref{tab:gnfwb16}. The halo has a mass  $m_\mathrm{200c}=3.3\times 10 ^{13} \, M_\odot$ and is at $z=0.55$.  The vertical lines indicate the truncation radius for the \textit{Adiabatic} and \textit{AGN} profiles. They are such that the enclosed mass is $m_\mathrm{200c}$.   The horizontal line indicates the gas mass of the halo, namely $f_\mathrm{b}\times3.3\,M_\odot\simeq 3.5\times 10^{12}\,M_\odot/h$.  \textit{Right:} Same as left but showing the scaled gas density. The critical density of the denominator of the $y$-axis label is evaluated at redshift $z$. The shaded area under the curves indicate the region below the truncation radius (vertical lines in the left plot). Other cosmological and halo model settings are set to our fiducial model (see Section \ref{s:fid}).}      
    \label{fig:gasprof}
\end{figure}

\section{Halo Model}\label{s:hm}
In this Section we describe the halo model approach. In Subsection \ref{ss:hybrid_bispec}  we use the halo model \citep[e.g.,][]{1991ApJ...381..349S,Mo_1996,Scoccimarro:2000gm,Seljak:2000gq,Cooray:2002dia} to estimate the hybrid bispectrum $B_{\delta_\mathrm{e}\delta_\mathrm{e}X}$ (Eq.~\eqref{eq:clpf}) and describe our two gas density profile parameterization choices. In Subsection \ref{ss:hmeqs} we explain how to compute the projected-field kSZ power spectrum with the halo model hybrid bispectrum.

\subsection{The Halo Model Code: \texttt{class\_sz}}\label{s:class_sz}

Along with this paper, we release version \verb|v1| of the halo model code \verb|class_sz|.\footnote{\href{https://github.com/borisbolliet/class_sz/releases/tag/v1.0.0}{https://github.com/borisbolliet/class\_sz/releases/tag/v1.0.0}} The code is written in C and was originally developed by \cite{Bolliet:2017lha}, specifically for the calculation of the thermal SZ power spectrum within the halo model and based on the Fortran code \verb|szfastdks| \citep{Komatsu:2002wc,Dolag_2016}. We have extended it to enable the calculation of  power spectra and bispectra of most LSS tracers, including galaxy number density, galaxy weak lensing, CMB weak lensing, the cosmic infrared background (CIB), and the kSZ effect. 

Version \verb|v1| of \verb|class_sz|  is built onto version \verb|v2.9.4| of the \verb|class| code\footnote{\href{https://github.com/lesgourg/class_public/releases/tag/v2.9.4}{https://github.com/lesgourg/class\_public/releases/tag/v2.9.4}} \citep{lesgourgues2011cosmic,blas2011}. This approach has two main advantages. First, with \verb|class_sz|, one can compute all the quantities available in \verb|class| such as the CMB temperature and polarization anisotropy power spectra or the matter power spectrum. Second, this makes most of the halo model quantities in \verb|class_sz| computable within all the cosmological models implemented in \verb|class|.
The  \verb|class_sz| code performs fast and accurate evaluations of the redshift and mass integrals (see Eq.~\ref{eq:massint} and \ref{eq:zint} hereafter) in the halo model using an adaptive Patterson scheme \citep{Patterson_1968} imported from  \verb|CosmoTherm| \citep{ct2011}. 
Where possible, the code's outputs have been checked with other halo-model codes, including \verb|ccl|  \citep{LSSTDarkEnergyScience:2018yem}, \verb|yxg| \citep{Koukoufilippas:2019ilu}, \verb|hmvec| \citep{Smith:2018bpn}, and \verb|HaloGen| \citep{Schaan_2018}.  Compared to other halo model codes, \verb|class_sz| has the unique property of full integration with \verb|class| and follows the same computational strategy as \verb|class| for the halo model quantities. 

The \verb|class_sz| code computes power spectra $P(k)$, bispectra $B(k,k^\prime,k^{\prime\prime})$, and angular power spectra $C_\ell$ within the Limber approximation (see Eq.~\ref{eq:limber}) on the flat sky.  These calculations are parallelized in a similar way as the transfer function calculations in the original \verb|class| code. In particular, each $k$- or $\ell$-mode of a given power spectrum or bispectrum, which requires its own redshift and mass integral, is computed by a single \textit{thread}.  Hence, the evaluation time of a typical calculation will generally benefit from setting the \verb|OpenMP|  environment variable \verb|OMP_NUM_THREADS| to the highest possible value given the computing platform architecture.\footnote{On an 8-core MacBook Pro, this value would be 16 since there are two computing threads available per core. On the Haswell nodes on NERSC, the value would be 64 since each node has 32 cores with two threads each.}  Quantities that are common to most $k$- and $\ell$-mode integrands, and which are computationally expensive, are pre-tabulated outside of the main parallel block. The tabulations themselves are parallelized, e.g., over a redshift and mass grid, when possible. 

The code borrows special functions and an integrator for oscillatory functions from \verb|gsl| \citep{galassi2018scientific}, interpolation and root finding routines from J.~Burkardt's scientific library,\footnote{\href{https://people.math.sc.edu/Burkardt/index.html}{https://people.math.sc.edu/Burkardt/index.html}} and the FFTLog algorithm \citep{Hamilton_2000} as implemented by A.~Slosar.\footnote{\href{https://github.com/slosar/FFTLog}{https://github.com/slosar/FFTLog}} Hence, in addition to the dependencies of the \verb|class| code, \verb|class_sz| requires \verb|gsl| and \verb|FFTW3| \citep{FFTW05}, properly installed and linked.  

The python wrapper \verb|classy_sz| can be called within any Python code. In particular, \verb|classy_sz| is easily interfaced with Markov Chain Monte Carlo (MCMC) samplers such as \verb|MontePython| \citep{Brinckmann:2018cvx,Audren:2012wb} and \verb|Cobaya| \citep{Torrado_2021} for Bayesian inference, as well as machine learning packages such as \verb|CosmoPower| \citep{Spurio_Mancini_2022}.

\newpage
\subsection{Halo Model for the Hybrid Bispectrum}\label{ss:hybrid_bispec}

The hybrid bispectrum, introduced in Eq.~(\ref{eq:c_approx}-\ref{eq:clpf}), is defined by
\begin{equation}
 \langle \delta_\mathrm{e}(\bm{k}_1) \delta_\mathrm{e}(\bm{k}_2) X(\bm{k}_3)\rangle=(2\pi)^3\delta_D(\bm{k}_1+\bm{k}_2+\bm{k}_3)B_{\delta_\mathrm{e}\delta_\mathrm{e}X}(k_1,k_2,k_3),
\end{equation}
where $\delta_\mathrm{e}$ is the electron density perturbation and $X$ is to be understood as the perturbation in the tracer field. 
In the halo model, the hybrid bispectrum is the sum of three terms
\begin{equation}
  B_{\delta_\mathrm{e}\delta_\mathrm{e}X} = B_{\delta_\mathrm{e}\delta_\mathrm{e}X}^\mathrm{1h}+B_{\delta_\mathrm{e}\delta_\mathrm{e}X}^\mathrm{2h}+B_{\delta_\mathrm{e}\delta_\mathrm{e}X}^\mathrm{3h},
\end{equation}
where the one-halo (1h) term corresponds to three points within the same halo, the two-halo term (2h) to three points within two halos, and the three-halo term (3h) to three points in three distinct halos \citep[see, e.g.,][for further details on halo model bispectra]{2000ApJ...543..503M,Scoccimarro:2000gm,Valageas_2011,Lacasa_2014,Lazanu_2016}. Then, each term is expressed in terms of ensemble averages over halos as (at a given redshift/comoving distance):
\begin{align}
          B_{\delta_\mathrm{e}\delta_\mathrm{e}X}^\mathrm{1h} &= \int \mathrm{d}n_1 \hat{u}_{k_1}^\mathrm{e}(m_1)\hat{u}_{k_2}^\mathrm{e}(m_1)\hat{u}_{k_3}^X(m_1)\label{eq:hybispec1h} \\
      B_{\delta_\mathrm{e}\delta_\mathrm{e}X}^\mathrm{2h} &= \int \mathrm{d}n_1b^{(1)}(m_1)\hat{u}_{k_1}^\mathrm{e}(m_1)\hat{u}_{k_2}^\mathrm{e} (m_1)\int \mathrm{d}n_2 b^{(1)}(m_2)\hat{u}_{k_3}^X(m_2)P_{L}(k_3)+\mathrm{perms}\label{eq:hybispec2h}\\
      B_{\delta_\mathrm{e}\delta_\mathrm{e}X}^\mathrm{3h} &= 2\int \mathrm{d}n_1b^{(1)}(m_1)\hat{u}_{k_1}^\mathrm{e}(m_1)P_{L}(k_1)\int\mathrm{d}n_2b^{(1)}(m_2)\hat{u}_{k_2}^\mathrm{e} (m_2)P_{L}(k_2)\int \mathrm{d}n_3b^{(1)}(m_3) \hat{u}_{k_3}^X(m_3)F_2(k_1,k_2,k_3)\nonumber\\
     &+\int \mathrm{d}n_1b^{(1)}(m_1)\hat{u}_{k_1}^\mathrm{e}(m_1)P_{L}(k_1)\int\mathrm{d}n_2b^{(1)}(m_2)\hat{u}_{k_2}^\mathrm{e} (m_2)P_{L}(k_2)\int \mathrm{d}n_3b^{(2)}(m_3) \hat{u}_{k_3}^X(m_3)+\mathrm{perms}\label{eq:hybispec3h}
\end{align}
where we did not write explicitly the permutations (see Appendix~\ref{ss:hm_bp} for details) and where $F_2$ is the kernel defined in Eq.~\eqref{eq:f2s} for the tree-level bispectrum. Here, $\hat{u}_{k}^\mathrm{e}(m)$ is the Fourier transform of the gas density profile (divided by $\rho_\mathrm{m,0}$, see below) and $\hat{u}_{k}^X(m)$ is the Fourier transform of the radial profile of $X$, e.g., the mass profile for weak lensing fields, or the galaxy HOD for galaxy density.

With $\delta_\mathrm{e}=X=\delta_\mathrm{m}$, this formula reduces to the halo model matter bispectrum \citep[see, e.g., Section IV.A of][]{Lazanu_2016}, which we plot in Figure \ref{fig:bks} (top row). We refer to \cite{Lazanu_2017} and \cite{phils2020} for a discussion on the accuracy of the halo model expressions and possible extensions.

\subsection{Gas Density Profile}\label{ss:gas_prof}

For the gas density profile, $\rho_\mathrm{gas,free}$, we consider two parameterizations. First, the NFW formula \citep{Navarro_1997} rescaled by the baryon fraction $f_\mathrm{b}$, i.e.,
\begin{equation}
    \rho_{\mathrm{gas,free}}(r)=f_\mathrm{b}f_\mathrm{free}\rho_{_\mathrm{NFW}}(r)
\end{equation}
where $\rho_{_\mathrm{NFW}}(r)$ is the usual NFW profile (see Appendix \ref{ss:app_lss} for details). Second, a generalized NFW (gNFW) formula, following   \cite{Battaglia_2016}:
\begin{equation}
    \rho_{\mathrm{gas,free}}(r)=f_\mathrm{b}f_\mathrm{free}\rho_\mathrm{crit}(z) C \left(\frac{r}{x_\mathrm{c}r_\mathrm{200c}}\right)^\gamma \left[1+\left(\frac{r}{x_\mathrm{c}r_\mathrm{200c}}\right)^\alpha\right]^{-\frac{\beta+\gamma}{\alpha}}, \label{eq:ugnfw}
\end{equation}
where $r_\mathrm{200c}$ is the characteristic radius associated with the overdensity mass $m_\mathrm{200c}$ (see Eq.~\ref{eq:rd}), with $x_c=0.5$ and $\gamma=-0.2$ kept fixed throughout the paper and with mass and redshift dependent parameters $C,\alpha,\beta,\gamma$, such that 
\begin{equation}
    p= A_0\left(\frac{m_\mathrm{200c}}{10^{14}M_\odot}\right)^{A_m}\left(1+z\right)^{A_z}\quad\mathrm{for}\quad p \in \{C,\alpha,\beta,\gamma\}.
\end{equation}
For $A_0,A_m,A_z$ we use the best-fit values from \cite{Battaglia_2016} reported in Table \ref{tab:gnfwb16}, corresponding to either the \textit{AGN feedback} model (that is our fiducial assumption) or the \textit{Adiabatic} model. Note that the NFW profile is a subcase of the gNFW formula, when parameters are set to $x_c=1/c_{200c}$, $\gamma=-1$, $\alpha=1$, $\beta=3$, and $C=\rho_{s}/\rho_\mathrm{crit}(z)$, where $c_{200c}$ is the concentration computed with the \cite{2013ApJ...766...32B} relation and $\rho_s$ is the normalization of the NFW profile defined in Eq.~\eqref{eq:rho_s}.
\begin{table}
\begin{centering}
\begin{tabular}{c|ccc|ccc}
 & \multicolumn{3}{c}{\textit{AGN feedback}} & \multicolumn{3}{c}{\textit{Adiabatic}}\tabularnewline
$p$ & $A_0$ & $A_{m}$ & $A_{z}$ & $A_0$ & $A_{m}$ & $A_{z}$\tabularnewline
\hline 
$C$ & $4\times10^{3}$ & $0.29$ & $-0.66$ & $1.9\times10^{4}$ & $0.09$ & $-0.95$\tabularnewline
$\alpha$ & $0.88$ & $-0.03$ & $0.19$ & $0.70$ & $-0.017$ & $0.27$\tabularnewline
$\beta$ & $3.83$ & $0.04$ & $-0.025$ & $4.43$ & $0.005$ & $0.037$\tabularnewline
\end{tabular}
\par\end{centering}
\caption{Best-fit values of the parameters of the generalized NFW gas density profile formula fit to simulations from  \cite{Battaglia_2016}. \textit{Adiabatic} corresponds to simulations whose sub-grid model has only gravitational heating. \textit{AGN feedback} corresponds to a sub-grid model with radiative cooling, star formation, supernova feedback, cosmic rays, and AGN feedback. The gas density profile is computed using these parameters in Eq.~\eqref{eq:ugnfw}. \citep[See][for details.]{Battaglia_2016}}
\label{tab:gnfwb16}
\end{table} 
With this, we compute the Fourier transforms entering Eqs.~(\ref{eq:hybispec1h}-\ref{eq:hybispec3h}) as 
\begin{equation}
\hat{u}^\mathrm{e}_k=4\pi \int_0^\infty \mathrm{d}r r^2 j_0(kr)\mathrm{H}(r_\mathrm{cut}-r)u^\mathrm{e}(r)\quad\mathrm{with}\quad j_0(x)=\frac{\sin(x)}{x}\quad\mathrm{and}\quad u^\mathrm{e}(r)=\frac{\rho_\mathrm{gas,free}(r)}{\rho_\mathrm{m,0}},\label{eq:hatub}
\end{equation}
where $\mathrm{H}$ is the Heaviside step function (which truncates the profile at $r_\mathrm{cut}$) and where we used the fact that the profiles are radially symmetric to write the Fourier transform as a Hankel transform. In general, it is necessary to truncate the density profiles because their volume integrals do not converge or may have support at unphysically large radii. For the NFW profile, we set the truncation radius to $r_\mathrm{cut}=r_\mathrm{200c}$. For the gNFW profile we require $r_\mathrm{cut}$ to be such that the enclosed gas mass is the same as in the NFW case, i.e., $f_\mathrm{b}m_{200c}$. We then find $r_\mathrm{cut}$ numerically with Brent's method \citep{brent2002algorithms}, solving
\begin{equation}
F(r_\mathrm{out};m_\mathrm{200c},z)=0\quad\mathrm{with}\quad F(r_\mathrm{out};m_\mathrm{200c},z)=4\pi \int_0^\mathrm{r_\mathrm{cut}}\mathrm{d}r r^2\rho_\mathrm{gas}(r;m_\mathrm{200c},z)-f_\mathrm{b}m_\mathrm{200c},\label{eq:brent}
\end{equation}
where we wrote the mass and redshift dependence explicitly to emphasize the fact that this operation is done at each mass and redshift. The left panel of Figure~\ref{fig:gasprof} illustrates the method. It shows the enclosed gas mass as a function of radius (in units of $r_\mathrm{200c}$).  The horizontal line indicates the gas mass of the halo, namely $f_\mathrm{b}\times3.3\,M_\odot$. By definition, it intersects the NFW curve at $r_\mathrm{200c}$, i.e., $x=1$. The blue and red vertical lines represent the truncation radii as obtained by solving Eq.~\eqref{eq:brent}, and consistently intersect the Adiabiatic and AGN curves at $f_\mathrm{b}\times3.3\,M_\odot$. The right panel of Figure~\ref{fig:gasprof} compares the NFW, Adiabatic and AGN profiles (rescaled by $x^2/f_\mathrm{b}\rho_\mathrm{crit}$). The area below each curve (which is proportional to the mass) is colored for $x$ corresponding to $r<r_\mathrm{cut}$.

We note that halo models based on the \cite{Battaglia_2016} gas density profile parameterization have been used in multiple previous analyses \citep[e.g.,][]{Smith:2018bpn, Munchmeyer:2018eey,Cayuso:2021ljq,Roy:2022muv},  which computed Fourier and harmonic space two-point functions. In principle, our results could be checked against these studies. One notable difference is that previous works often truncate the gas density profile at $r_\mathrm{200c}$, and rescale its amplitude by a factor such that the enclosed mass is $m_\mathrm{200c}$.  We argue that our truncation method is more consistent, as it preserves the total gas mass but does not alter the density as a function of radius.

In Figure~\ref{fig:bks}, we show the matter hybrid bispectrum for $kSZ^2\times\delta_\mathrm{g}$ cross-correlation at $z=1$ computed according to the different approaches discussed above. We show three different triangle configurations, parameterized via $(k,k^\prime,k^{\prime\prime})=(k,\lambda k,\lambda^\prime k)$, with $\lambda=\lambda^\prime=1$ for the \textit{Equilateral} configuration,  $\lambda=1,\lambda^\prime=0.1$  for the \textit{Squeezed} configuration and  $\lambda=0.5,\lambda^\prime=0.5$ for the \textit{Flattened} configuration. For the hybrid bispectrum, we show the dimensionless combinations  $b_\mathrm{g}(z)W^\mathrm{kSZ}(\chi)^2 f_\mathrm{b}^2 f_\mathrm{free}^2 B_\mathrm{TL}$ for the Tree-Level line, $b_\mathrm{g}(z)W^\mathrm{kSZ}(\chi)^2 f_\mathrm{b}^2 f_\mathrm{free}^2 B_\mathrm{eff}^{NL}$ for the effective approach lines (where $b_\mathrm{g}(z)$ is given in Eq.~\ref{eq:ngbar}), and $W^\mathrm{kSZ}(\chi)^2 B_{\delta_\mathrm{e}\delta_\mathrm{e}\delta_\mathrm{g}}^{i\mathrm{h}}$ with $i=1,2,3$ for the halo-model lines (see Eq.~\ref{eq:hybispec1h}-\ref{eq:hybispec3h}).

\begin{figure}
    \includegraphics[width=1.\columnwidth]{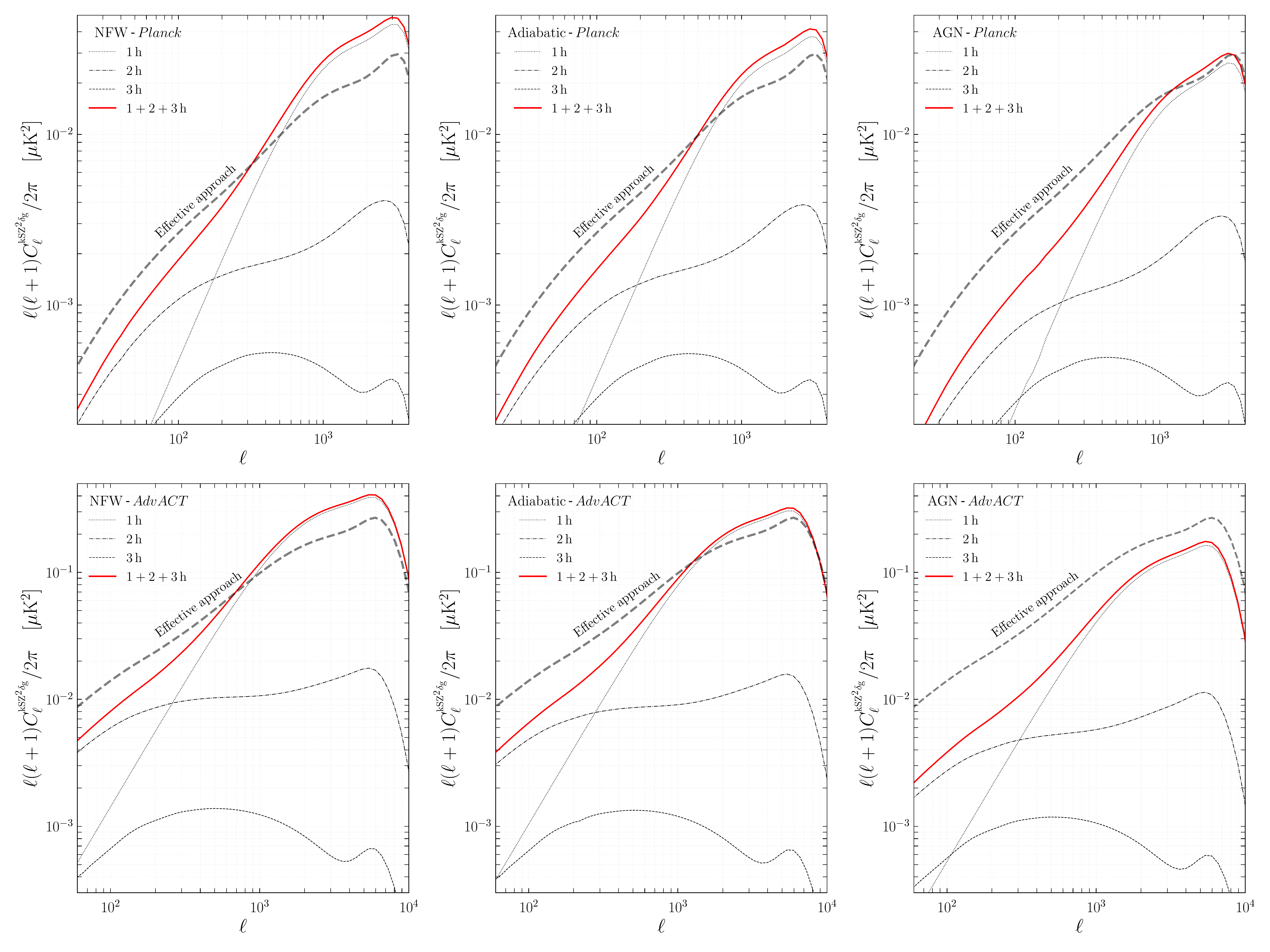}
    \caption{\textbf{(Galaxy density)} Projected-field kSZ  power spectrum in cross-correlation with galaxies, assuming our fiducial model (see Section \ref{s:fid}). The halo model predictions (Section \ref{s:hm}) are the thick red lines and thin dashed and dotted lines. The top row shows the three different profiles, i.e., \textit{NFW}, \textit{Adiabatic}, and \textit{AGN Feedback} (see Table~\ref{tab:gnfwb16} and Eq.~\ref{eq:ugnfw})  assuming a \textit{Planck} CMB map and the bottom row shows the same but assuming an AdvACT CMB map (see Table \ref{tab:exps}). The \textit{effective approach} predictions (Subsection \ref{ss:eff}) are the grey dashed lines -- they do not change from column to column. Furthermore, they are multiplied by the redshift dependent bias $b_\mathrm{g}(z)$ of Eq.~\ref{eq:ngbar}. The galaxy HOD parameters are for an  \textit{unWISE}-like catalogue and are given below Eq.~\ref{eq:ngbar}. The overall signal is much larger in the bottom row because the AdvACT filter preserves much more of the small-scale kSZ signal than that of \emph{Planck} (see Figure~\ref{fig:filter_figure}).} 
    \label{fig:fls_cls_ttg}
\end{figure}

\begin{figure}
    \includegraphics[width=1.\columnwidth]{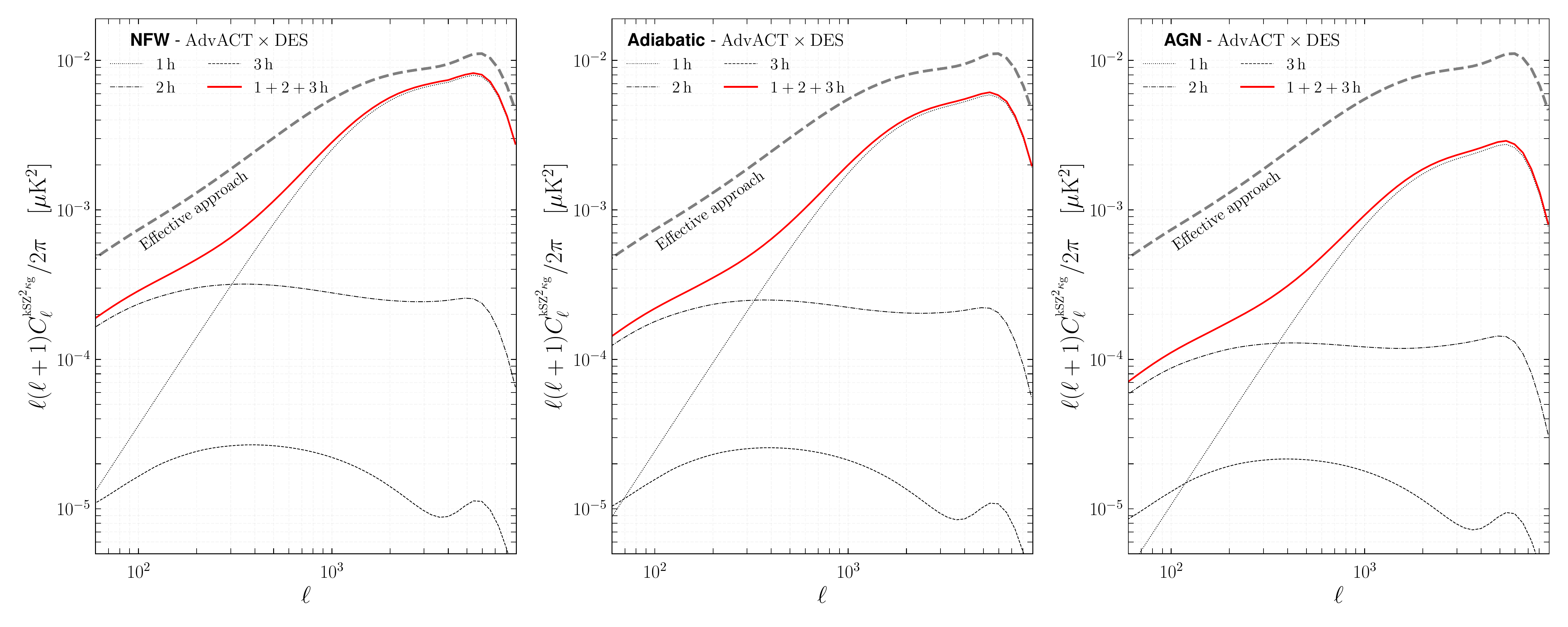}
    \caption{\textbf{(Galaxy lensing)} Projected-field kSZ  power spectrum in cross-correlation with galaxy lensing, assuming our fiducial model (see Section \ref{s:fid}). The \textit{effective approach} prediction (see Subsection \ref{ss:eff}) is the thick dashed line and is the same in each panel. The halo model predictions (Section \ref{s:hm}) are the thick red lines and thin dashed and dotted lines. Each panel shows a different profile, i.e., \textit{NFW}, \textit{Adiabatic}, and \textit{AGN Feedback} (from left to right -- see Table~\ref{tab:gnfwb16} and Eq.~\ref{eq:ugnfw}), and they all assume an AdvACT CMB map (see Table \ref{tab:exps}). For galaxy weak lensing we assume a DES-like survey (see Table~\ref{tab:exps_lss}).} 
    \label{fig:fls_cls_ttkg}
\end{figure}

\begin{figure}
    \includegraphics[width=1.\columnwidth]{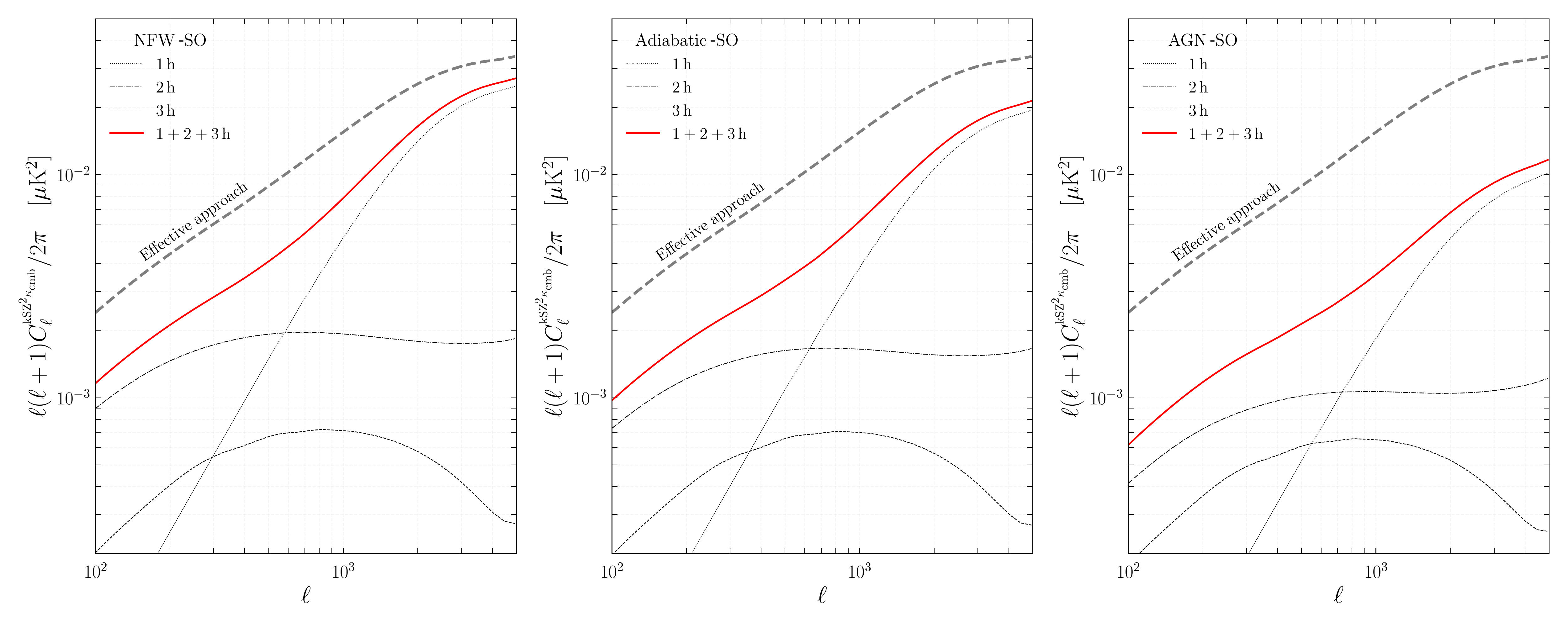}
    \caption{\textbf{(CMB lensing)} Projected-field kSZ  power spectrum in cross-correlation with CMB lensing, assuming our fiducial model (see Section \ref{s:fid}). The \textit{effective approach} prediction (see Subsection \ref{ss:eff}) is the thick dashed line and is the same in each panel. The halo model predictions (Section \ref{s:hm}) are the thick red lines and thin dashed and dotted lines. Each panel shows a different profile, i.e., \textit{NFW}, \textit{Adiabatic}, and \textit{AGN Feedback} (from left to right -- see Table~\ref{tab:gnfwb16} and Eq.~\ref{eq:ugnfw}), and they all assume an SO CMB map (see Table \ref{tab:exps}). For CMB weak lensing we assume an SO map as well (see Table~\ref{tab:exps_lss}). } 
    \label{fig:fls_cls_ttkc}
\end{figure}

\subsection{The Projected-field kSZ Power Spectrum and its Numerical Implementation}\label{ss:hmeqs}

To compute the projected-field kSZ power spectrum with the halo model, we use Eq.~\eqref{eq:clpf} where we replace $B_{\delta_\mathrm{e}\delta_\mathrm{e}X}$ by the halo model hybrid bispectrum discussed in the previous section. The three terms of the halo model hybrid bispectrum (see Eq.~\ref{eq:hybispec1h}-\ref{eq:hybispec3h}) yield three terms for the projected-field kSZ power spectrum, which we can write as 
\begin{equation}
C_\ell^{\mathrm{kSZ}^2X}= C_\ell^{\mathrm{kSZ}^2X,1\mathrm{h}}+C_\ell^{\mathrm{kSZ}^2X,2\mathrm{h}}+C_\ell^{\mathrm{kSZ}^2X,3\mathrm{h}}.
\end{equation}
At this point, we know how to compute the halo model projected-field kSZ power spectrum. The challenge is to find an efficient numerical implementation. The most straightforward option is to carry out the multiple integrals one after the other, as we described for the effective approach implementation (see Subsection~\ref{ss:eff}). The difference is that the halo model hybrid bispectrum requires integration over halo mass. Hence for the one-halo term, we need to compute a four-dimensional integral (over redshift, $\ln \ell^\prime$, $\theta$ and $m_1$), for the two-halo term a five-dimensional integral (same as one-halo, plus $m_2$ integral) and for the three-halo term a six-dimensional integral (same as for the two-halo term, plus $m_3$ integral). This is numerically tractable, however the computation is time-consuming, taking $\sim\mathcal{O}(10^3\mathrm{s})$ on a laptop.

Fortunately, there exists a way to accelerate the computation because the halo model enables us to take a shortcut when evaluating the 2D integral over  $\bm{\ell}^\prime$ in Eq.~\eqref{eq:clpf}. It relies on the fact that the halo model hybrid bispectrum terms are all separable with respect to scale/wavenumber. Hence, the triangle power spectrum (Eq.~\ref{eq:clpf}) is a 2D convolution which can be evaluated rapidly using the \verb|FFTLog| algorithm \citep{Hamilton_2000}. Let us describe the procedure in more detail. Starting from Eq.~\eqref{eq:clpf}, we first split the triangle power spectrum into the three halo-model terms and treat them separately: $T(\ell,\chi)=T^\mathrm{1h}(\ell,\chi)+T^\mathrm{2h}(\ell,\chi)+T^\mathrm{3h}(\ell,\chi)$.  Using the expression of the one-halo term of the hybrid bispectrum (Eq.~\ref{eq:hybispec1h}) and re-arranging, we can write the one-halo term of the triangle power spectrum as
\begin{equation}
    T^\mathrm{1h}(\ell,\chi)=\frac{1}{(2\pi)^2}\int\mathrm{d}n \hat{u}^X_{\ell}\int\mathrm{d}^2\bm{\ell}^\prime  \left[w(\ell^\prime)\hat{u}^\mathrm{e}_{\ell^\prime}\right]\left[w(\left|\bm{\ell}+\bm{\ell}^\prime\right|)\hat{u}^\mathrm{e}_{|\bm{\ell}+\bm{\ell}^\prime|}\right].
\end{equation}
From the convolution theorem, we can replace the integral over $\mathrm{d}^2\bm{\ell}^\prime$ by Fourier transform operations as
\begin{equation}
    T^\mathrm{1h}(\ell,\chi)=\frac{1}{(2\pi)^2}\langle \hat{u}^X_{\ell}\varphi(\ell)\rangle_n,\quad\mathrm{with}\quad\varphi^\mathrm{1h}\equiv\mathcal{F}^{-1}\left\{\mathcal{F}\left\{w\hat{u}^\mathrm{e}\right\}\cdot\mathcal{F}\left\{w\hat{u}^\mathrm{e}\right\}\right\},\label{eq:fft1h}
\end{equation}
where $\mathcal{F}$ is the Fourier transform (from harmonic space to angular space), and $\mathcal{F}^{-1}$ is its inverse. Note that here, the Fourier transform operations are performed inside the mass integral (the integral over $\mathrm{d}n$, denoted $\langle\cdots\rangle_n$). Eventually, we integrate over redshift to get $C_\ell^{\mathrm{kSZ}^2X,1\mathrm{h}}$. 

For the two-halo term, there are two mass integrals. Moreover, due to the cyclic permutation we have three terms to deal with. For one of these terms, which we denote $T_\mathrm{tog.}^\mathrm{2h}$ , we have two $\hat{u}^\mathrm{e}$'s within the same mass integral. For the other two terms, each $\hat{u}^\mathrm{e}$ is in a separate mass integral, and we denote their sum  $T_\mathrm{sep.}^\mathrm{2h}$. In fact, both terms in $T_\mathrm{sep.}^\mathrm{2h}$ contribute equally, because of the invariance of the expression under the transformation $\ell\leftrightarrow|\bm{\ell}+\bm{\ell}^\prime|$.

Using the expression of the two-halo term of the hybrid bispectrum (Eq.~\ref{eq:hybispec2h}), and re-arranging, we get 
\begin{align}
    T^\mathrm{2h}_{\mathrm{tog.}}(\ell,\chi)&= \frac{1}{(2\pi)^2}\langle b^{(1)} \varphi^\mathrm{2h}_\mathrm{tog.}(\ell)\rangle_n\langle b^{(1)}\hat{u}^X_{\ell}\rangle_nP_{L},\quad\mathrm{with}\quad\varphi^\mathrm{2h}_\mathrm{tog.}\equiv \mathcal{F}^{-1}\left\{\mathcal{F}\left\{w\hat{u}^\mathrm{e}\right\}\cdot\mathcal{F}\left\{w\hat{u}^\mathrm{e}\right\}\right\},
    \label{eq:fft2htog}\\
    T^\mathrm{2h}_{\mathrm{sep.}}(\ell,\chi)
    &=\frac{1}{2\pi^2}\varphi^\mathrm{2h}_\mathrm{sep.}(\ell),\quad\mathrm{with}\quad\varphi^\mathrm{2h}_\mathrm{sep.}\equiv \mathcal{F}^{-1}\{\mathcal{F}\{w\langle b^{(1)}\hat{u}^\mathrm{e} \rangle_nP_L\}\cdot \mathcal{F}\{w\langle b^{(1)}\hat{u}^\mathrm{e} \hat{u}^X_\ell\rangle_n\}\},
    \label{eq:fft2hsep}
\end{align}
where $P_L$ in Eq.~\eqref{eq:fft2htog} is evaluated at $k=(\ell+1/2)/\chi$.  Note that, as for the one-halo term, the Fourier transform operations in Eq.~\eqref{eq:fft2htog} are carried out inside the mass integral, while in Eq.~\eqref{eq:fft2hsep} the Fourier transform operations are carried out \textit{after} the mass integrals. Then we have $T^\mathrm{2h}=T^\mathrm{2h}_{\mathrm{tog.}}+T^\mathrm{2h}_{\mathrm{sep.}}$, which we can integrate over redshift according to Eq.~\eqref{eq:clpf} to get $C_\ell^{\mathrm{kSZ}^2X,2\mathrm{h}}$.

For the three-halo term, let us first notice that the hybrid bispectrum splits into terms proportional to the second order bias $b^{(2)}$ and terms proportional to the $F_2$-kernel (see Eq.~\ref{eq:hybispec3h}). Hence, for the triangle power spectrum we have $T^\mathrm{3h}=T^\mathrm{3h}_{b^{(2)}}+T^\mathrm{3h}_{F_2}$. Accounting for permutations, there are three terms proportional to $b^{(2)}$ in the hybrid bispectrum, but two of them contribute equally to $T^\mathrm{3h}_{b^{(2)}}$ due to the symmetry $\ell\leftrightarrow|\bm{\ell}+\bm{\ell}^\prime |$. After re-arranging we get
\begin{equation}
T_{b^{(2)}}^\mathrm{3h}(\ell,\chi)
=\frac{1}{(2\pi)^2}\varphi^{\mathrm{3h,1}}_{b^{(2)}}\langle b^{(2)}\hat{u}^X_\ell\rangle_{n}+\frac{1}{2\pi^2}\varphi^{\mathrm{3h,2}}_{b^{(2)}}\langle b^{(1)}\hat{u}^X_\ell\rangle_{n}P_L,\label{eq:Il3hb2}
\end{equation}
where here $P_L$ is evaluated at $k=(\ell+1/2)/\chi$ and where we introduced the definitions
\begin{equation}
\varphi^{\mathrm{3h,1}}_{b^{(2)}}\equiv \mathcal{F}^{-1}\{\mathcal{F}\{w\langle b^{(1)}\hat{u}^\mathrm{e}\rangle_{n}P_L\}\cdot\mathcal{F}\{w\langle b^{(1)}\hat{u}^\mathrm{e}\rangle_{n}P_L\}\}\quad\mathrm{and}\quad\varphi^{\mathrm{3h,2}}_{b^{(2)}}\equiv\mathcal{F}^{-1}\{\mathcal{F}\{w\langle b^{(1)}\hat{u}^\mathrm{e}\rangle_{n}\}\cdot\mathcal{F}\{w\langle b^{(2)}\hat{u}^\mathrm{e}\rangle_{n}P_L\}\}.
\end{equation}
Note that here, the Fourier transform operations are carried out \textit{after} the mass integrals.

To compute $T^\mathrm{3h}_{F_2}$, we start by expanding $F_2$ as 
\begin{align}
F_2(k_1,k_2,k_3)=\frac{5}{14}+\frac{3}{28}x_{31}^2+\frac{3}{28}x_{32}^2-\frac{5}{28}x_{12}^2-\frac{5}{28}x_{21}^2+\frac{1}{14}x_{31}^2x_{32}^2\quad\mathrm{with}\quad x_{ij}\equiv k_i/k_j,\,\,i,j\in\{1,2,3\}\label{eq:f2expanded}
\end{align}
Since this form of $F_2$ has six terms and accounting for the cyclic permutation, we can write the hybrid bispectrum terms proportional to $F_2$ as a sum of eighteen terms, namely
\begin{align}
      B^{\mathrm{3h}}_{F_2}=\sum_{i=1}^{18}\omega_i\left[\langle b^{(1)}\hat{u}^\mathrm{e}_{k_1}\rangle_{n}P_L(k_1)^s k_1^p\right] \left[\langle b^{(1)}\hat{u}^\mathrm{e}_{k_2}\rangle_{n}P_L(k_2)^t k_2^q\right]\left[\langle  b^{(1)}\hat{u}^X_{k_3}\rangle_{n}P_L(k_3)^u k_3^r\right]\label{eq:b3hf2t}
\end{align}
where $\omega_i$ are rational prefactors determined by the prefactors in Eq.~\eqref{eq:f2expanded}, where $p,q,r\in\{0,2,-2,4\}$ such that $p+q+r=0$ and where $s,t,u\in\{0,1\}$ such that $s+t+u=2$. (The combination of indices $p,q,r,s,t,u$ is different for each of the 18 terms.) Here we used wavenumbers rather than multipoles --- we assume the same correspondence as in Eq.~\eqref{eq:clpf}. Due to the symmetry $\ell\leftrightarrow|\bm{\ell}+\bm{\ell}^\prime|$, these eighteen terms yield ten different terms in $T^\mathrm{3h}_{F_2}$ (twelve of the eighteen terms contribute equally). After re-arranging, we get
\begin{equation}
T_{F_2}^\mathrm{3h}(\ell,\chi)
= \frac{1}{(2\pi)^2}\sum_{i=1}^{10} c_i  \mathcal{F}^{-1}\left\{\mathcal{F}\{w\langle b^{(1)}\hat{u}^\mathrm{e}\rangle_{n}P_L(k)^s k^p\}\cdot\mathcal{F}\{w\langle b^{(1)}\hat{u}^\mathrm{e}\rangle_{n}P_L(k)^t k^q\}
\right\}\left[\langle  b^{(1)}\hat{u}^X_{k_3}\rangle_{n}P_L(k_3)^u k_3^r\right]\label{eq:il3hf2}
\end{equation}
where $k_3=(\ell+1/2)/\chi$ and where $c_i$ are numerical constants, $c_1=c_5=5/7,c_2=2c_8=2c_9=-5/7,c_3=2c_6=2c_7=3/7,c_4=c_{10}=1/7$, with $(u,s,t)=(0,1,1)$ for $1\le i\le4$ and $(u,s,t)=(1,1,0)$ for $5\le i\le10$. Eventually, we obtain $C_\ell^{\mathrm{kSZ}^2X,\mathrm{3h}}$ by summing Eqs.~\eqref{eq:Il3hb2} and \eqref{eq:il3hf2}, and integrating over redshift according to Eq.~\eqref{eq:clpf}.

The halo model prediction is shown in detail in Figures~\ref{fig:fls_cls_ttg}, \ref{fig:fls_cls_ttkg} and \ref{fig:fls_cls_ttkc}, where we plot it against the effective approach prediction.  Naively, one would expect that on large scales the projected-field power spectrum should always hold to the same limit, irrespective of gas profile assumptions (determined by the fact that the enclosed mass is always the same). However, in our case it is not true because of the convolution which mixes contributions from \textit{all} scales at each $\ell$ (see Eq.~\eqref{eq:cldef}). One way to recover the intuitive large-scale behavior is to artificially filter out the small scales. 

\begin{figure}
    \includegraphics[width=1.\columnwidth]{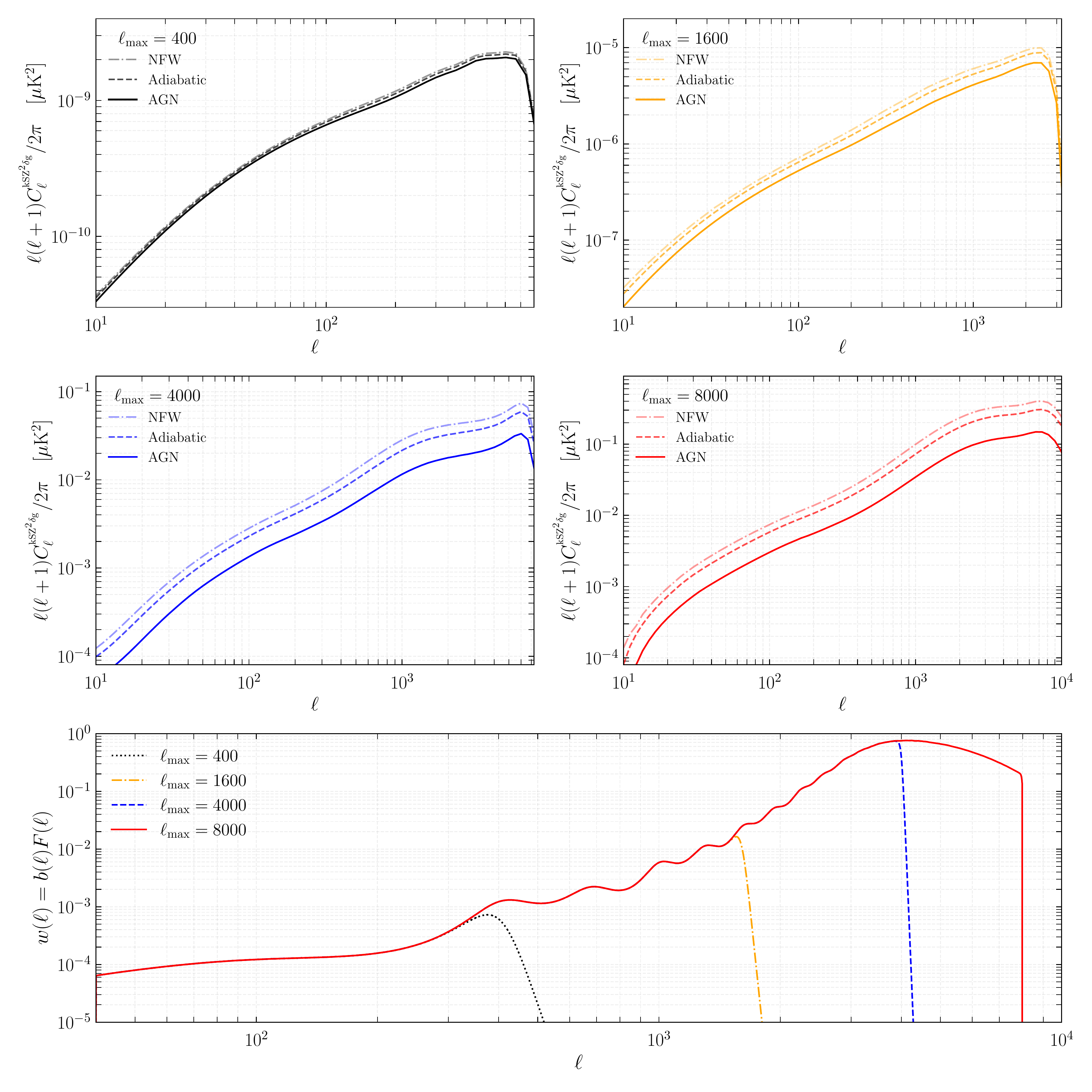}
    \caption{Projected-field kSZ power spectrum in cross-correlation with galaxy number density (assuming the case of \textit{unWISE} galaxies, see Table~\ref{tab:exps_lss}). Here, we assume an SO-like CMB map (see Table~\ref{tab:exps}). The bottom panel shows four filter choices with different $\ell_\mathrm{max}$ (multiplied by the beam --- see Eq.~\ref{eq:bwf}). The top panels show the corresponding predictions for the projected-field power spectrum. Each of the top panels shows the three different gas profile assumptions: \textit{NFW}, \textit{Adiabatic}, and \textit{AGN Feedback} (see Table~\ref{tab:gnfwb16} and Eq.~\ref{eq:ugnfw}).  This figure illustrates that when small scales are removed (case $\ell_\mathrm{max}=400$), the three different profile assumptions become almost indistinguishable.}
    \label{fig:fls_noss}
\end{figure}

In Figure~\ref{fig:fls_noss}, we show the halo model kSZ$^2$-galaxy projected-field power spectrum computed without contributions from small scales. We remove the small-scale contributions by defining a filter in harmonic space which vanishes for $\ell>\ell_\mathrm{max}$, using the taper function of footnote \ref{fn:taper}. The filters (multiplied by the beam) are shown on the same plot in the bottom panel.  The top panels are the resulting power spectra for four different values of $\ell_\mathrm{max}$ between 400 and 8000. We see that for $\ell_\mathrm{max}=8000$ and 4000, the different density profile assumptions (NFW, Adiabatic, and AGN) are easily distinguishable since the amplitudes of the power spectra significantly differ. However, for $\ell_\mathrm{max}=1600$ the difference is smaller and for $\ell_\mathrm{max}=400$ the three profile assumptions yield nearly the same projected-field power spectrum. The fact that the power spectra converge to the same amplitude is a good sanity check. It shows that our implementation is consistent with the fact that on large scales the projected-field power spectrum is sensitive to the overall gas mass rather than the details of the shape of the profiles. Although this figure shows results only for the kSZ${}^2$--galaxy cross-correlation, we expect the same results for other LSS tracers.  

\begin{figure}
    \includegraphics[width=1\columnwidth]{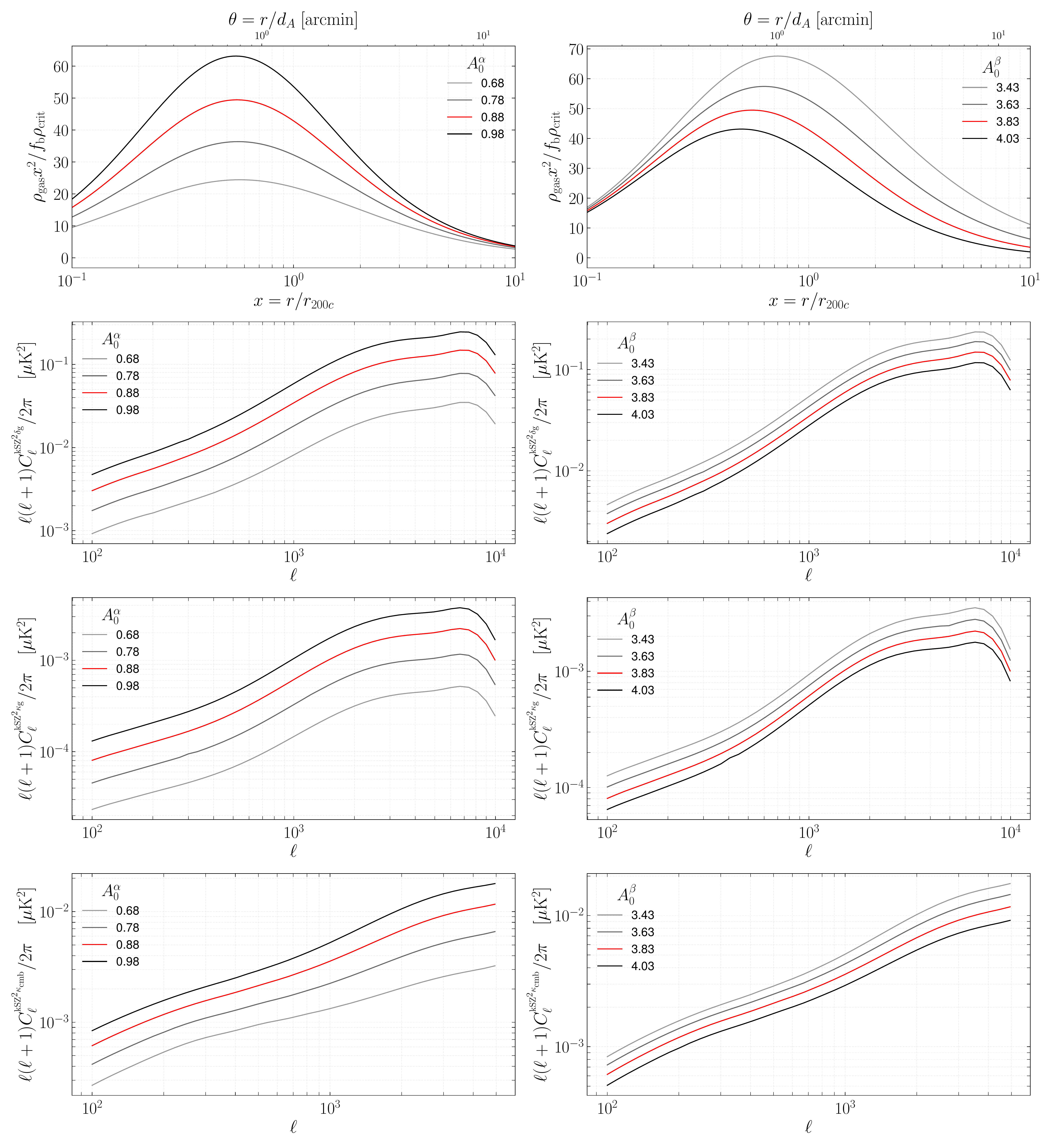} 
    \vspace{-0.2cm}
    \caption{Radial gas density profile (top row) for the same dark matter halo as in Figure~\ref{fig:gasprof} and projected-field kSZ power spectrum predictions for different values of the inner slope parameter $\alpha$ (left) and outer slope parameter $\beta$ (right). See Eq.~\eqref{eq:ugnfw} for the parameter definitions. When one parameter is varied, the others are kept fixed to their fiducial values, i.e., \textit{AGN Feedback}. The second, third, and fourth rows respectively assume an SO CMB map (see Table~\ref{tab:exps}) cross-correlated with galaxy number density (assuming \textit{unWISE}), galaxy weak lensing (assuming a VRO/\textit{Euclid}-like survey), and CMB weak lensing (assuming SO). See Table \ref{tab:exps_lss} for the LSS survey specifications. In all panels, our fiducial prediction (AGN feedback) is in red.}    
    \label{fig:varying_alpha_beta}
\end{figure}

The halo-model implementation allows us to study the projected-field power spectrum for arbitrary shapes of the gas density profile beyond the AGN, Adiabatic, and NFW models. In this paper we implement the gNFW formula from \cite{Battaglia_2016}. (Of course, one could easily extend our implementation to other gas models.)  Thus, we can study the projected-field power spectrum predictions for different values of the parameters entering the gNFW formula. In the top row of Figure~\ref{fig:varying_alpha_beta}, we consider the same dark matter halo as in Figure~\ref{fig:gasprof}, i.e., with a mass of $3.3\times 10^{13}\,M_\odot$ at $z=0.55$, and show the scaled density profile for different values of the slope parameters $\alpha$ and $\beta$ around the fiducial AGN model (red line). In the formalism of \cite{Battaglia_2016} these slope parameters have a mass and redshift dependence. Here, for simplicity, we keep the mass and redshift dependence fixed to the fiducial AGN feedback values, and only change the overall amplitude determined by $A_0^\alpha$ and $A_0^\beta$. On the left plot, we see that $\alpha$ mainly determines the gas density in the inner part of the halo --- $\alpha$ is referred to as the inner slope. On the right plot we see that $\beta$ mainly determines the gas density in the outer part of the profile --- $\beta$ is referred to as the outer slope. In the bottom rows of Figure~\ref{fig:varying_alpha_beta}, we show the projected-field power spectrum predictions associated with these different gas density profiles determined by the specific values of $A_0^\alpha$ and $A_0^\beta$. The fiducial AGN feedback predictions are shown in red and each row corresponds to a different LSS tracer: galaxies (second row), galaxy lensing (third row), and CMB weak lensing (fourth row). The same conclusion holds in all cases: changing the gas profile shape appears to amount to an overall shift of the projected-field power spectrum. This behavior differs from what is obtained for the kSZ power spectrum, where different gas density profiles yield the same large-scale power (see Figure~\ref{fig:cls_ksz}). Again, it is due to the fact that in the projected-field estimator, small scales (sensitive to the gas profile shape rather than the integrated mass) contribute at all multipoles including the low-$\ell$ part, because of the convolution of the profiles in harmonic space (see Eq.~\eqref{eq:cldef}). This is entirely due to the ``squaring operation'' on the CMB map, irrespective of the LSS tracer, as illustrated in the figure.

\subsection{Analytical Covariance Matrix}\label{ss:covmat}

In order to forecast results for upcoming CMB observations and LSS survey data, we need an estimate of the covariance matrix associated with a measurement of the projected-field kSZ power spectrum. We follow \cite{Dore2004} and \cite{Ferraro2016}, assuming that the Gaussian contribution dominates so that the covariance matrix reads:

\begin{equation}
    M_{\ell\ell^\prime}=\frac{1}{(2\ell+1)f_\mathrm{sky}}\{C_\ell^{\Theta_f^2\Theta_f^2}(C_\ell^{XX}+N_\ell^{XX})+(C_\ell^{\mathrm{kSZ}^2 X})^2\}\delta_{\ell\ell^\prime},\label{eq:covmat}
\end{equation}
where $f_\mathrm{sky}$ is the sky fraction spanned by the overlap between the CMB map and the LSS survey, $\delta_{\ell\ell^\prime}$ is the identity matrix,  $C_\ell^{\mathrm{kSZ}^2 X}$ is the projected-field kSZ power spectrum of Eq.~\eqref{eq:clpf}, and where 
\begin{equation}
C_\ell^{\Theta_f^2\Theta_f^2} = \frac{1}{2\pi^2}\int\mathrm{d}^2\bm{\ell}^\prime C_{\ell^\prime}^{\Theta_f \Theta_f}C_{|\bm{\ell}-\bm{\ell}^\prime|}^{\Theta_f \Theta_f }\quad\mathrm{with}\quad C_{\ell}^{\Theta_f \Theta_f} = w(\ell)^2\{C_\ell^{\Theta\Theta,\mathrm{lensed}}+C_\ell^{\mathrm{kSZ}}+N_\ell^{\Theta\Theta}\}.\label{eq:cltftf}
\end{equation}
Here $C_\ell^{\Theta\Theta,\mathrm{lensed}}$ is the \textit{lensed} primary CMB anisotropy power spectrum, which we compute with \verb|class/class_sz| in the fiducial model; $C_\ell^{\mathrm{kSZ}}$ is the kSZ anisotropy power spectrum which we compute according to Eq.~\eqref{eq:clksz2} in the fiducial model\footnote{Note that this is an underestimate of the total kSZ power, since that approximation is only accurate at high-$\ell$ (see Subsection \ref{sec:kszani}).}; and $N_\ell^{\Theta\Theta}$ is the noise power spectrum of the CMB map. It can be computed from the pixel noise level and beam as $N_\ell^{\Theta\Theta}=\Delta_T^2b(\ell)^{-2}$ with $b(\ell)$ from Eq.~\eqref{eq:bwf} for the case where other foregrounds are neglected, or extracted from Internal Linear Combination analyses (ILC) which includes the contribution from all other foregrounds. We consider several configurations summarized in Table~\ref{tab:exps}. (See also Figure~\ref{fig:cls_ksz_tt} for the CMB noise curves.) Note that in \verb|class_sz| we use Fourier transform methods, i.e., the convolution theorem, to evaluate Eq.~\eqref{eq:cltftf} efficiently. The remaining terms in Eq.~\eqref{eq:covmat} are the two-point contributions from tracer $X$: the noise $N_\ell^{XX}$ and the auto-power spectrum $C_\ell^{XX}$. They take on different expressions depending on the tracer. For galaxy number density, $N_\ell^{XX}$ is the \textit{shot noise}, i.e., $N_\ell^\mathrm{\delta_g\delta_g}=1/\bar{n}$ where $\bar{n}$ is the galaxy density per steradian; for galaxy weak lensing $N_\ell^{XX}$ is the \textit{shape noise}, i.e., $N_\ell^{\kappa_\mathrm{g}\kappa_\mathrm{g}} = \sigma_\gamma^2/n_s$ where  $n_s$ is the source galaxy number density, and $\sigma_\gamma^2$ is the intrinsic ellipticity dispersion (per shear component); for CMB lensing we use lensing noise estimates plotted in Figure~\ref{fig:cls_kk} and computed with the usual minimum-variance quadratic estimator \citep[see, e.g.,][]{Hu_Oka_2002}. In all cases, the power spectra $C_\ell^{XX}$ are computed within the halo model with \verb|class_sz| (see Appendix~\ref{ss:app_lss} for details). The specifications of the LSS surveys are summarized in Table~\ref{tab:exps_lss}.

The covariance matrix is binned with the same binning scheme as the one adopted for the measurement of $C_\ell^{\mathrm{kSZ}^2X}$. Since it is diagonal, and we assume it is slowly varying in each bin, we have $M_{a{a}} \approx M_{\ell_a\ell_{a}}/n_a$
where the effective multipole $\ell_a$ is at the center of the bin and $n_a$ is the number of multipoles in the bin. 

In the right panels of Figure \ref{fig:fls_cls_ksz2kg_cov}, we show the different contributions to the covariance matrix for the three types of cross-correlations. In all cases, the high-$\ell$ regime is dominated by noise from the LSS tracer. This suggests that future surveys, beyond the ones considered here, will improve the sensitivity.

\subsection{CMB Lensing Contribution}\label{ss:cmb_lens_cont}

The procedure to measure the projected-field kSZ power spectrum  relies on squaring the CMB temperature map. Inevitably, this implies that the measurement picks up correlations from the lensing field  whose leading-order contribution is of the form $\langle \psi X\rangle\langle \Theta_u \Theta_u\rangle $ where $\Theta_u$ is the unlensed CMB temperature field and $\psi$ is the CMB lensing potential. This simply arises from the fact that the lensed CMB temperature field is $\Theta = \Theta_u +\nabla \psi \cdot \nabla \Theta_u + \cdots$. Hence, the leading-order lensing contribution to $\bar{C}_\ell^{\mathrm{kSZ}^2 X}$ is \citep[see][for details]{Ferraro2016}:
\begin{equation}
    C_\ell^{\Theta_u^2\psi X} = -2\ell\frac{C_\ell^{\psi X}}{(2\pi)^2}\int \mathrm{d}\bm{\ell}^\prime \ell^\prime w(\ell^\prime) C_{\ell^\prime}^{\Theta_u\Theta_u}w(|\bm{\ell}^\prime+\bm{\ell}|)\cos\theta,\label{eq:lenscon}
\end{equation}
where $\theta=\hat{(\bm{\ell},\bm{\ell}^\prime)}$, $C_{\ell}^{\Theta_u\Theta_u}$ is the unlensed primary CMB power spectrum, and $C_\ell^{\psi X}$ is the cross-power spectrum between the lensing potential and tracer $X$, which we write in terms of the CMB lensing convergence as $C_\ell^{\psi X} = [{2}/{\ell(\ell+1)}]C_\ell^{X\kappa_{_\mathrm{CMB}}}$. We compute $C_\ell^{X\kappa_{_\mathrm{CMB}}}$ within the halo model (see Appendix~\ref{sss:useful_cls}). Note also that since the cosine can be expanded as 
$\cos \theta = (|\bm{\ell}^\prime+\bm{\ell}|^2-\ell^2 - \ell^{\prime 2})/2\ell \ell^\prime$, we can compute Eq.~\eqref{eq:lenscon} efficiently with FFT methods. 

\begin{figure}
    \includegraphics[width=0.5\columnwidth]{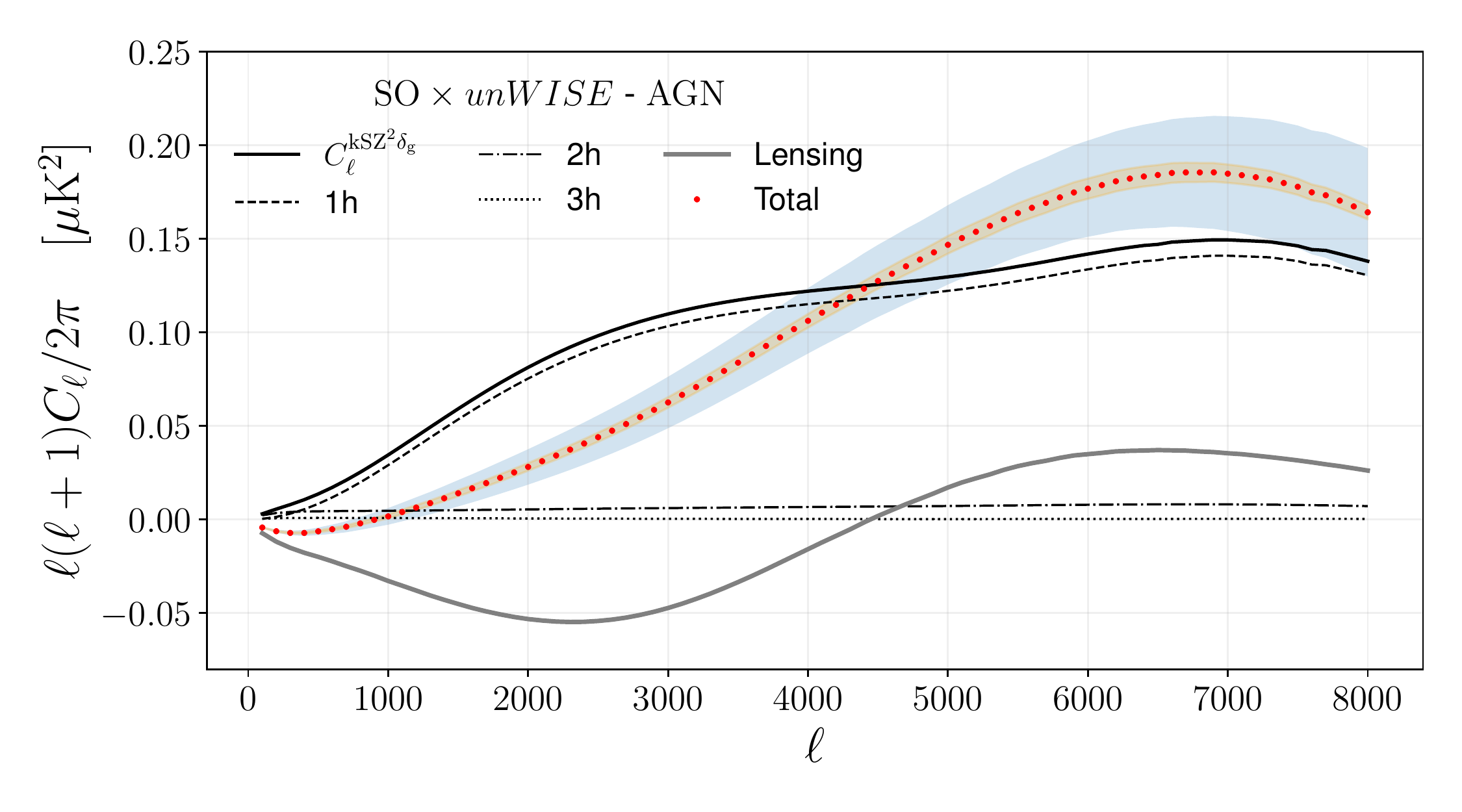} 
    \includegraphics[width=0.5\columnwidth]{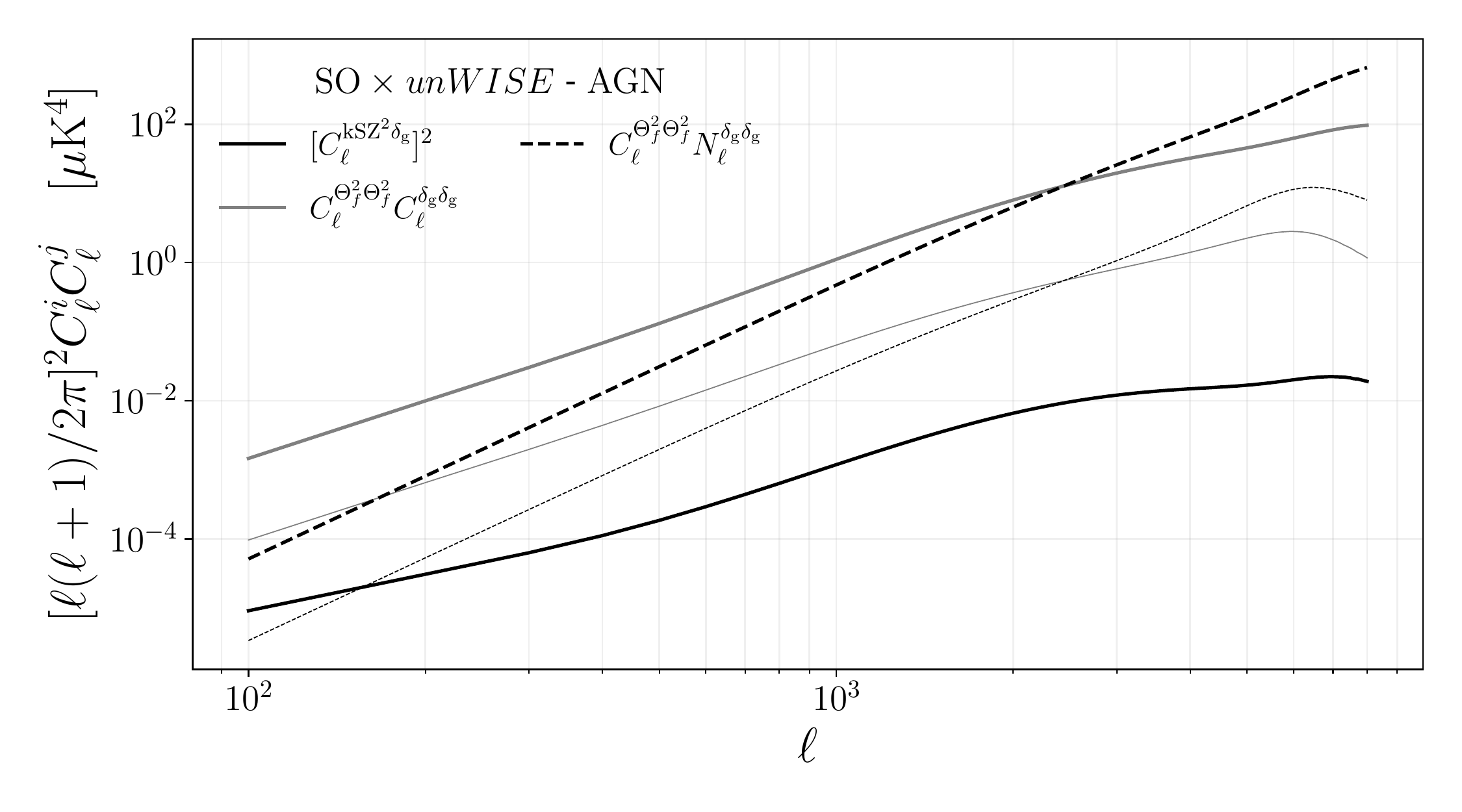} 
    \includegraphics[width=0.5\columnwidth]{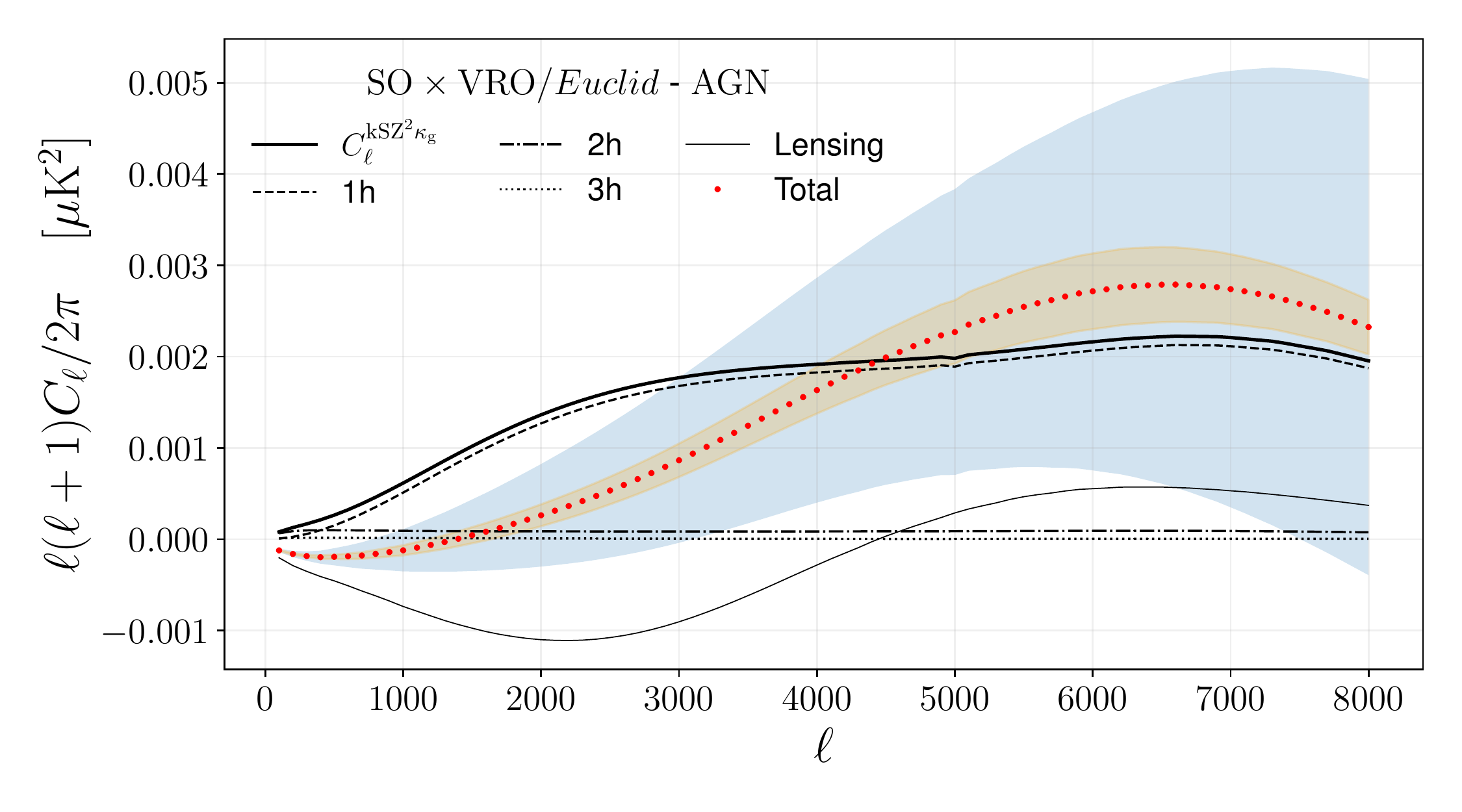} 
    \includegraphics[width=0.5\columnwidth]{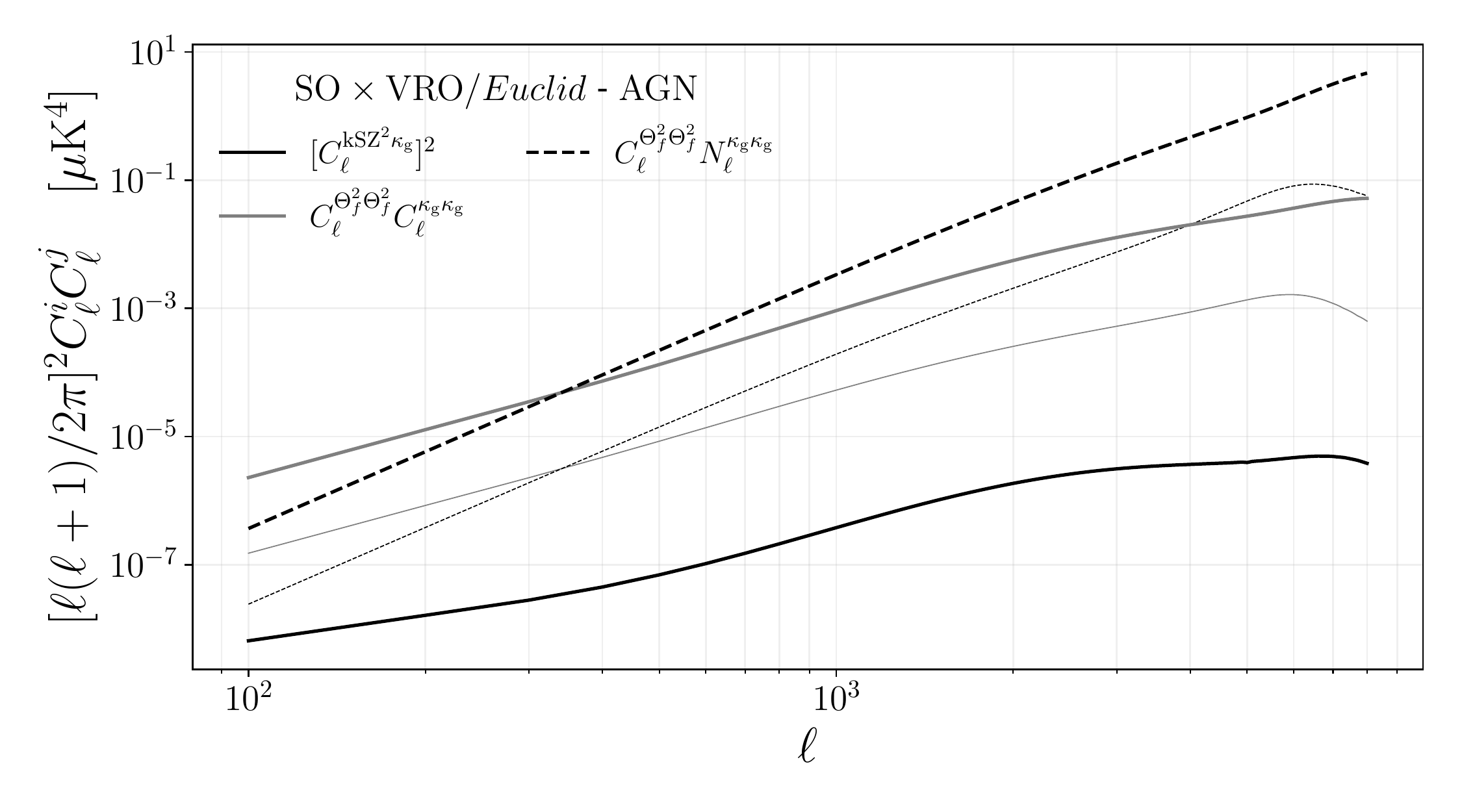}
    \includegraphics[width=0.5\columnwidth]{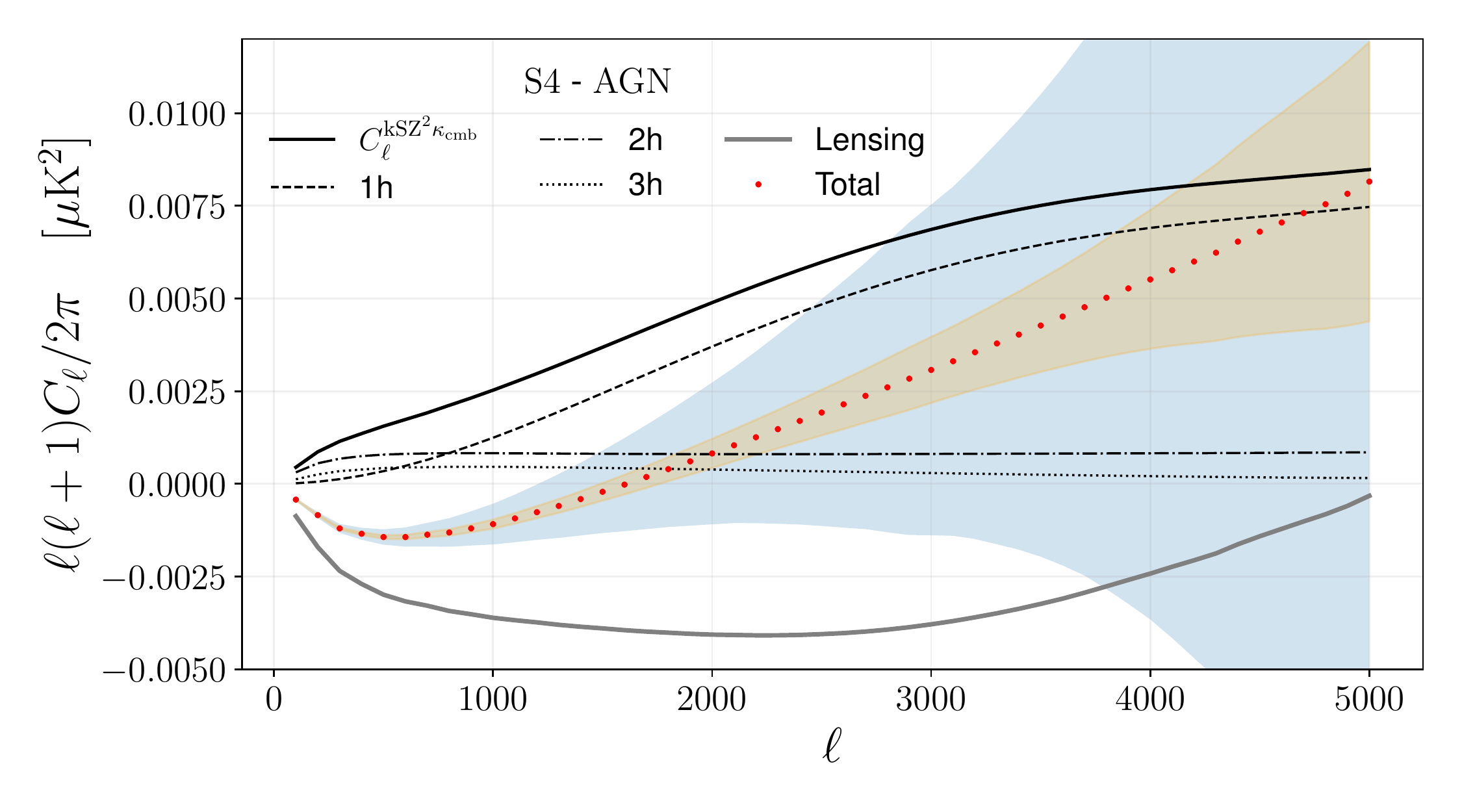} 
    \includegraphics[width=0.5\columnwidth]{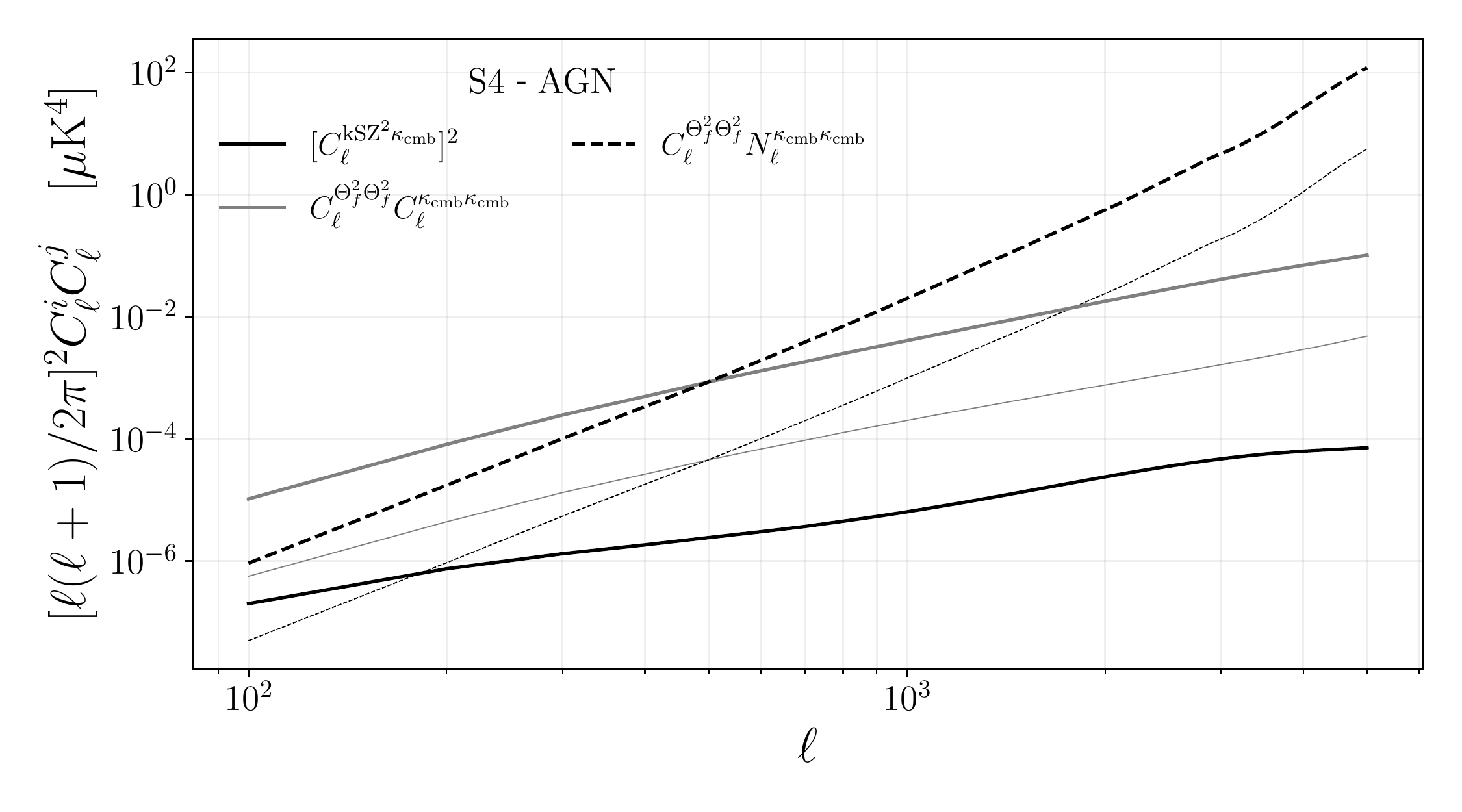}
    \vspace{-0.2cm}
    \caption{Contributions to the projected-field kSZ power spectrum measurements (left column) and covariance matrix (right column) in cross-correlation with galaxy number density (top, assuming \textit{unWISE}), galaxy weak lensing (middle, assuming a VRO/\textit{Euclid}-like survey), and CMB weak lensing (bottom, assuming CMB-S4 CMB lensing). For the CMB map specifications see Table~\ref{tab:exps}, and for the LSS survey specifications see Table~\ref{tab:exps_lss}. In the left plots, the lensing contribution of Eq.~\eqref{eq:lenscon} is the thick grey line; the shaded areas show the statistical uncertainty computed with the covariance matrix of Eq.~\eqref{eq:covmat}. In the right panels, the thin lines show the contributions without the CMB temperature map noise. In the left panels, the orange area indicates the size of the error bars assuming no CMB temperature map noise, i.e., the cosmic-variance-limited CMB measurement. The CMB noise levels are shown in Figure \ref{fig:cls_ksz_tt}. Note that for this spectra we used a linear binning with $\Delta\ell=100$. }
    \label{fig:fls_cls_ksz2kg_cov} 
\end{figure}

The lensing contribution is shown as the thick grey line in the left panels of Figure~\ref{fig:fls_cls_ksz2kg_cov}. At low-$\ell$ this contribution is negative and generally larger than the kSZ term. At larger $\ell$ the lensing contribution remains smaller than the kSZ contribution for all the cases shown here. In the \emph{Planck}-based analyses of \cite{Hill2016} and \cite{Kusiak:2021hai} the lensing contribution dominated over most of the $\ell$-range. The difference is that here we are showing predictions for SO and CMB-S4 CMB maps, which probe scales beyond $\ell\sim3000$ where the kSZ effect dominates the anisotropy (see Figure~\ref{fig:cls_ksz_tt}), unlike \textit{Planck}, for which the lensed CMB dominates over the kSZ signal on the relevant scales.

\begin{table}
\begin{centering}
\begin{tabular}{c|c|lcc}
Experiment& Tracer & Specifications
\tabularnewline
\hline 
\textit{unWISE} & $\delta_\mathrm{g}$&$f_\mathrm{sky}=0.6$; \textit{blue} HOD from \cite{Kusiak:2022xkt}; $N_\ell^{\delta_\mathrm{g}\delta_\mathrm{g}}=8.94\times 10^{-8}$ ($\bar{n}=3409\,\mathrm{deg}^{-2}$)\tabularnewline
DES-like&$\kappa_\mathrm{g}$ & $f_\mathrm{sky}=0.12$; $n_s=8\,\mathrm{arcmin}^{-2}$, $\sigma_\gamma^2=0.3$; $n(z)$= sources bin 3 \citep[][]{za2021}  \tabularnewline
VRO/\textit{Euclid}-like & $\kappa_\mathrm{g}$ & $f_\mathrm{sky}=f_\mathrm{sky}^{_\mathrm{CMB}}$; $n_s=40\,\mathrm{arcmin}^{-2}$, $\sigma_\gamma^2=0.3$; $n(z)$= sources bin 3 \citep[][]{za2021}  \tabularnewline
SO CMB Lensing&$\kappa_\mathrm{cmb}$& truncation at $\ell_\mathrm{max}=5000$; noise curve from online repository (see footnote \ref{fn:solens}).\tabularnewline
CMB-S4 CMB Lensing&$\kappa_\mathrm{cmb}$& truncation at $\ell_\mathrm{max}=5000$; noise curve from online repository (see footnote \ref{fn:s4lens}).
\end{tabular}
\par\end{centering}
\caption{Specifications for the LSS survey experiments considered in our forecast analysis. For VRO/\textit{Euclid},  $f_\mathrm{sky}=f_\mathrm{sky}^{_\mathrm{CMB}}$ assumes that the sky area is limited by the sky coverage and mask of the CMB map (see Table~\ref{tab:exps} for the $f_\mathrm{sky}^{_\mathrm{CMB}}$ values).}
\label{tab:exps_lss}
\end{table}

\begin{figure}
    \includegraphics[width=1.\columnwidth]{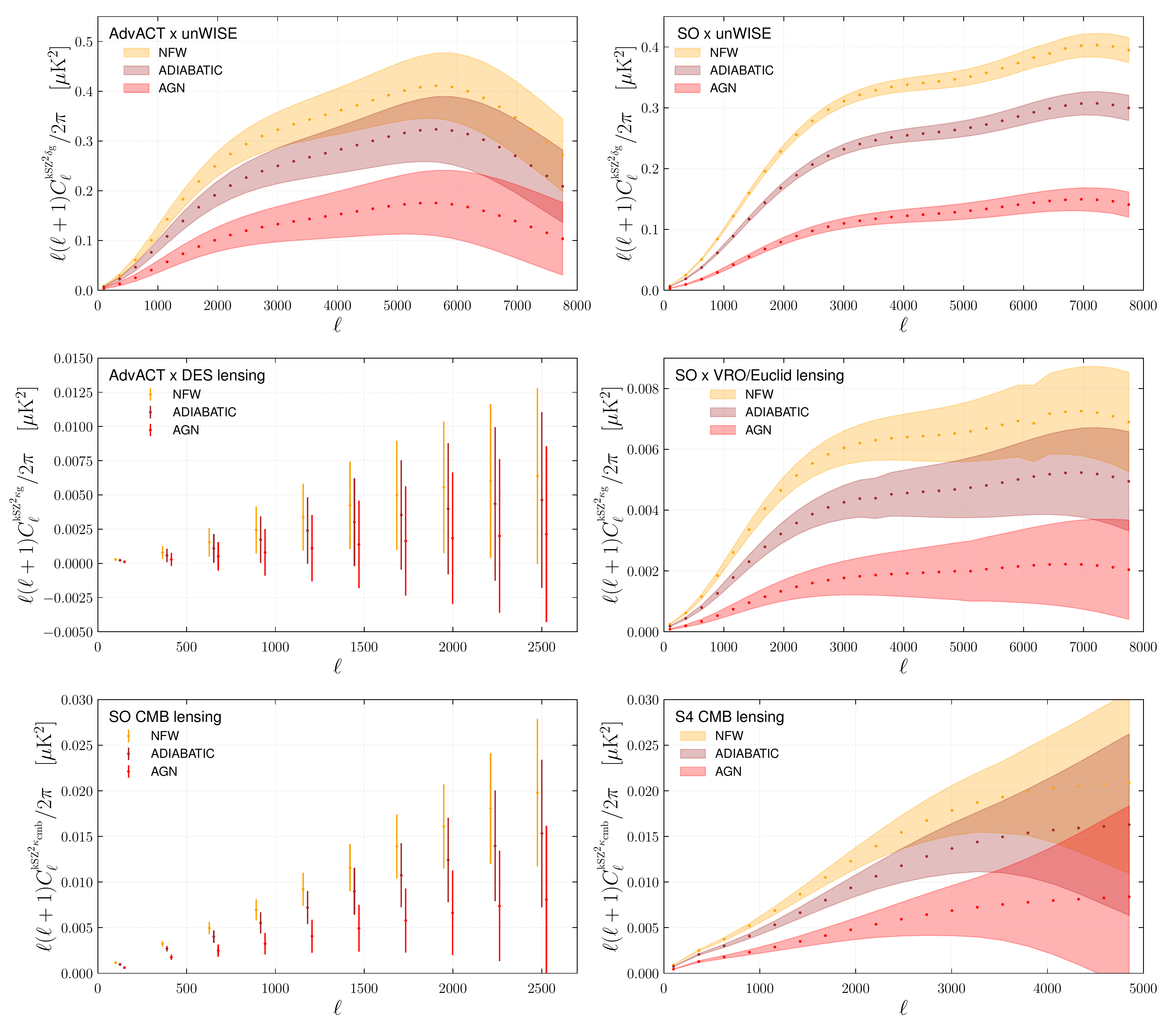}
    \vspace{-0.8cm}
    \caption{Projected-field kSZ power spectrum in cross-correlation with galaxy number density (top, assuming \textit{unWISE}), galaxy weak lensing (middle, assuming a DES-like and a VRO/\textit{Euclid}-like survey), and CMB weak lensing (bottom, assuming SO and CMB-S4). For the CMB map specifications see Table~\ref{tab:exps} and for the LSS survey specifications see Table~\ref{tab:exps_lss}. Each color corresponds to a different gas profile assumption: \textit{NFW}, \textit{Adiabatic}, or \textit{AGN Feedback} (see Table~\ref{tab:gnfwb16} and Eq.~\ref{eq:ugnfw}). The shaded areas (or error bars) represent the statistical uncertainty computed with the covariance matrix of Eq.~\eqref{eq:covmat}. The spectra in this figure are computed with a bin size of $\Delta \ell = 264$.} 
    \label{fig:fls_cls_varprof_varhod}
\end{figure}
\section{Forecasts}\label{s:forecasts}

In this section we forecast the detection signal-to-noise for the projected-field kSZ power spectrum signal and for some of the gas profile parameters, including its normalization (determined by $f_\mathrm{b}f_\mathrm{free}$), the inner slope $\alpha$, and the outer slope $\beta$ (see Subsection \ref{ss:hybrid_bispec}). Our forecasts are obtained with the Fisher matrix formalism \citep[e.g.,][]{Tegmark_1997}. The results depend on the characteristics of the CMB and LSS tracer maps which enter the covariance matrix calculation (resolution, noise, sky coverage and overlap, etc). We consider four CMB maps:
\begin{itemize}
\item\textbf{\textit{Planck}} \citep[][]{Planck2018}, as in \cite{Hill2016} and \cite{Kusiak:2021hai}. In our fiducial calculations we compute the noise with $\Delta_T=47\,\mu\mathrm{K}$-arcmin, matching that in the LGMCA CMB map~\cite{Bobin2016}, and use $f_\mathrm{sky}=0.6$. For the beam we use $\Theta_\mathrm{FWHM}=5$ arcmin. In the Wiener filter we set $\ell_\mathrm{max}=3000$.

\item \textbf{AdvACT} \citep[][]{Henderson_2016} for upcoming maps from ACT \citep[e.g.,][]{Aiola2020}. We use $\Delta_T=20\,\mu\mathrm{K}$-arcmin and $f_\mathrm{sky}=0.3$ for the noise and covariance matrix, with $\Theta_\mathrm{FWHM}=1.5$ arcmin and $\ell_\mathrm{max}=8000$ for the beam and Wiener filter. The noise level here has been inflated over the raw temperature map noise in order to account for the effects of component separation, using the same ILC methodology as applied in the SO and CMB-S4 calculations below.

\item \textbf{Simons Observatory} \citep[SO;][]{SO2019}, for next-generation CMB maps. We use $f_\mathrm{sky}=0.4$ in the covariance matrix, with $\Theta_\mathrm{FWHM}=1.4$ arcmin and $\ell_\mathrm{max}=8000$ for the beam and Wiener filter. For the noise, we use the post-ILC component-separated noise curves constructed in \cite{SO2019} (see Sec.~2.4-2.5 of that work), which are available online.\footnote{\href{https://github.com/simonsobs/so_noise_models/tree/master/LAT_comp_sep_noise/v3.1.0}{https://github.com/simonsobs/so\_noise\_models/tree/master/LAT\_comp\_sep\_noise/v3.1.0}\\(filename: SO\_LAT\_Nell\_T\_atmv1\_goal\_fsky0p4\_ILC\_CMB.txt --- We use the deproj-0 case, i.e., standard ILC.)} These noise curves are constructed using a harmonic ILC method applied to simulated sky maps containing all relevant Galactic and extragalactic foregrounds.  For simplicity, we use the standard minimum-variance ILC noise curves here, although in an actual analysis it may be necessary to apply constrained ILC methods to deproject contamination arising from the thermal SZ effect or the cosmic infrared background.

\item \textbf{CMB-Stage 4} \citep[CMB-S4;][]{CMBS4DSR}, with the same assumptions as SO, but different post-ILC component-separated noise curves. The CMB-S4 noise curves are also publicly available online\footnote{\url{https://sns.ias.edu/~jch/S4_190604d_2LAT_Tpol_default_noisecurves.tgz} or \href{https://github.com/msyriac/orphics/tree/master/data}{https://github.com/msyriac/orphics} (filename: S4\_190604d\_2LAT\_T\_default\_noisecurves\_deproj0\_SENS0\_mask\_16000\_ell\_TT\_yy.txt). See also the wiki page \url{https://cmb-s4.uchicago.edu/wiki/index.php/Survey\_Performance\_Expectations} for further information.} and were computed with the same methodology as the SO post-ILC noise curves (see Appendix A.3 of~\cite{CMBS4DSR}).
\end{itemize}
These specifications are summarized in Table~\ref{tab:exps} and the CMB temperature map noise curves are shown in Figure~\ref{fig:cls_ksz_tt}. For the LSS survey data, we consider five configurations for different tracers:
\begin{itemize}
    \item \textbf{\textit{unWISE}} \citep[][]{Schlafly_2019} galaxy number density, i.e., $X=\delta_\mathrm{g}$,  as in \cite{Alex,Kusiak:2021hai}. For halo-model calculations, we characterize the galaxy-halo connection using the HOD results for the \textit{blue sample} obtained in \cite{Kusiak:2022xkt}. See Appendix~\ref{sss:hod} for details on the HOD parameters. For the shot noise we set $N_\ell^{\delta_\mathrm{g}\delta_\mathrm{g}}=8.94\times 10^{-8}$, which corresponds to $1/\bar{n}$ with $\bar{n}=3409\,\mathrm{deg}^{-2}$ \citep[from][]{Alex} converted to steradians.
    
    \item \textbf{DES}-like galaxy weak lensing, i.e., $X=\kappa_\mathrm{g}$, assuming a galaxy weak lensing survey similar to DES-Y3 \citep[e.g.,][]{Sevilla_Noarbe_2021}, with $f_\mathrm{sky}=0.12$ (i.e., 5000 deg$^2$ sky area) a source galaxy number density $n_s=8\,\mathrm{arcmin}^{-2}$ and shape noise $\sigma_\gamma^2=0.3$. For the source galaxy redshift distribution, we take the distribution of the third redshift bin of the Redmagic sample peaking at $z\approx0.7$ \citep[the dashed red line in the top panel of Figure 3 of ][]{za2021}.
    
    \item \textbf{VRO/\textit{Euclid}}-like galaxy weak lensing, i.e., $X=\kappa_\mathrm{g}$, assuming a galaxy weak lensing survey similar to VRO \citep{LSSTDarkEnergyScience:2018yem} or \textit{Euclid} \citep[][]{EC2020}, with the same source distribution and shape noise as above, but a galaxy source density five times larger, $n_s=40\,\mathrm{arcmin}^{-2}$, and full overlap with the CMB maps.
    
    \item \textbf{SO} CMB lensing, i.e., $X=\kappa_\mathrm{cmb}$, with the reconstructed CMB lensing noise curve from the online repository, which is built by applying the lensing quadratic estimator to the same post-ILC noise curves described above.\footnote{\href{https://github.com/simonsobs/so_noise_models/blob/master/LAT_lensing_noise/lensing_v3_1_1/nlkk_v3_1_0_deproj0_SENS2_fsky0p4_it_lT30-3000_lP30-5000.dat}{(SO-Goal) https://github.com/simonsobs/so\_noise\_models/blob/master/LAT\_lensing\_noise/lensing\_v3\_1\_1}. (filename: ../nlkk\_v3\_1\_0\_deproj0\_SENS2\_fsky0p4\_it\_lT30-3000\_lP30-5000.dat)\label{fn:solens}} Note that we truncate the power spectra at $\ell_\mathrm{max}=5000$ (since there is no SNR beyond this multipole).
    
    \item \textbf{CMB-S4} CMB lensing, i.e., $X=\kappa_\mathrm{cmb}$, with the reconstructed CMB lensing noise curve from the online repository and truncating at $\ell_\mathrm{max}=5000$.\footnote{ \href{https://github.com/toshiyan/cmblensplus/tree/master/example/data}{(CMB-S4) https://github.com/toshiyan/cmblensplus/tree/master/example/data}.\label{fn:s4lens}} 
\end{itemize}
These specifications are summarized in Table~\ref{tab:exps_lss} and the SO and CMB-S4 CMB lensing noise curves are shown in Figure~\ref{fig:cls_kk} of the Appendix.

The Fisher matrix is computed using the analytical covariance matrix (see Subsection \ref{ss:covmat})  and the halo model predictions for the projected-field kSZ power spectrum (see Subsection \ref{ss:hybrid_bispec}). With the lensing contribution included (see Subsection \ref{ss:cmb_lens_cont}), the total signal is $\bar{C}_\ell^{\mathrm{kSZ}^2X}=C_\ell^{\mathrm{kSZ}^2X}+ C_\ell^{\Theta_u^2\psi X}$. (Note the over-bar to distinguish the total signal from the kSZ signal only.) Then the Fisher matrix elements are
\begin{equation}
    F_{ij}=\sum_{aa^\prime}\frac{\partial \bar{C}_{a}^{\mathrm{kSZ}^2X}}{\partial p_i} [M^{-1}]_{a{a^\prime}}\frac{\partial \bar{C}_{a^\prime}^{\mathrm{kSZ}^2X}}{\partial p_j}\label{eq:fishmat}
\end{equation}
where $p$ denotes the vector of varied parameters, $i,j$ are the varied parameter indices, and $a$ is the multipole bin index. With this, the marginalized 1$\sigma$ uncertainty on a parameter $p_i$ is given by $\sqrt{[F^{-1}]_{ii}}$. The binning of the covariance matrix is discussed in Subsection~\ref{ss:covmat}. For the binned power spectrum we simply assume that is it well approximated by the value at the center of the bin. Note that we use a linear binning with $\Delta\ell=564$ and $\ell_\mathrm{min}=100$ for the Fisher forecasts.  

To start, we compute the total detection signal-to-noise ratio, SNR$_\mathrm{tot}$. We can define a single parameter, $A_\mathrm{kSZ^2}$, such that  $C_\ell^{\mathrm{kSZ}^2X}=A_\mathrm{kSZ^2}C_\ell^{\mathrm{kSZ}^2X,fid.}$ where  $C_\ell^{\mathrm{kSZ}^2X,fid.}$ is the fiducial prediction, corresponding to $A_\mathrm{kSZ^2}=1$. In this case, the derivatives in Eq.~\eqref{eq:fishmat} are straightforward and the Fisher calculation yields
\begin{equation}
     \mathrm{SNR}_\mathrm{tot} = \left(\frac{\Delta{A}_{\mathrm{kSZ^2}}}{{A}_{\mathrm{kSZ^2}}}\right)^{-1}=\sqrt{F}=\left[\sum_a \frac{(C_a^{\mathrm{kSZ}^2X,fid.})^2}{M_{aa}}\right]^{1/2}.
\end{equation}
Results are reported in the first column of Table~\ref{tab:forecasts_SNRtot_hm}. The highest SNR values are obtained for galaxy number density cross-correlation. For \textit{Planck} $\times$ \textit{unWISE} we find $\mathrm{SNR}_\mathrm{tot}=1.7$, consistent with the measurement in \cite{Kusiak:2021hai}, which reported $\mathrm{SNR}_\mathrm{tot}=1.35$ (for the \textit{unWISE} \textit{blue} galaxy sample).
With data that is already on hand, e.g., AdvACT CMB maps and \textit{unWISE} galaxies, we forecast a detection SNR over 10$\sigma$. In the next decade, with CMB-S4 $\times$ \textit{unWISE}, the SNR should reach $\gtrsim 100\sigma$. For galaxy-density cross-correlation, we note that $A_\mathrm{kSZ^2}$  is degenerate with the galaxy bias, $b_\mathrm{g}$ (which also determines the amplitude of the lensing term). Nonetheless, the galaxy bias can be tightly constrained by galaxy clustering ($C_\ell^{\delta_\mathrm{g}\delta_\mathrm{g}}$) and galaxy-lensing cross-correlation ($C_\ell^{\delta_\mathrm{g}\kappa_\mathrm{g}}$) measurements. If we set a Gaussian prior on $b_\mathrm{g}$ corresponding to the measurements of \cite{Alex}, we find that the $\mathrm{SNR}_\mathrm{tot}$ forecasts only change marginally. 

Galaxy weak lensing cross-correlation based on DES data ($\mathrm{SNR_{tot}}\approx10$ with CMB-S4) do not reach SNR values as high as with galaxy density (due to the influence of shape noise), but with VRO/\textit{Euclid} we expect clear detections of this cross-correlation as well ($\mathrm{SNR_{tot}}\approx30$ with CMB-S4). A detection with CMB lensing cross-correlation will certainly require next-generation CMB maps such as those from SO and CMB-S4, for which we find $\mathrm{SNR_{tot}}\approx16$ and $\mathrm{SNR_{tot}}\approx35$, respectively.

With such high SNR detections, we can ask how well different gas profile shapes could be distinguished. In Figure \ref{fig:fls_cls_varprof_varhod} we see that galaxy density cross-correlation using \textit{unWISE} will be able to exclude particular gas profile models at high significance. It appears to be more challenging with galaxy weak lensing or CMB lensing cross-correlations, but forecasts with SO $\times$ VRO/\textit{Euclid} (galaxy weak lensing) and CMB-S4 (CMB weak lensing) are promising. 

\begin{figure}
    \includegraphics[width=1.\columnwidth]{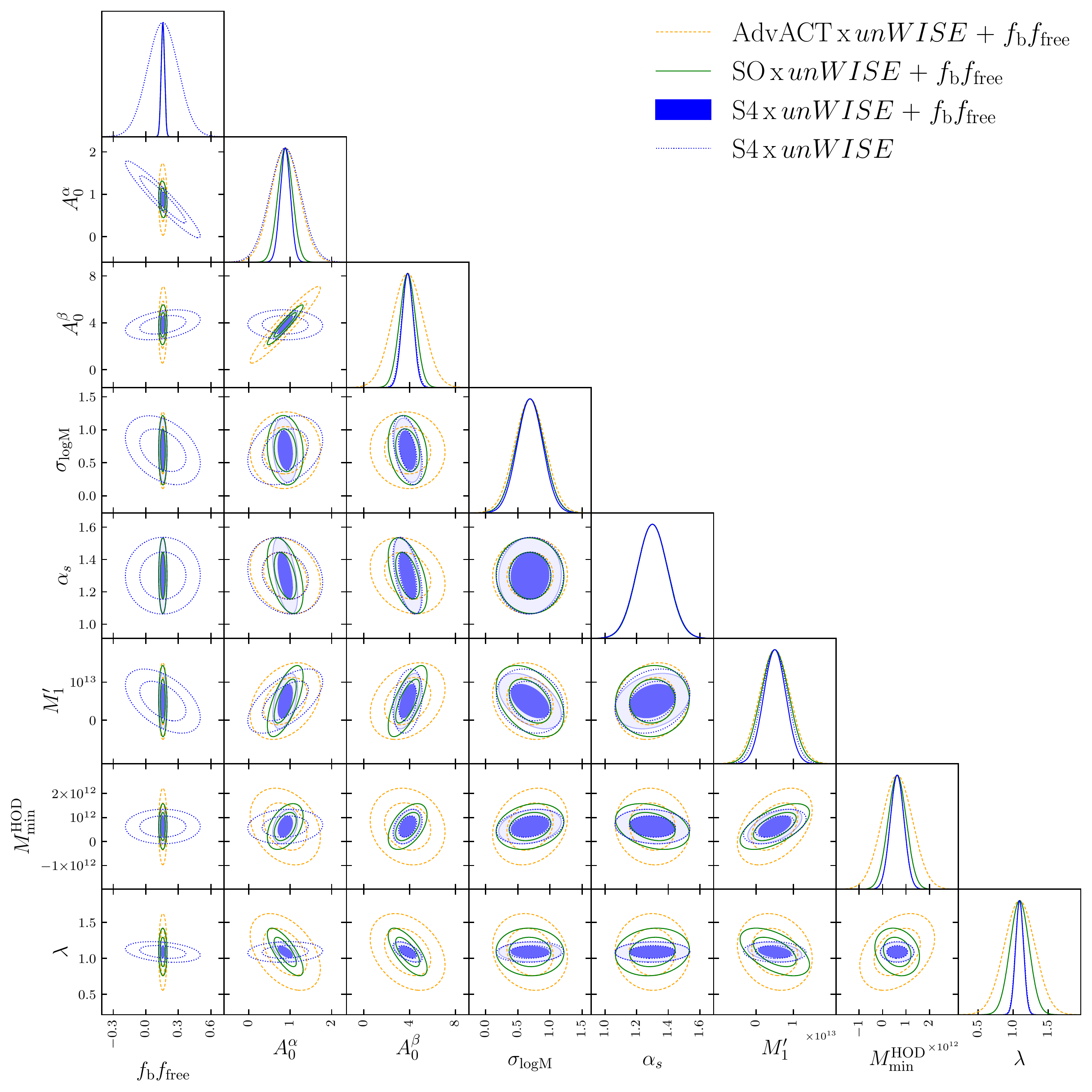}
    \vspace{-0.6cm}
    \caption{2D marginalized posterior probability distribution from Fisher matrix calculations for kSZ$^2$-galaxy density cross-correlation. Galaxies are assumed to be the \textit{unWISE blue} sample. The orange and green lines assume AdvACT and SO CMB maps, respectively, as well as a 10\% prior on $f_\mathrm{free}$. The blue contours are with CMB-S4 CMB maps with (filled) and without (empty) the prior on $f_\mathrm{b}f_\mathrm{free}$.}  
    \label{fig:TTG-forecast}
\end{figure}

\begin{figure}
    \includegraphics[width=0.5\columnwidth]{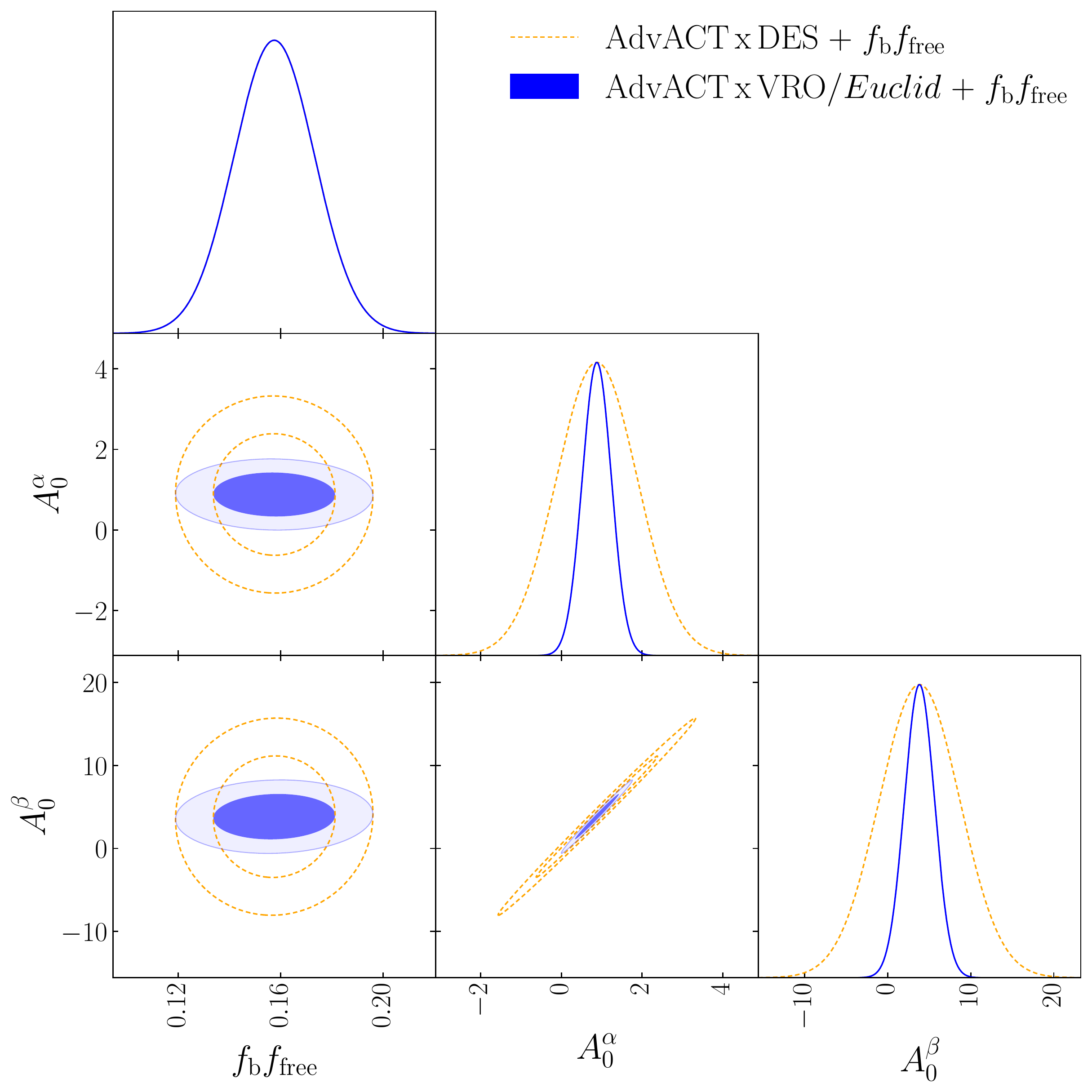}\includegraphics[width=0.5\columnwidth]{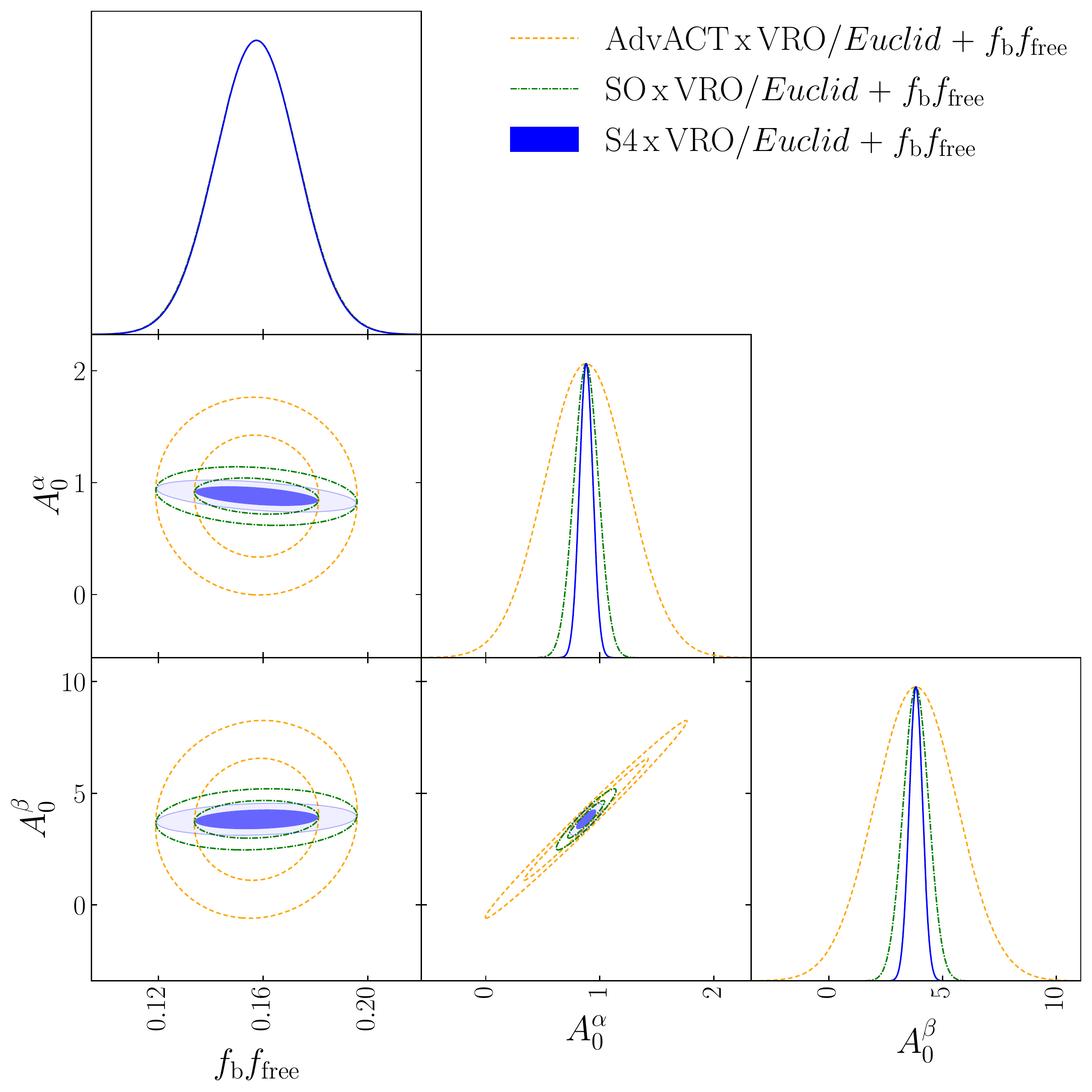}
    \includegraphics[width=0.5\columnwidth]{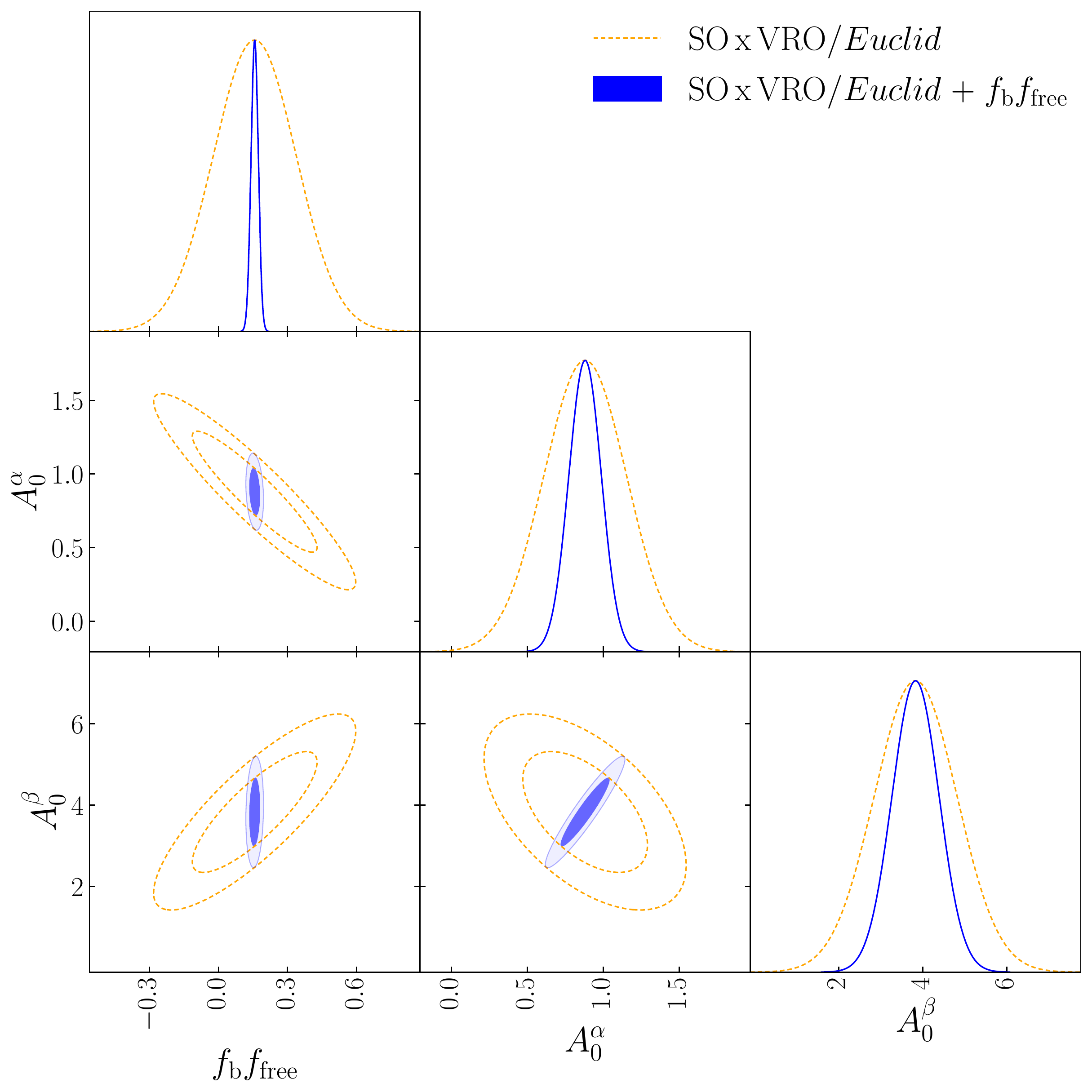}\includegraphics[width=0.5\columnwidth]{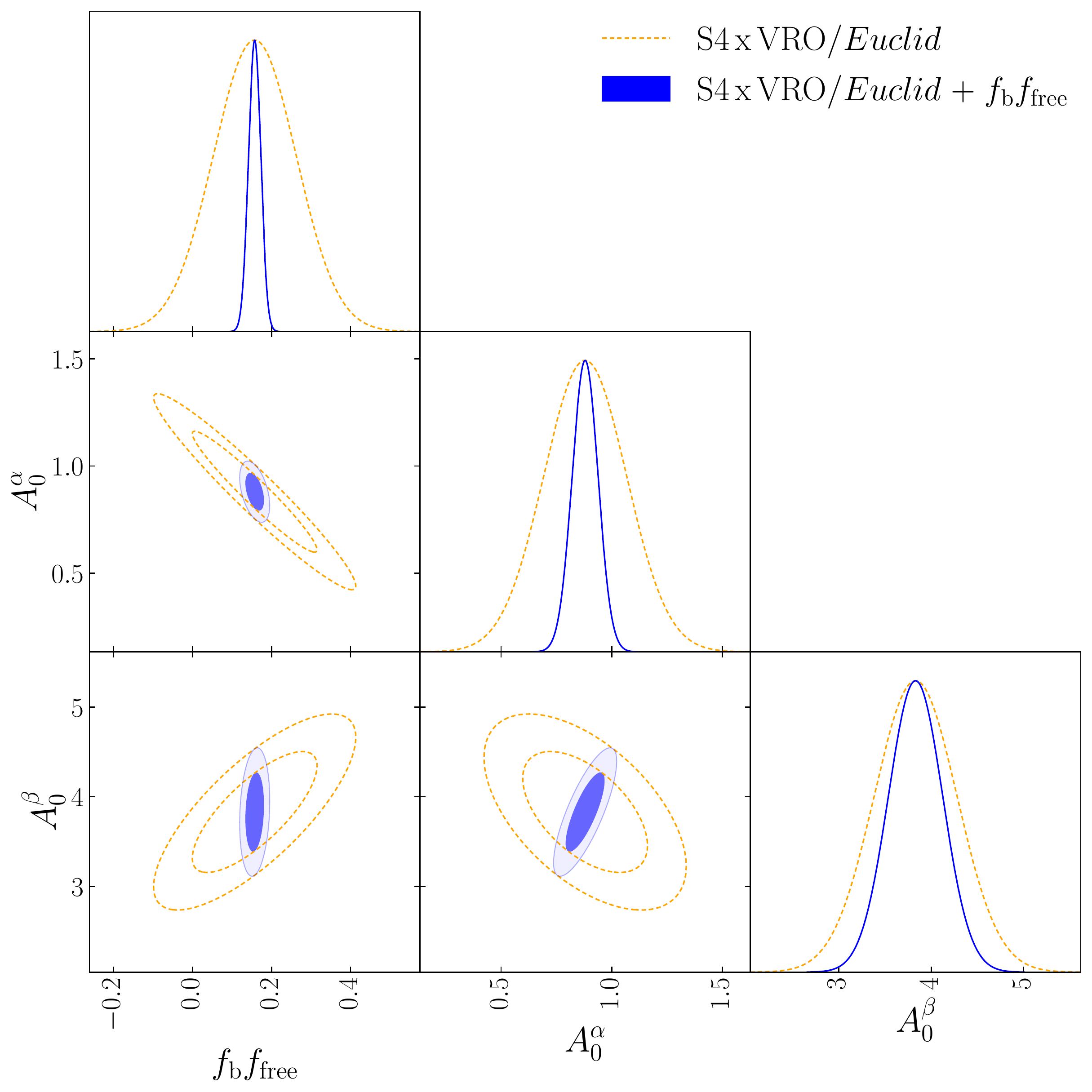}
    \vspace{-0.2cm}
    \caption{2D marginalized posterior probability distributions from Fisher matrix calculations for kSZ$^2$-galaxy weak lensing cross-correlation. (\textit{Top Left:}) AdvACT CMB maps with DES (orange) and VRO/\textit{Euclid} (blue) galaxy weak lensing, with a 10$\%$ prior on $f_\mathrm{free}$ in both cases.  (\textit{Top right:}) AdvACT (orange), SO (green), and CMB-S4 (blue) CMB maps with VRO/\textit{Euclid} galaxy weak lensing, with a 10$\%$ prior on $f_\mathrm{b}f_\mathrm{free}$ in all cases. (\textit{Bottom Left:}) SO CMB maps with VRO/\textit{Euclid} galaxy weak lensing, with (blue) and without (orange) a 10$\%$ prior on $f_\mathrm{free}$. (\textit{Bottom Right:}) Same as bottom left but with CMB-S4 CMB maps.}  
    \label{fig:TTKg-forecast}
\end{figure}

\begin{figure}
    \begin{center}
    \includegraphics[width=0.6\columnwidth]{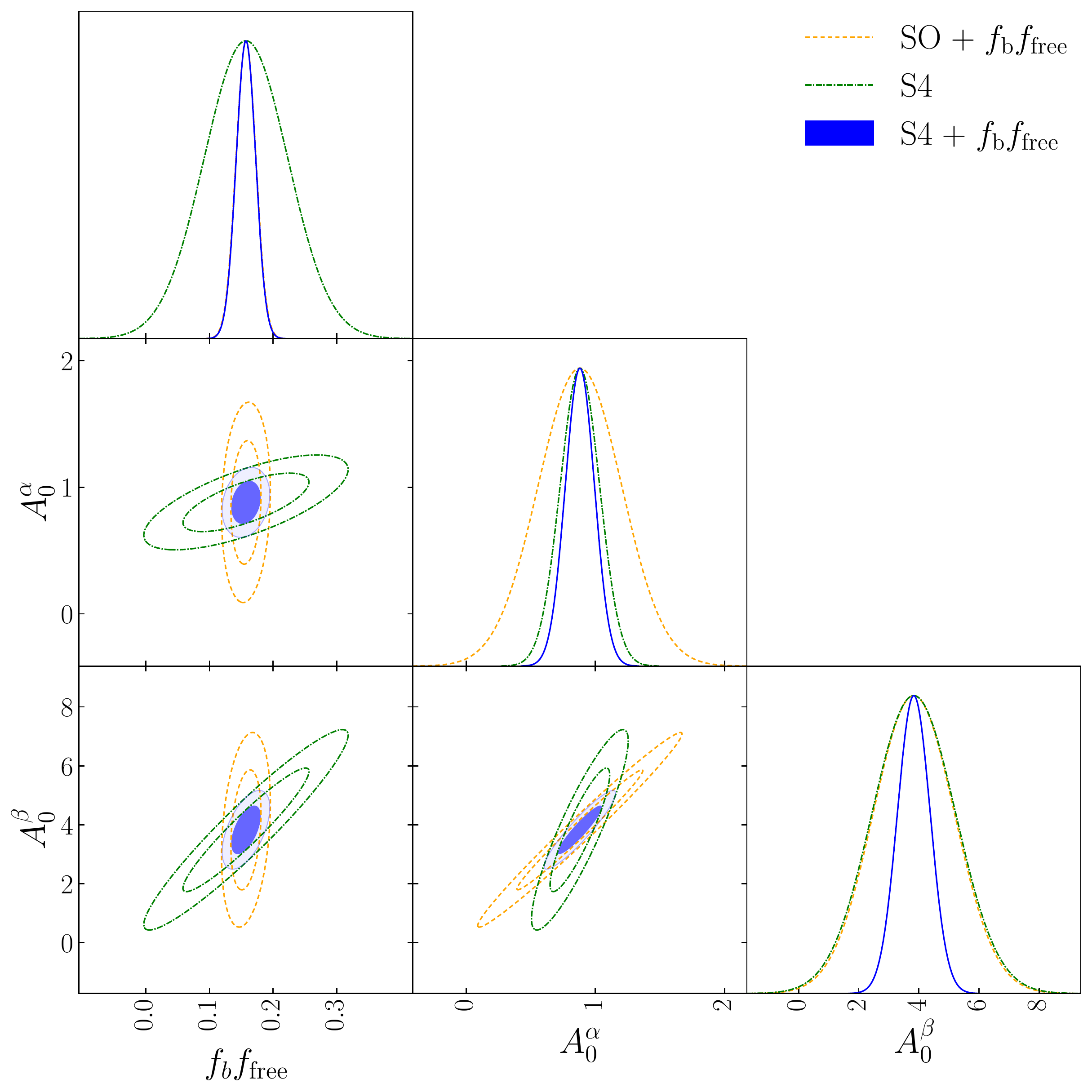}
    \end{center}
    \vspace{-0.1cm}
    \caption{2D marginalized posterior probability distribution from Fisher matrix calculations for kSZ$^2$-CMB lensing cross-correlation. The orange line assumes SO CMB maps and a 10\% prior on $f_\mathrm{b}f_\mathrm{free}$. The blue and green lines assume CMB-S4 CMB maps with and without a 10\% prior on $f_\mathrm{b}f_\mathrm{free}$, respectively.}  
    \label{fig:TTKcmb-forecast}
\end{figure}

A more refined way to answer this question is to study forecasts on the gas profile parameters. There are three parameters: first, the overall amplitude of the gas density profile (i.e., the parameter $C$ in Eq.~\ref{eq:ugnfw}  which is completely degenerate with the product of the free electron and baryon fractions, $f_\mathrm{b}f_\mathrm{free}$; second, the parameter $\alpha$ controlling the inner slope of the profile; third, the parameter $\beta$ controlling the outer slope of the profile. We compute the Fisher matrix for these gas profile parameters. To do so, we evaluate the derivatives in Eq.~\eqref{eq:fishmat} around the fiducial model. The Fisher confidence ellipses are shown in Figure~\ref{fig:TTG-forecast} for cross-correlation with galaxy number density, in Figure~\ref{fig:TTKg-forecast} for galaxy weak lensing, and in Figure~\ref{fig:TTKcmb-forecast} for CMB weak lensing. The marginalized constraints are reported in Table~\ref{tab:forecasts_SNRtot_hm}. Since the baryon fraction at large halo-centric radii should be close to the cosmological value $\Omega_\mathrm{b}/\Omega_\mathrm{m}$ and the free electron fraction close to unity because the universe is fully reionized at low redshift, we can safely assume an external constraint on $f_\mathrm{b}f_\mathrm{free}$. In Table~\ref{tab:forecasts_SNRtot_hm} the forecasts where we have imposed a 10\% prior on $f_\mathrm{b}f_\mathrm{free}$ are reported in parentheses. 

\begin{table}
\begin{centering}
\begin{tabular}{rlc|ccc}
\tabularnewline
  & &  SNR$_\mathrm{tot}$ & $(\frac{\Delta A_0^\mathrm{\beta}}{A_0^\mathrm{\beta}})^{-1}$ & $(\frac{\Delta A_0^\mathrm{\alpha}}{A_0^\mathrm{\alpha}})^{-1}$ & $(\frac{\Delta f_\mathrm{free}}{f_\mathrm{free}})^{-1}$\\[1ex] 
\hhline{======}
\ldelim\{{4}{5mm}[$\delta_\mathrm{g}$]& \textit{Planck} $\times$ \textit{unWISE} \dotfill & 1.7 & 0.18 (0.37) & 0.29 (0.38) & 0.19 (10)\tabularnewline
& AdvACT $\times$ \textit{unWISE} \dotfill & 17.8 & 1.72 (2.87) & 2.22 (2.54)& 0.71 (10)\tabularnewline
& SO $\times$ \textit{unWISE} \dotfill & 61.9 & 3.70 (5.51) & 2.07 (4.98) & 0.78 (10)\tabularnewline
& CMB-S4 $\times$ \textit{unWISE} \dotfill & 102.9 & 7.32 (7.83) & 2.38 (7.18) &  1.12 (10)\tabularnewline
\hhline{======}
\ldelim\{{6}{5mm}[$\kappa_\mathrm{g}$] & AdvACT $\times$ DES  \dotfill & 2.24 & 0.28 (0.79) & 0.59 (0.88) &  0.09 (10)\tabularnewline
& AdvACT $\times$ VRO/\textit{Euclid} \dotfill & 5.98 & 0.92 (2.11) & 1.72 (2.44) & 0.31 (10)\tabularnewline
& SO $\times$ DES \dotfill & 6.14 & 1.03 (2.75) & 0.93 (2.34) & 0.23 (10)\tabularnewline
& SO $\times$ VRO/\textit{Euclid}  \dotfill & 18.81 & 3.89 (6.84) & 3.24 (8.22) & 0.88 (10)\tabularnewline
& CMB-S4 $\times$ DES  \dotfill & 9.71 & 2.19 (4.36) & 1.33 (5.23) & 0.40 (10)\tabularnewline
& CMB-S4 $\times$ VRO/\textit{Euclid} \dotfill & 29.72 & 8.57 (13.07) & 4.71 (15.08) & 1.51 (10)\tabularnewline
\hhline{======}
\ldelim\{{2}{8mm}[$\kappa_\mathrm{cmb}$] & SO  \dotfill & 16.39 & 0.92 (2.84) & 1.72 (2.72) & 0.94 (10)\tabularnewline
& CMB-S4 \dotfill & 34.52 & 2.76 (7.01) & 5.75 (7.79) & 2.4 (10)\tabularnewline
\end{tabular}
\par\end{centering}
\caption{Forecasts for projected-field kSZ power spectrum detection significance (first column), gNFW parameters for the gas profile (second and third column), and the free electron fraction (last column). Numbers quoted in parentheses are computed assuming a 10\% prior on $f_\mathrm{free}$. Note that we assume our fiducial \textit{AGN feedback} model in all calculations here (see Table \ref{tab:gnfwb16}). Note that $\delta_g$ forecasts have been marginalized over HOD parameters, while this is not necessary for the lensing cross-correlation forecasts (since they do not depend on galaxy HOD's). }
\label{tab:forecasts_SNRtot_hm}
\end{table} 

For cross-correlation with galaxy density, we also take into account the uncertainties on the HOD parameters (see Appendix~ \ref{sss:hod}) by using priors corresponding to the marginalized 1$\sigma$ uncertainty found in \cite{Kusiak:2022xkt}. In this case, the peculiar scale dependence of the lensing contribution, which depends on HOD parameters, also helps to break degeneracies with the gas profile parameters (see Figure~\ref{fig:TTG-forecast} for the contour plots including HOD parameters). 

For all the cross-corelations (galaxy density, galaxy lensing, and CMB lensing) we see that $\alpha$ and $\beta$ are always degenerate. This is because a larger $\alpha$ can be compensated by a larger $\beta$, consistent with the results of Figure \ref{fig:varying_alpha_beta}, which shows that these parameters mainly change the amplitude of the projected field power spectrum in a rather scale-independent manner. That means that the slope parameters are also strongly degenerate with $f_\mathrm{b}f_\mathrm{free}$ (see dotted blue contours in Figure~\ref{fig:TTG-forecast},  dotted orange contours in Figure~\ref{fig:TTKg-forecast}, and dotted green contours in Figure~\ref{fig:TTKcmb-forecast}). 
These degeneracies can be  significantly reduced with a prior on $f_\mathrm{b}f_\mathrm{free}$. Still, it is worth noting that even without external information on $f_\mathrm{b}f_\mathrm{free}$, CMB-S4 CMB maps should allow high significance measurements of $\alpha$ and $\beta$. Indeed, for CMB-S4 $\times$ \textit{unWISE} (galaxy density cross-correlation) we forecast a 7$\sigma$ measurement of $\beta$. For CMB-S4 $\times$ VRO/\textit{Euclid} (galaxy weak lensing cross-correlation) we forecast a 9$\sigma$ measurement of $\beta$ and a 5$\sigma$ measurement of $\alpha$. And for the CMB-S4 kSZ${}^2$--CMB lensing cross-correlation we find that, unlike galaxy density or galaxy lensing cross-correlation, the inner slope is better measured ($6\sigma$) than the outer slope ($3\sigma$).    

As can be seen in Figure \ref{fig:TTG-forecast}, and in the bottom plots of Figure \ref{fig:TTKg-forecast}, the addition of an external prior on $f_\mathrm{b}f_\mathrm{free}$ allows us to constrain the slope parameters with forthcoming SO CMB maps before CMB-S4 will become available. For SO $\times$ \textit{unWISE} (galaxy density cross-correlation) we forecast 5$\sigma$ measurements of both $\alpha$ and $\beta$. With SO $\times$ VRO/\textit{Euclid} (galaxy weak lensing cross-correlation) we find 7$\sigma$ for $\beta$ and 8$\sigma$ for $\alpha$. With CMB-S4 maps and the prior on $f_\mathrm{b}f_\mathrm{free}$ the significance of the forecasted measurements of $\alpha$ and $\beta$ is roughly two times larger than without the external prior. 

With readily available AdvACT CMB maps, although we forecast high SNR detections of the projected-field power spectrum, a measurement of the slope parameters appears to be more challenging owing to the degeneracies discussed above. With the external prior on $f_\mathrm{b}f_\mathrm{free}$, we forecast measurements of $\alpha$ and $\beta$ at a significance of $\sim 2\sigma$ in cross-correlation with \textit{unWISE} galaxies and VRO/\textit{Euclid} weak lensing data. Nonetheless, we emphasize that our forecasts are conservative in the sense that they rely on the AGN feedback gas density model, which yields a low projected-field power spectrum compared to the NFW and Adiabatic models (see Figure \ref{fig:fls_cls_ttg}, \ref{fig:fls_cls_ttkg}, \ref{fig:fls_cls_ttkc}, and \ref{fig:fls_cls_varprof_varhod} for comparisons of the respective predictions). It is possible that the spatial distribution of the gas around real halos is significantly different than the AGN feedback model and could be measured with the the projected-field power spectrum using AdvACT or SPT CMB maps.

\section{Conclusion}\label{s:conclusions}

Our main results are the first calculation of the projected-field kSZ power spectrum using the halo model formalism and its numerical implementation in \verb|class_sz| (Section~\ref{s:hm}) for galaxy density, galaxy weak lensing, and CMB weak lensing cross-correlations. Improving upon previous works by \cite{Dore2004}, \cite{DeDeo}, and \cite{Ferraro2016} that were based on an \textit{effective approach} for the computation of the hybrid bispectrum (Subsection \ref{ss:eff}), the halo model formalism allows us to access the scale-dependent information on the gas distribution around halos. Since the projected-field power spectrum is an integral of a three-point function (i.e., the hybrid bispectrum), the halo model expressions are a sum of a 1-halo, 2-halo, and 3-halo term. We find that the 1-halo term largely dominates for $\ell\gtrsim 500$ (see Figure~\ref{fig:fls_cls_ttg}, \ref{fig:fls_cls_ttkg}, and \ref{fig:fls_cls_ttkc}).  

An important aspect of the numerical implementation is to take advantage of the separable form of the expressions so we can evaluate integrals as products in Fourier space, using FFTLog methods for the Fourier transforms (Subsection \ref{ss:hmeqs}). This speeds up the computation by a factor of $\sim100$ compared to trapezoidal or quadrature rules. Still, the evaluation of a projected-field power spectrum takes $\mathcal{O}(10\mathrm{s})$ on a laptop. This may be too time-consuming for a fast Monte-Carlo-Markov-Chain parameter inference analysis. In order to accelerate MCMCs, it will be profitable to develop emulators using existing codes such as \verb|cosmopower| \citep{Spurio_Mancini_2022}.  

Projected-field power spectrum measurements require Wiener filtering of the CMB temperature map in order to mitigate foreground contamination and maximize the detection significance (Subsection~\ref{sec:pf_estimator}). As a secondary improvement, we found a Wiener filter that is more optimal than the previous ansatz and justified it with a heuristic derivation (Appendix~\ref{app:filters}). We leave for future work the task of finding a rigorous derivation of the optimal filter.

We parameterized the gas density profile using a gNFW formula with parameters calibrated on hydrodynamical simulations from \cite{Battaglia_2016}. We computed halo model predictions for three different shapes of the gas profile, motivated by different physical assumptions on feedback mechanisms, namely an NFW-like model, an Adiabatic model, and an AGN feedback model (Subsection \ref{ss:gas_prof}). Regarding the subtle treatment of the truncation of the gas density profile, necessary to obtained a converged Fourier transform, we proposed a procedure that preserves the total gas mass without altering the scale dependence of the profile (see text around Eq.~\ref{eq:brent}). For a fixed total gas mass, the gas profile extends toward larger radius in the Adiabatic model compared to the NFW model and even more so in the AGN feedback model (Figure \ref{fig:gasprof}), in accordance with the fact that energetic mechanisms tend to push the gas from the center towards outer regions of halos. Because of this, the projected-field power spectrum is the lowest in the AGN feedback model.  Although we focused on the \cite{Battaglia_2016} parameterization, the formalism presented here can be easily extended to other gas density profiles. For instance, motivated by recent results from simulations \citep{will_leandre} and observations \citep{Pandey_21}, it will be interesting to include a more general dependence on halo concentration and allow for a broken power law mass dependence of the gas density profile. As cosmological inference is moving into a stage where it relies heavily on non-analytical models trained on results from simulations, it will also be beneficial to extend our numerical implementation so that it can accomodate emulators for the gas density profiles \citep[e.g.,][]{Moser_2022} and the halo mass function \citep[e.g.,][]{bocquet_mira_titan}. 

We found that the halo model predictions roughly match the previous \textit{effective approach} predictions for galaxy density cross-correlation (Figure~\ref{fig:fls_cls_ttg}). However, they are significantly lower for weak lensing cross-correlation (Figure~\ref{fig:fls_cls_ttkg} and \ref{fig:fls_cls_ttkc}).

The low-$\ell$ behavior of the projected-field power spectrum when we change the gas density profile shape is not straightforward to interpret. The naive expectation that the low-$\ell$ limit should be independent of the scale dependence of the gas density profile because it is determined by the total gas mass does not hold here. In fact, the low-$\ell$ limit also receives contributions from small scales probing the inner part of halos. This is because of the convolution in harmonic space that arises from the real-space squaring operation of the CMB temperature field. We found that variations of the inner and outer slopes of the profiles mainly amount to a scale-independent rescaling of the projected field power spectrum at the sensitivity level of near-term data (Figure~\ref{fig:varying_alpha_beta}), although there is a noticeable scale-dependent effect associated with the outer slope for CMB weak lensing cross-correlation. 

We also used the halo model to estimate the covariance matrix of the projected-field power spectrum (Subsection~\ref{ss:covmat}) and the CMB lensing contribution to the measured signal (Subsection~\ref{ss:cmb_lens_cont}). 

We assumed that the covariance matrix is dominated by the Gaussian contribution and neglected contributions from higher-point functions, i.e., in this case a connected 6-point function, combinations of 3-point functions, and combinations of 4- and 2-point functions. As an example, we neglect terms like a $\langle TTTT\rangle\langle XX\rangle$ term, i.e., the connected 4-pt function of $T$ generated by lensing and the trispectrum of the foregrounds and kSZ, which are expected to be small. At some point these higher order terms should be investigated, but the Gaussian contributions certainly dominate for the forecasts considered here \citep[see, e.g.,][for an analysis of non-Gaussian covariance for the primordial bispectrum from CMB observations]{Coulton_2020}.

For the covariance matrices we considered four classes of CMB maps (\textit{Planck}, AdvACT, SO, and CMB-S4) characterized by different resolutions and noise properties (Table~\ref{tab:exps}) in combination with five LSS survey configurations, namely: \textit{unWISE} for galaxy density, DES and VRO/\textit{Euclid} for galaxy weak lensing, and SO and CMB-S4 for CMB weak lensing (Table~\ref{tab:exps_lss}). 

With these experimental specifications we obtained Fisher forecasts on the total detection SNR of the projected-field power spectrum and on measurements of the inner slope $\alpha$ and outer slope $\beta$ of the gas density profile  (Table \ref{tab:forecasts_SNRtot_hm} in Section~\ref{s:forecasts}), accounting for degeneracies with HOD parameters present in galaxy density cross-correlation (Figure~\ref{fig:TTG-forecast}). 
We found that galaxy density cross-correlation will be the easiest to detect with current datasets ($\mathrm{SNR}_\mathrm{tot}=17.8$ with AdvACT $\times$ \textit{unWISE}) although it will be challenging to probe the radial shape of the profiles. A first measurement of the slopes of the density profile, using galaxy density cross-correlation should be achievable with SO CMB maps.  For galaxy weak lensing cross-correlation, we forecast a first robust detection (above $5\sigma$) with AdvACT $\times$ VRO/\textit{Euclid} and high-significance measurements of the profile slopes with SO CMB maps. CMB lensing cross-correlation detections should be possible with SO CMB maps and measurements of the slopes with CMB-S4 maps. Interestingly, CMB lensing cross-correlation seems to be more sensitive to the inner slope of the profile than the outer slope, while galaxy density and galaxy lensing cross-correlations appear to probe both parameters equally well. Whether this is an artefact of our approximations or a real feature will be assessed in future work.

Although current SNR forecasts are often higher for cross-correlation with galaxy number density, studying cross-correlations with CMB or galaxy weak lensing is a particularly appealing avenue because it is free of degeneracies with HOD parameters or galaxy bias. Moreover, cross-correlation with CMB lensing can yield a measurement of the gas profile from \textit{all} halos up to $z\approx 2-3$ (without selection effects), while cross-correlation with galaxy weak lensing should enable gas tomography depending on the redshift distribution of the source galaxies. 

We note that the halo model formalism we presented here can be extended to other cross-correlations. For instance, kSZ$^2$-21cm cross-correlation has been shown to open a unique window on patchy reionization  \citep[see, e.g.,][]{2018MNRAS.476.4025M,LaPlante2020,LaPlante:2021ced}. Our formalism can also be used to compute the full three-point function (the hybrid bispectrum) which should carry more information than the projected-field power spectrum. 

A priority for future work will be to establish the robustness of the modeling choices based on further analytical investigations \citep[e.g.,][]{raagini} and comparison with hydrodynamical simulations, building on the initial comparisons in~\cite{Hill2016} and \cite{Ferraro2016}.  State-of-the-art simulations that could be used for such investigations include cosmological hydrodynamical simulations like Illustris TNG \citep{Springel_10.1093/mnras/stx3304,tng_18} and BAHAMAS \citep{McCarthy_2016}, or baryon pasting simulations such as the ones presented in \cite{2022arXiv220102632O}.

This work continues to pave the way for kinetic SZ measurements with upcoming CMB observations and LSS surveys to become a major source of information on the ICM and CGM gas thermodynamics \citep[e.g.,][]{astro20}. This is an increasingly important research topic, as current cosmological analyses show signs of inconsistencies which could be due to our misunderstanding of the behavior of baryonic matter in dense regions of the universe \citep[e.g.,][]{Amon:2022azi}. Understanding baryons will be crucial in order to maximize cosmological information extraction from ongoing and upcoming LSS surveys.

\section*{Data Availability}
The code \verb|class_sz| is public.\footnote{\href{https://github.com/borisbolliet/class_sz}{https://github.com/borisbolliet/class\_sz}} All numerical computations presented in this paper are  reproducible using a Jupyter notebook online.\footnote{\href{https://github.com/borisbolliet/class_sz/blob/master/notebooks/projected_fields_KSZ2X_2022.ipynb}{https://github.com/borisbolliet/class\_sz/blob/master/notebooks/projected\_fields\_KSZ2X\_2022.ipynb.}} We used \verb|getdist| \citep{Lewis:2019xzd} for computing the Fisher contours. 

\section*{Acknowledgments}

We thank the Aspen Center for Physics for hospitality during the preparation of this work, and the participants at the workshop on CMB secondary anisotropies held in September 2021. We thank Mathew Madhavacheril for crucial advice on FFT methods and Emmanuel Schaan for consultation on halo model consistency relations. We thank Fiona McCarthy and Oliver Philcox for many helpful discussions on the halo model formalism. We thank Raagini Patki and Nick Battaglia  for many helpful conversations on the kSZ effect in astrophysics and cosmology. We thank Abhishek Maniyar and Blake Sherwin for helpful conversations about CMB lensing. Finally, we thank Matthew Johnson and Tony Mroczkowski for discussions on kSZ detection methods.  JCH and AKK acknowledge support from NSF grant AST-2108536.  JCH acknowledges support from NASA grant 21-ATP21-0129.  The Flatiron Institute is supported by the Simons Foundation. SF is funded by the Physics Division of Lawrence Berkeley National Laboratory. Research at Perimeter Institute is supported in part by the Government of Canada through the Department of Innovation, Science and Economic Development and by the Province of Ontario through the Ministry of Colleges and Universities. AK thanks the AMTD Foundation for support.

\appendix
\section{Velocity dispersion}\label{s:veldisp}
\begin{figure}
    \begin{center}
    \includegraphics[width=0.6\columnwidth]{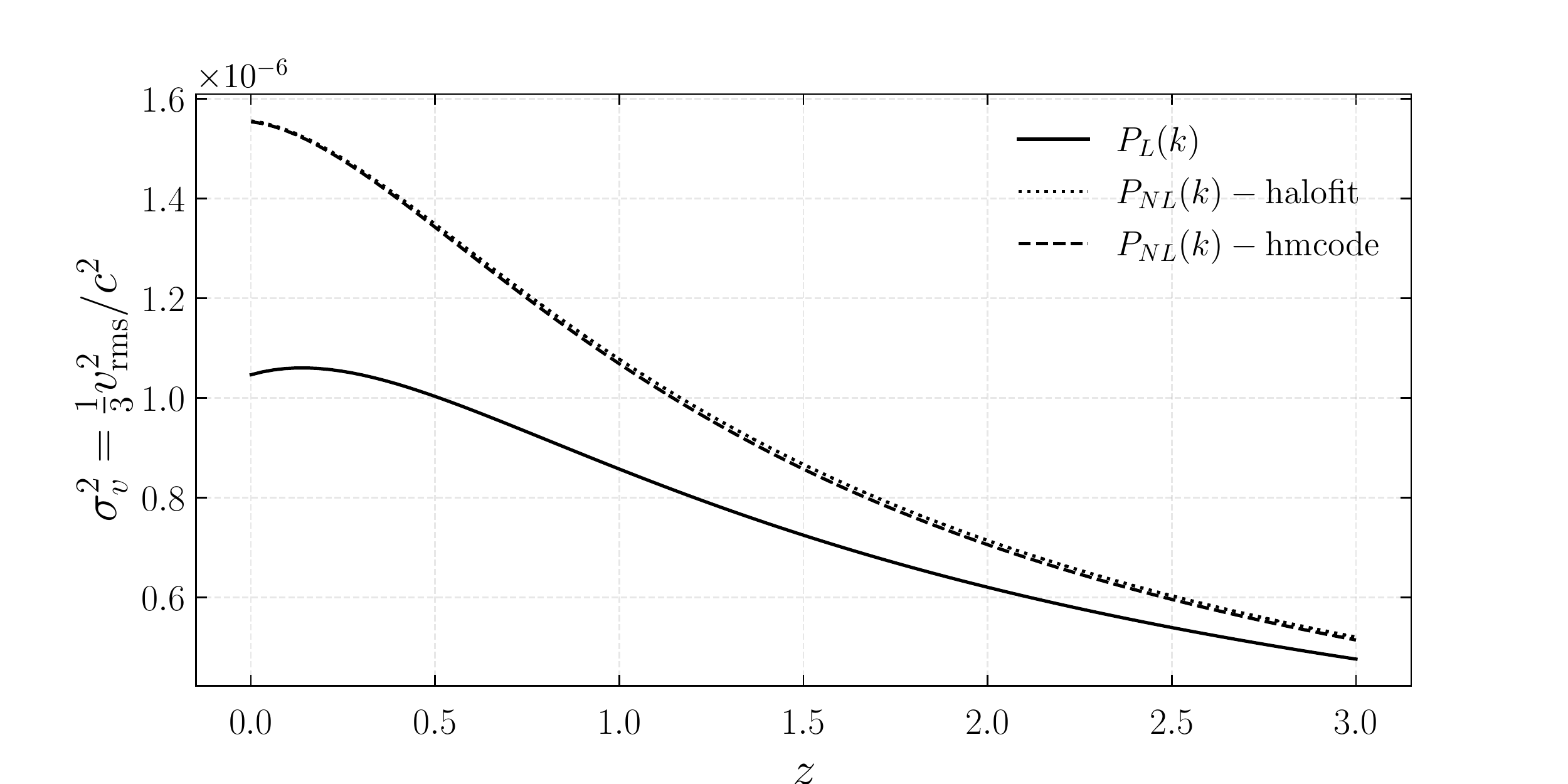}
    \end{center}
    \vspace{-0.1cm}
    \caption{Velocity dispersion (Eq.~\ref{eq:vrms2}) with different matter power spectrum choices, computed with {\texttt{class\symbol{95}sz}}. We assumed our fiducial cosmology.}  
    \label{fig:vrms2}
\end{figure}

In Figure~\ref{fig:vrms2} we show the velocity dispersion, as computed in Eq.~\eqref{eq:vrms2} with different choices for the matter power spectrum. In our fiducial model \citep[as in, e.g.,][]{Kusiak:2021hai}, we use the non-linear matter power spectrum computed with \verb|halofit| (dotted line). With this, we get $\sigma_v^2\simeq1.55\times10^{-6}$ of $v_\mathrm{rms}^2\simeq 648\,\mathrm{km/s}$. The computation with the \verb|hmcode| non-linear matter power spectrum is nearly identical to the \verb|halofit| one. If we use the linear matter power spectrum instead \citep[as in, e.g.,][]{Ferraro2016,Hill2016}, the velocity dispersion is $\approx 30\%$ lower at $z=0$ -- we get  $\sigma_v^2\simeq1.05\times10^{-6}$ of $v_\mathrm{rms}^2\simeq 531\,\mathrm{km/s}$. This difference would propagate linearly into the projected-field kSZ power spectrum, since it is proportional to $\sigma_v^2$ (in our approximation, see Eq.~\ref{eq:clpf}).

\section{Useful Halo Model Tools}\label{app:hm}
The halo model \citep{1991ApJ...381..349S,Mo_1996,Seljak:2000gq,Scoccimarro:2000gm,2000ApJ...543..503M,Peacock_2000,Cooray:2002dia} is an analytic model to compute power spectra, bispectra and higher order statistics of LSS tracers. 
\subsection{Halo Mass Function}\label{app:hmf}
The model assumes that matter is distributed within distinct spherical halos whose abundance is determined by the linear matter power spectrum through the halo mass function (HMF). The HMF determines the comoving number density of haloes of mass $M$ at redshift $z$ via \citep[e.g,][]{1974ApJ...187..425P,1991ApJ...379..440B,Tinker_2008,Tinker_2010}
\begin{equation}
    \frac{\mathrm{d}n}{\mathrm{d}m} = \nu f(\nu)\frac{\rho_\mathrm{m,0}}{m}\frac{\mathrm{d}\ln\sigma^{-1}}{\mathrm{d}m}\label{eq:hmf}
\end{equation}
where $\nu=\delta_c/\sigma$\footnote{Note that \cite{Tinker_2010} use the \textit{peak-height} definition $\nu \equiv \delta_c/\sigma(m,z)$ while {\texttt{class\symbol{95}sz}} uses $\nu \equiv (\delta_c/\sigma)^2$ as in E. Komatsu's {\texttt{szfast}} code. Also,  \cite{Tinker_2008} do not use the peak height explicitly, but  $\sigma^{-1}$ instead.} is the \textit{peak height} in the linear density field with $\delta_c=(3/20)(12\pi)^{2/3}\approx1.686$ the spherical collapse density threshold \citep[see][for the $\Omega_\mathrm{m}$ correction - not used here]{nm1997},  $\rho_\mathrm{m,0}$ is the mean matter density at $z=0$ and 
\begin{equation}
    \sigma^2(m,z) =\frac{1}{2\pi^2}\int\mathrm{d}kk^2\hat{\mathrm{W}}(kR)^2 P_{L}(k,z)
\end{equation}
is the variance of the matter density field smoothed in region of radius $R=(3m/4\pi\rho_\mathrm{m,0})^{1/3}$ using the Fourier transform of the real-space top-hat window function $\hat{\mathrm{W}}(x)=3j_1(x)/x$ where $j_1(x) =  \left[\sin(x)-x\cos(x)\right]/x^2$ is the first-order spherical Bessel function. Here, $P_L$ is the linear matter power spectrum.
In this paper, we use the \cite{Tinker_2008} formula (see their Eq. 3) for the function $f(\nu)$ rather than the  \cite{Tinker_2010} formula for the same reasons as explained in Appendix B of \cite{Kusiak:2022xkt}. One should keep in mind that these fitting formulas are calibrated on simulations with a limited mass and redshift range. Namely $0.25\lesssim \sigma^{-1}\lesssim 2.5$, which corresponds to masses $\sim 10^{10}-10^{15}\,\mathrm{M}_\odot /h$ at $z=0$) and $0<z\lesssim 2$. \cite{Tinker_2008} suggests to use $f(\sigma,z=2.5)$ for all $z>2.5$, while \cite{Tinker_2010} suggests to use $f(\sigma,z=3)$ for all $z>3$.

By consistency, the HMF must be such that
\begin{equation}
    \int \mathrm{d}\nu f(\nu)=1,\quad\int \mathrm{d}\nu b^{(1)}(\nu)f(\nu)=1,\quad\int \mathrm{d}\nu b^{(n)}(\nu)f(\nu)=0\,\,\mathrm{for}\,\,n>1,
\end{equation}
where $b^{(1)}$ is the linear bias (see Eq.~\ref{eq:b1tink}) and $b^{(n)}$ are are higher order biases. These constraints enforce that all matter is within halos and that it is not bias with respect to itself \citep[e.g.,][]{Tinker_2010}.

\subsection{Consistency Conditions}\label{s:hmcds}

To impose the consistency conditions numerically, the mass integrals are approximated as follows \citep{Schmidt_2016}:
\begin{align}
\int_0^{+\infty} \mathrm{d}n\hat{u}^X(m,z)\hat{u}^Y(m,z) &= \int_{m_\mathrm{min}}^{m_\mathrm{max}}\mathrm{d}n \hat{u}^X(m,z)\hat{u}^Y(m,z)+N_\mathrm{min}(z)\hat{u}^X(m_\mathrm{min},z)\hat{u}^Y(m_\mathrm{min},z)\label{eq:hmct1}\\
\int_0^{+\infty} \mathrm{d}nb^{(i)}(m,z)\hat{u}^X(m,z) &=\int_{m_\mathrm{min}}^{m_\mathrm{max}}\mathrm{d}nb^{(i)}(m,z)\hat{u}^X(m,z) + b^{(i)}_{m_\mathrm{min}}(z)[\rho_\mathrm{m,0}/m_\mathrm{min}]\hat{u}^X(m_\mathrm{min},z)\label{eq:hmct2}
\end{align}
with $i=1,2$ for the first and second order bias \citep[see also][]{phils2020,mead2021}. The \textit{counter-terms} on the RHS account for the low-mass part of HMF that cannot be parameterized using current N-body simulations. With this implementation,  ``halo model predictions do not depend on any properties of low-mass halos that are smaller than the scales of interest" \citep{Schmidt_2016}. The counter-terms require a mass integral at each redshift that we pretabulate as 
\begin{align}
N_\mathrm{min}(z) &= [1-I_0(z)]\rho_\mathrm{m,0}/m_\mathrm{min}\quad\mathrm{with}\quad I_0(z)=\int_{m_\mathrm{min}}^{m_\mathrm{max}}\mathrm{d}n m/\rho_\mathrm{m,0}\label{eq:ct1}\\
b^{(1)}_\mathrm{min}(z) &= 1-I_1(z)\quad\mathrm{with}\quad I_1(z)=\int_{m_\mathrm{min}}^{m_\mathrm{max}}\mathrm{d}nb^{(1)}(m,z)m/\rho_\mathrm{m,0}\label{eq:ct2}\\
b^{(2)}_\mathrm{min}(z) &= -I_2(z)\quad\mathrm{with}\quad I_2(z)=\int_{m_\mathrm{min}}^{m_\mathrm{max}}\mathrm{d}nb^{(2)}(m,z)m/\rho_\mathrm{m,0}.\label{eq:ct3}
\end{align}
One can check that Eq. \eqref{eq:hmct1}-\eqref{eq:hmct2} amounts to substituting the HMF $\mathrm{d}n/\mathrm{d}m$ with $\mathrm{d}n/\mathrm{d}m+N_\mathrm{min}\delta(m-m_\mathrm{min})$ in all mass integrals and setting a cut-off at $m_\mathrm{min}$. Eq. \eqref{eq:ct1}-\eqref{eq:ct3} are then equivalent to the consistency conditions
\begin{equation}
\int \mathrm{d}n m =\rho_\mathrm{m,0},\quad
\int \mathrm{d}n m b^{(1)}(m,z) = \rho_\mathrm{m,0}, \quad
\int \mathrm{d}nm b^{(2)}(m,z) = 0 ,\label{eq:hmc}
\end{equation}
ensuring that all mass is within halos and that matter is unbiased with respect to itself. 

As shown on the left panel of Fig.~\ref{fig:pkshm}, the correction of the matter power spectrum associated with these consistency condition can be very significant.

\subsection{Halo Model Power Spectra}\label{sec:hmps}
Let $X$ and $Y$ be two LSS tracers with radial profiles  $u^X$ and $u^Y$.  Their 3D power spectrum is defined via
\begin{equation}
    \langle X(\bm{k}_1)Y(\bm{k}_2)\rangle=(2\pi)^2\delta(\bm{k}_1+\bm{k}_2)P_{XY}(k_1)
\end{equation}
The halo model power spectrum for the RHS is $P_{XY}^\mathrm{hm} = P_{XY}^\mathrm{1h}+P_{XY}^\mathrm{2h}$ where the  1-halo term, $P_{XY}^\mathrm{1h}$, accounts for correlations between points within the same halo, and the 2-halo term, $P_{XY}^\mathrm{2h}$, accounts for correlations between points in distinct halos. Each term can be expressed  using the 3D Fourier transforms of the profiles. Note that the Fourier transforms reduce to Hankel transforms since the profiles are radially symmetric. The Fourier transform of a radial profile is given by
\begin{equation}
    \hat{u}(k)=4\pi \int_0^\infty \mathrm{d}r r^2 j_0(kr)\mathrm{H}(r_\mathrm{out}-r)u(r)\quad\mathrm{where}\quad j_0(x)=\sin(x)/x\label{eq:hatu}
\end{equation}
is the spherical Bessel function of order 0 and where we added the Heaviside step function $\mathrm{H}$ in order to truncate the profile at some radius $r_\mathrm{out}$. Note that in the $k\rightarrow 0$ limit, $\hat{u}$ is the volume average of $u$ within a sphere of radius $r_\mathrm{out}$.
Explicitly, the 1- and 2-halo terms are 
\begin{equation}
    P_{XY}^\mathrm{1h} = \langle \hat{u}^X\hat{u}^Y\rangle_n\quad\mathrm{with}\quad
    P_{XY}^\mathrm{2h} = \langle b^{(1)}\hat{u}^X\rangle_n\langle b^{(1)}\hat{u}^Y\rangle_n P_{L}
\end{equation}
where $b^{(1)} = b^{(1)}(m,z)$ is the first-order halo bias \citep[e.g.,][]{1999MNRAS.308..119S,Tinker_2010}. Here, we use
\begin{equation}
    b^{(1)}(\nu)=1-A\frac{\nu^a}{\nu^a+\delta_c^a}+B\nu^b+C\nu^c\label{eq:b1tink}
\end{equation}
with parameters fixed to the values in Table 2 of \cite{Tinker_2010}. 
  In general, for two fields $X$ and $Y$ there is a contribution to the 1-halo power spectrum coming from  correlated fluctuations so that $\langle\hat{u}^X\hat{u}^Y\rangle =(1+r) \langle\hat{u}^X\rangle\langle\hat{u}^Y\rangle$ with $r\neq0$ (here the angle brackets are to be understood as ensemble-average at fixed mass and redshift). Although, we can often assume $r\ll1$ since it is unlikely that two different fields $X$ and $Y$ would fluctuate in a correlated way. See e.g. \cite{Koukoufilippas:2019ilu} where they took this effect into account.

\subsection{Halo Model Bispectra}\label{ss:hm_bp}

 Let $X,Y,Z$ be three LSS tracers. Their bispectrum is defined by 
\begin{equation}
    \langle X(\bm{k}_1) Y(\bm{k}_2) Z(\bm{k}_3)\rangle=(2\pi)^3\delta(\bm{k}_1+\bm{k}_2+\bm{k}_3)B(k_1,k_2,k_3).
\end{equation}
Its halo model expression is the sum of three terms,  $B^\mathrm{hm} = B^\mathrm{1h}+B^\mathrm{2h}+B^\mathrm{3h}$, associated with correlations between triplets within 1,2 and 3 halos, respectrively. The halo model terms expressions are \citep{Scoccimarro:2000gm,Valageas_2011,Lazanu_2016}:
\begin{align}
         B^\mathrm{1h} &= \langle \hat{u}_{k_1}^X\hat{u}_{k_2}^Y\hat{u}_{k_3}^Z\rangle_n +\,\mathrm{perm}(X,Y,Z)\label{eq:bispec1h}\\
     B^\mathrm{2h} &= \langle \hat{u}_{k_1}^X\hat{u}_{k_2}^Y\rangle_n\langle \hat{u}_{k_3}^Z\rangle_n P_{L}(k_3)+\langle \hat{u}_{k_3}^X\hat{u}_{k_1}^Y\rangle_n\langle \hat{u}_{k_2}^Z\rangle_n P_{L}(k_2)+\langle \hat{u}_{k_2}^X\hat{u}_{k_3}^Y\rangle_n\langle \hat{u}_{k_1}^Z\rangle_n P_{L}(k_1)+\,\mathrm{perm}(X,Y,Z)\label{eq:bispec2h}\\
     B^\mathrm{3h}&= 2\langle b^{(1)}\hat{u}_{k_1}^X\rangle_n P_{L}(k_1)\langle b^{(1)}\hat{u}_{k_2}^Y\rangle_n P_{L}(k_2)\langle b^{(1)} \hat{u}_{k_3}^Z\rangle_nF_2(k_1,k_2,k_3)\nonumber+2\mathrm{cyc}\\
     &+\langle b^{(1)}\hat{u}_{k_1}^X\rangle_n P_{L}(k_1)\langle b^{(1)}(m_2)\hat{u}_{k_2}^Y \rangle_n P_{L}(k_2)\langle b^{(2)}(m_3) \hat{u}_{k_3}^Z\rangle_n+2\mathrm{cyc}\\
     &+\mathrm{perm}(X,Y,Z)
     \label{eq:bispec3h}
\end{align}
where $F_2$ is given in Eq. \eqref{eq:f2s}
and $b^{(2)}$ is the second order halo bias. We compute $b^{(2)}$ with the peak background split formula using Eq.~(8) of \cite{10.1093/mnras/stv702}. Here, `$2\mathrm{cyc}$' denotes the cyclic permutations with respect to scales/wavenumbers (as explicitly written in the 2-halo term equation), whereas, `$\mathrm{perm}(X,Y,Z)$' denotes the cyclic permutations of the tracers. 

In this paper, we are interested in a case where $X=Y=\delta_\mathrm{e}$ and where $Z$ is always evaluated at the scale $k_3$ (so is the hybrid bispectrum in Eq.~\ref{eq:clpf}). Hence, for the 1-halo term, there is only one permutation to evaluate. For the 2-halo term there are three permutations (those of Eq.~\ref{eq:bispec2h} where $Z$ and $k_3$ are together). Similarly, for the 3-halo term there are three permutations proportional to $F_2$ and three other permutations proportional to $b^{(2)}$. In \cite{Hill_2018}, the 2-halo term of the kSZ-kSZ-ISW bispectrum was computed including the nine  permutations (see their Eq.~30).

\subsection{Angular Power Spectra and Correlation Functions}\label{ss:cls_hm}

With the Limber approximation in flat-sky \citep[e.g., Appendix A of  ][and references therein]{hp2013}, angular power spectra are obtained by integrating the 3D power spectra, evaluated at $k=(\ell+1/2)/\chi$, over comoving volume 
\begin{equation}
    C_\ell^{XY}=\int\mathrm{d}\mathrm{v}  W^X(\chi)W^Y(\chi)P_{XY}^\mathrm{hm}\left(\frac{\ell+1/2}{\chi}\right)\label{eq:clxy}
\end{equation}
where $W^X$ and $W^Y$ are projection kernels. From the angular power spectra we get the angular 2-point correlation functions (2PCF) as
\begin{equation}
    \xi^{XY}(\theta) = \frac{1}{2\pi}\int \mathrm{d}\ell \ell J_i(\ell\theta)C_\ell^{XY}\label{eq:xi}
\end{equation}
where $J_i$ is the $i$th  order Bessel function of the first kind and $i$ depends on the spin of the field. For instance, $i=2$ for  $X=\delta_\mathrm{g}$ and $Y=\kappa_\mathrm{g}$. In this case, the angular 2PCF is the so-called galaxy tangential shear $\gamma_t(\theta)$.  For $X=Y=\delta_\mathrm{g}$, we have $i=0$ and the angular 2PCF is the galaxy clustering correlation function, often denoted $w(\theta)$. For $X=Y=\kappa_\mathrm{g}$ we have $i=0/4$ and the angular 2PCF is the so-called shear $\xi_{+/-}(\theta)$  \citep[see,e.g.,][and references therein]{DES12018}. Numerically, the integral in Eq. \eqref{eq:xi} can be evaluated efficiently with FFTLog routines.

\subsection{Navarro-Frenk-White  Density Profile}\label{sss:nfw}
The Navarro-Frenk-White density profile is defined as $\rho_{_\mathrm{NFW}}(r)=\rho_{\mathrm{m},0} u^\mathrm{_{NFW}}(r)$ with 
\begin{equation}
    u^\mathrm{_{NFW}}(r) =\frac{\rho_s}{\rho_{\mathrm{m},0}} \frac{1}{\frac{r}{r_s}\left(1+\frac{r}{r_s}\right)^2}\quad\mathrm{where}\quad r_s = r_\Delta/c_\Delta.
\end{equation}
Here, the scale radius $r_s$ is defined in terms of characteristic radius and concentrations $r_\Delta$ and $c_\Delta$. These depend on the halo mass $m_\Delta$. The concentration is often computed with a relation calibrated on simulations \citep[e.g.,][]{Duffy_2008,2013ApJ...766...32B}. In this paper, we use the \cite{2013ApJ...766...32B} relation. It is common to take $r_\Delta$ as the radius of the spherical region of mass $m_\Delta$ within which the density is $\Delta$ times the critical density,
at redshift $z$. Thus,
\begin{equation}
    r_\Delta = \left[3 m_\Delta/(4\pi  \Delta \rho_\mathrm{crit}(z)) \right]^{1/3}.\label{eq:rd}
\end{equation}
Common values for $\Delta$ are $180$, $200$ and $500$. Instead of using the critical density as a reference, one can use the matter density which means replacing $\Delta$ by $\Delta\Omega_\mathrm{m}(z)$, where $\Omega_\mathrm{m}(z)=\rho_\mathrm{m}(z)/\rho_\mathrm{crit}(z)$. Another common choice for these definitions are the virial mass and radius, which amount to taking $\Delta=\Delta_c(z)$ given in \cite{1998}. By consistency, we  have $m_\Delta = \int_0^{r_\Delta}\mathrm{d}r4\pi r^2 \rho_{_\mathrm{NFW}}(r)$, which yields 
\begin{equation}
    \rho_s = \frac{m_\Delta}{4\pi r_s^3 }f_{_\mathrm{NFW}}(c_\Delta)\quad\mathrm{with}\quad f_{_\mathrm{NFW}}(x)=[\ln(1+x)-x/(1+x)]^{-1}.\label{eq:rho_s}
\end{equation}

The Fourier transform of $u^{_\mathrm{NFW}}$ truncated at $r_\mathrm{cut}=\lambda r_\Delta$ (see Eq. \ref{eq:hatu}) has an analytical expression given by \citep{Scoccimarro:2000gm}:
\begin{equation}
    \hat{u}^\mathrm{_{NFW}}(k) = \frac{m_{\lambda r_\Delta}}{\rho_{\mathrm{m},0}}    \left([\mathrm{Ci}((1+\lambda c_\Delta)q)-\mathrm{Ci}(q)] \cos q
    +    [\mathrm{Si}((1+\lambda c_\Delta) q)-\mathrm{Si}(q)]\sin q
    -\frac{\sin(\lambda c_\Delta q)}{(1+\lambda c_\Delta)q}\right)f_{_\mathrm{NFW}}(\lambda c_\Delta)\label{eq:nfwtrunc}
\end{equation}
where $m_{\lambda r_\Delta}$ is the mass within $\lambda r_\Delta$ (i.e., $m_\Delta$ for $\lambda=1$) and where $\mathrm{Ci}(x)=\int_x^{\infty}\mathrm{d}t\cos(t)/t$ and $\mathrm{Si}(x)=\int_0^x\mathrm{d}t\sin (t)/t$ are the cosine and sine integrals, and $q = (1+z)kr_s = \ell/\ell_s$.\footnote{In the last equality we defined  $\ell_s=d_A/r_s$ with angular diameter distance $d_A = \chi/(1+z)$, and traded wavenumber for multipole according to $k\chi=\ell+1/2$.}
Noting that $q\propto m_\Delta^{1/3}$, the asymptotic behaviors of $\hat{u}^\mathrm{_{NFW}}$ when $k\rightarrow0$ or $m_\Delta\rightarrow0$ are the same, namely
\begin{equation}
    \lim_{q\rightarrow 0}\hat{u}^\mathrm{_{NFW}}_k = \frac{m_{\lambda r_\Delta}}{\rho_\mathrm{m,0}}.\label{eq:q0limnfw}
\end{equation}
This is an important property which implies that in the low-$k$ regime $\langle \hat{u}^\mathrm{_{NFW}}_k\rangle_n\approx1$ (when $\lambda =1$), as a consequence of the consistency conditions. 

\begin{figure}
    \includegraphics[width=1.\columnwidth]{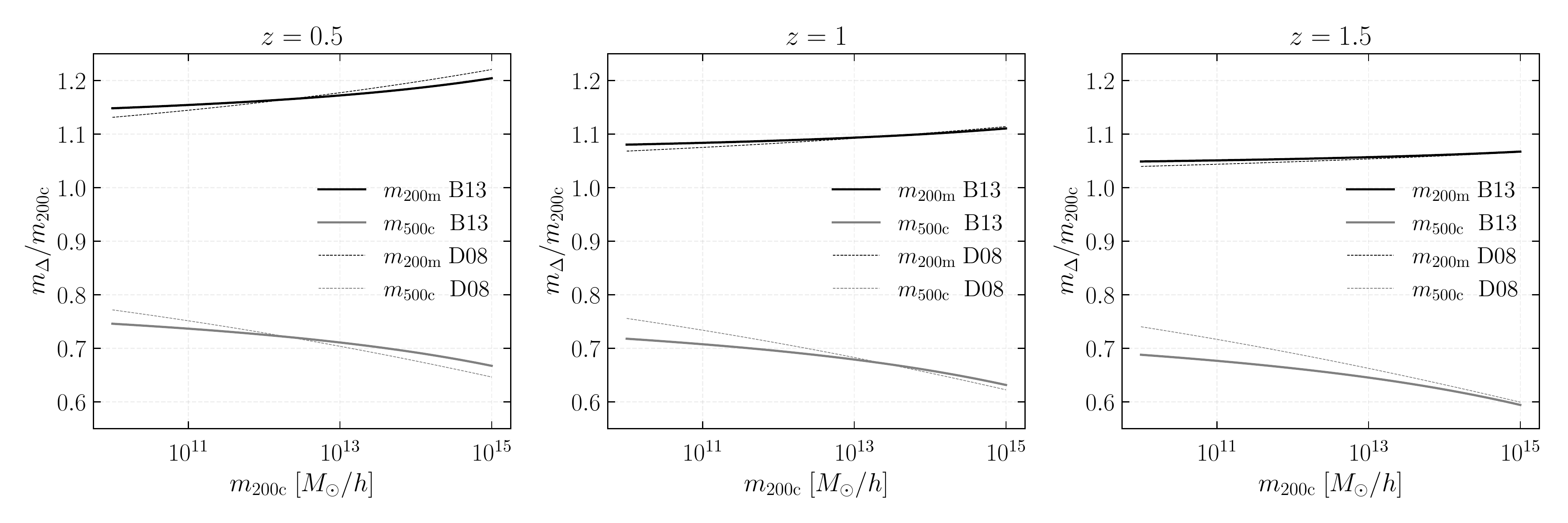}
    \caption{Ratios between mass definitions at $z=0.5$ (left), $z=1$ (middle) and $z=1.5$ (right) using the \cite{2013ApJ...766...32B} relation (B13) and the \cite{Duffy_2008}  relation (D08).} 
    \label{fig:mconv}
\end{figure}

\subsection{Mass Conversions}\label{ss:mass_conv}
 Although here we exclusively used the $m_{200\mathrm{c}}$ mass definition, we explain how to convert between  mass definition as it can be useful for comparison with other analyses or to implement different mass functions, HOD's and tracer profiles. To convert between $m_{\Delta}$ and $m_{\Delta^{\prime}}$, we  compute $m_{\Delta^{\prime}}=\int_0^{r_{\Delta^\prime}}\mathrm{d}r4\pi r^2 \rho_{_\mathrm{NFW}}(r)$ with the NFW profile defined in terms of $r_s=r_\Delta/c_\Delta$. Its expression is equivalent to
\begin{equation}
\frac{m_{\Delta^{\prime}}}{m_{\Delta}}-\frac{f_{_\mathrm{NFW}}(c_{\Delta})}{f_{_\mathrm{NFW}}(c_{\Delta}r_{\Delta^\prime}/r_{\Delta})}=0\quad\mathrm{with}\quad r_{\Delta^\prime} = \left[3 m_\Delta/(4\pi  \Delta^\prime \rho_\mathrm{crit}(z)) \right]^{1/3}
\end{equation}
which can be solved for $m_{\Delta^{\prime}}$ with a root-finding algorithm. In \verb|class_sz| we use Brent's method \citep{brent2002algorithms}. For reference, we show the conversions between $m_\mathrm{200c}$, $m_\mathrm{200m}$ and $m_\mathrm{500c}$ at three redshifts for the \cite{2013ApJ...766...32B} and \cite{Duffy_2008} concentration-mass relations in Figure~\ref{fig:mconv}. Overall, $m_\mathrm{200m}$ is $\approx 5-20\%$ larger than  $m_\mathrm{200c}$, while $m_\mathrm{500c}$ is $\approx 20-40\%$ lower than  $m_\mathrm{200c}$. The \cite{2013ApJ...766...32B} and \cite{Duffy_2008} agree well at high masses but differ substantially at low masses.

\subsection{Galaxy Halo Occupation Distributions}\label{sss:hod}

Galaxies populate dark matter halos in complicated ways. Its simple and faithful description, galaxy Halo Occupation Distributions (HOD) was proposed in  \cite{zz2007}. Here we use a slightly different parameterization, to match that  of \cite{za2021}. The expectation value for number of central galaxies in a halo of mass $m$ is given by 
\begin{equation}
  N_\mathrm{cent}(m) = \frac{1}{2}\left(1+\mathrm{erf}\left[\frac{\log_{10}(m/m_\mathrm{min})}{\sigma_{\mathrm{log}_{10}m}}\right]\right)\label{eq:ncent}
\end{equation}
where $m_\mathrm{min}$ is a pivot mass above which, on average, halos have a central galaxy. Here,  $\sigma_{\mathrm{log}_{10}m}$ controls the steepness of the transition in mass from no galaxy to having at least one galaxy in the halo. The expectation value for number of satellite galaxies is a power law with an exponent $\alpha_s$,
\begin{equation}
 N_\mathrm{sat}(m) =  N_\mathrm{cent}(m)\left(\frac{m}{m_1}\right)^{\alpha_s},\label{eq:nsat}
\end{equation}
where $m_\mathrm{1}$ is a pivot mass above which the number of satellites increases steeply. In our fiducial model, corresponding to the best-fit to \textit{blue} \textit{unWISE} galaxies, these parameters are set to  $m_\mathrm{min} = 1.01\times 10^{13} {M_{\odot}}/h$, $m_1 = 1.18\times10^{14} {M_{\odot}}/h$, $\alpha_s=2.08$ and $\sigma_{\mathrm{log}_{10}m}=0.76$ \citep[see][for details]{Kusiak:2022xkt}. 

Given a specific HOD (Eq.~\ref{eq:ncent} and \ref{eq:nsat}) we can compute the galaxy number density and galaxy bias at $z$ as 
\begin{equation}
    \bar{n}_{\mathrm{g}}(z) =\langle  N_\mathrm{cent}+ N_\mathrm{sat} \rangle_n,\quad\mathrm{with}\quad b_\mathrm{g}(z)=\frac{1}{\bar{n}_{\mathrm{g}}(z)}\langle b^{(1)}(N_\mathrm{cent}+ N_\mathrm{sat})\rangle_n,\label{eq:ngbar}
\end{equation}
where $b^{(1)}$ is the linear bias of Eq.~\eqref{eq:b1tink}. As for the spatial distribution, central galaxies are naturally assumed to be located at the center of halos (their \textit{density profile} is a Dirac delta function) and satellite galaxies are assumed to be randomly distributed along an NFW-like radial profile. Thus, the Fourier transform of  the \textit{galaxy density profile} is
\begin{equation}
    \hat{u}^\mathrm{g}_k = \frac{1}{\bar{n}_\mathrm{g}}\left(N_\mathrm{cent}+N_\mathrm{sat}\hat{u}^\mathrm{sat}_k\right)\label{eq:ugk}
\end{equation}
where $\hat{u}^\mathrm{sat}_k$ is the same Eq.~\eqref{eq:nfwtrunc} with $\lambda=1$ and without the ${m_{\lambda r_\Delta}}/{\rho_{\mathrm{m},0}}$ prefactor. In addition, $c_\Delta$ is often replaced by a free parameter $c_\mathrm{sat}$ to allow for more freedom in the radial distribution. Another important HOD quantity is the Fourier transform of the second moment of the satellites galaxy distribution:
\begin{equation}
\hat{u}^\mathrm{gg}_k=\frac{1}{\bar{n}_{\mathrm{g}}^2}\left[N_\mathrm{sat}^2(\hat{u}^{\mathrm{sat}}_k)^2+2N_\mathrm{sat}\hat{u}^{\mathrm{sat}}_k\right],\label{eq:uggk}
\end{equation}
\citep[see, e.g.,][]{van_den_Bosch_2013,Koukoufilippas:2019ilu}, as it determines the 1-halo term of the galaxy power spectrum.

\begin{figure}
    \includegraphics[width=1.\columnwidth]{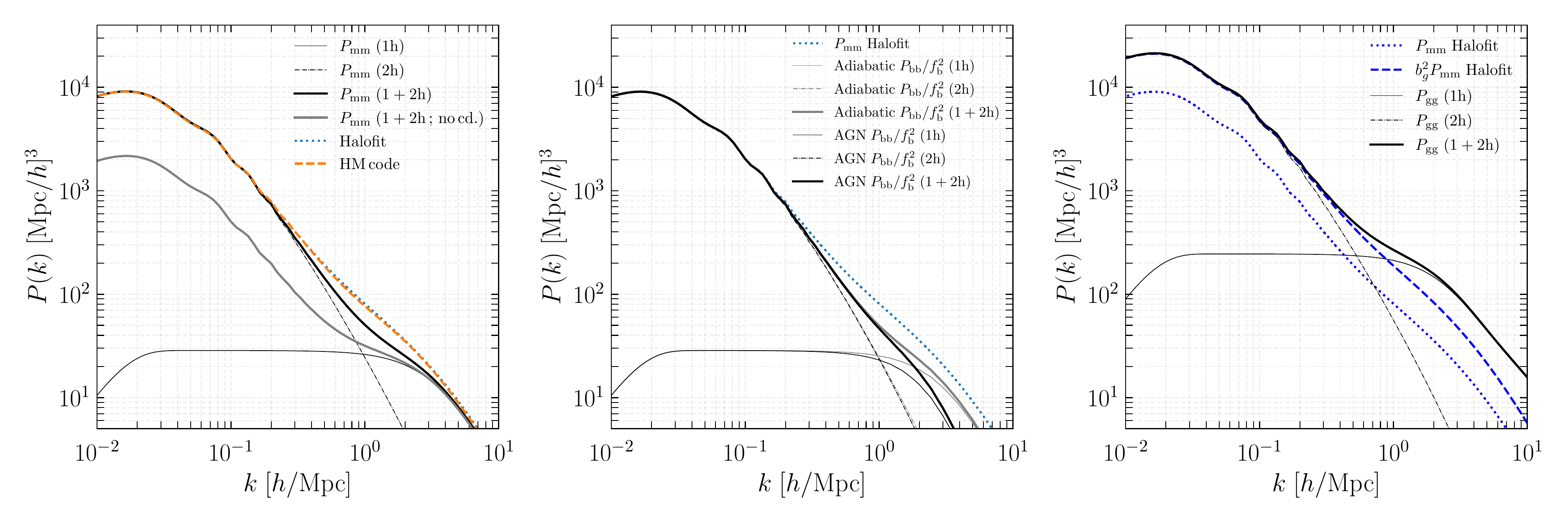}
    \vspace{-0.4cm}
    \caption{\textit{Left:} The matter power spectrum at $z=1$ computed in the halo model (1-halo term is the solid black line, 2-halo term is the dash-dotted black line, and sum is the thick solid line) compared with the  {\texttt{halofit}} formula (dotted blue line) \cite{Takahashi_2012} and the {\texttt{hmcode}} prediction (dashed orange) \cite{mead2021}. The thick grey line shows the halo model prediction when the consistency prescriptions of Subsection\,\ref{s:hmcds} are not taken into account. \textit{Middle:} The power spectrum of baryons in the halo model for two different density profile types (\textit{Adiabatic} and \textit{AGN Feedback}) and rescaled by $1/f_b^2$ compared with the {\texttt{halofit}} matter power spectrum. \textit{Right:} Galaxy power spectrum in the halo model (black lines) at $z=1$ compared with the matter power spectrum computed with {\texttt{halofit}} (dotted blue line) and the {\texttt{halofit}} power spectrum times $b_{\mathrm{g}}^2$ squared (dashed blue) with $b_{\mathrm{g}}$ from Eq.~\eqref{eq:ngbar}.}  
    \label{fig:pkshm}
\end{figure}

\subsection{Application to Tracers}\label{ss:app_lss}

In this Section we present explicit expressions for 3D auto-power spectra of tracers in the halo-model. Cross-power spectra expressions are easily obtained in a similar fashion. 

\subsubsection{Cold Dark Matter}\label{sss:matter}

 When CDM density is assumed to follow the NFW profile (see Appendix~\ref{sss:nfw}), the halo model matter power spectrum at $z$ is $P^\mathrm{hm}_{\delta_\mathrm{m}\delta_\mathrm{m}}=P_{\delta_\mathrm{m}\delta_\mathrm{m}}^\mathrm{1h}+P_{\delta_\mathrm{m}\delta_\mathrm{m}}^\mathrm{2h}$ with 
 \begin{equation}
      P_{\delta_\mathrm{m}\delta_\mathrm{m}}^\mathrm{1h}(k,\chi) = \langle \hat{u}^{_\mathrm{NFW}}_{k}\hat{u}^{_\mathrm{NFW}}_{k}\rangle_n\quad\mathrm{and}\quad P_{\delta_\mathrm{m}\delta_\mathrm{m}}^\mathrm{2h}(k,\chi) = \langle b^{(1)}\hat{u}^{_\mathrm{NFW}}_{k} \rangle_n^2 P_{L}(k,\chi)\label{eq:pkhm}
 \end{equation}
where $P_L$ is the linear matter power spectrum. In the low-$k$ limit we have $\langle b^{(1)}\hat{u}^{_\mathrm{NFW}}_{k} \rangle_n\rightarrow 1$ (by construction and consistency) so that $P_{\delta_\mathrm{m}\delta_\mathrm{m}}^\mathrm{2h}\sim P_L$, whereas  $P_{\delta_\mathrm{m}\delta_\mathrm{m}}^\mathrm{1h}\sim \langle m_\Delta^2/\rho_\mathrm{m,0}^2\rangle$ which is independent of $k$.\footnote{On ultra-large scales the power spectrum should grow as $k^4$. Hence, we follow \cite{mead2015}  and add an exponential damping to the 1-halo term. In principle, the 2-halo term should also be damped in the non-linear regime due to perturbative effects. We did not account for that subtlety here and refer to \cite{mead2021} and \cite{phils2020} for details on these aspects.} Hence, at low-$k$ the 2-halo term dominates and we have $P^\mathrm{hm}_{\delta_\mathrm{m}\delta_\mathrm{m}}\sim P_L$. The halo model matter power spectrum is plotted on the left panel of Figure~\ref{fig:pkshm} against the \verb|halofit| and \verb|hmcode| (which are nearly identical). The mismatch between the halo model power spectrum and the N-body calibrated formulas (\verb|halofit| and \verb|hmcode|) in the transition regime between the 2-halo and 1-halo term is a well-known short-coming of the halo model. This issue has been addressed in several manners. For instance, \cite{mead2015} suggest using $P^{\mathrm{hm}}_{\delta_\mathrm{m}\delta_\mathrm{m}}=[(P^{\mathrm{1h}}_{\delta_\mathrm{m}\delta_\mathrm{m}})^\alpha+(P^{\mathrm{2h}}_{\delta_\mathrm{m}\delta_\mathrm{m}})^\alpha)]^{1/\alpha}$ where $\alpha$ is a free parameter. Another approach, proposed in \cite{phils2020} is to use perturbation theory at one-loop order in the modeling of the 2-halo term, amounting to replace $P_{L}$ in Eq.~\eqref{eq:pkhm} by 
    $P_{NL}=P_{L}+P_\mathrm{spt}+P_\mathrm{ct}$
where $P_\mathrm{spt}=P_{22}+P_{13}$ with $P_{22}$ and $P_{13}$ resulting from higher-order terms associated with the $F_2$ and $F_3$ coupling kernels  \citep[see, e.g.,][]{bernardeau2002} and $P_\mathrm{ct}(k)=-c_\mathrm{s}^2k^2P_{L}(k)$ with $c_\mathrm{s}$ a free parameter of the model. What these approaches have in common is inclusion of the  \textit{nuisance} parameters to the model. The extra nuisance parameters then need to be calibrated on simulations or marginalized over. Here we do not study these extra pieces of modelling for the transition regimes and leave them for future work.

\subsubsection{Electrons}\label{sss:baryons}
The electron power spectrum is computed in the halo model using the Fourier transform of the gas density profile, $\hat{u}^\mathrm{e}$ of Eq.~\eqref{eq:hatub}, as 
 \begin{equation}
      P_{\delta_\mathrm{e}\delta_\mathrm{e}}^\mathrm{1h}(k,\chi) = \langle \hat{u}^{\mathrm{e}}_{k}\hat{u}^{\mathrm{e}}_{k}\rangle_n\quad\mathrm{and}\quad P_{\delta_\mathrm{e}\delta_\mathrm{e}}^\mathrm{2h}(k,\chi) = \langle b^{(1)}\hat{u}^{\mathrm{e}}_{k} \rangle_n^2 P_{L}(k,\chi).\label{eq:pkb}
 \end{equation}
The gas density profile is normalized such that 
\begin{equation}
\lim_{k\rightarrow 0}  \hat{u}^{\mathrm{e}}_{k} =  f_\mathrm{b} f_\mathrm{free}\frac{m_\Delta}{\rho_\mathrm{m,0}},
\end{equation}
and there for in the low-$k$ limit we have $P_{\delta_\mathrm{e}\delta_\mathrm{e}}^\mathrm{hm}\approx P_{\delta_\mathrm{e}\delta_\mathrm{e}}^\mathrm{2h}\approx f_\mathrm{b}^2P_L$, irrespective of the gas density profile assumption. In the high-$k$ regime, the difference between the gas density profile and the NFW profile can be significant and therefore the scale dependence of the baryon power spectrum can depart from that of the non-linear matter power spectrum. This is illustrated in the middle panel of Figure~\ref{fig:pkshm}.

\subsubsection{Galaxies}\label{sss:galaxies}

The galaxy power spectrum is $P_{\delta_\mathrm{g}\delta_\mathrm{g}}^\mathrm{hm}=P_{\delta_\mathrm{g}\delta_\mathrm{g}}^\mathrm{SN}+P_{\delta_\mathrm{g}\delta_\mathrm{g}}^\mathrm{1h}+P_{\delta_\mathrm{g}\delta_\mathrm{g}}^\mathrm{2h}$,
where $P_{\delta_\mathrm{g}\delta_\mathrm{g}}^\mathrm{SN}$ is the shot-noise contribution, whose expression is 
\begin{equation}
P_{\delta_\mathrm{g}\delta_\mathrm{g}}^\mathrm{SN}=\frac{1}{\bar{n}_\mathrm{g}},
\end{equation}
where $\bar{n}_\mathrm{g}$ is defined in Eq.~\eqref{eq:ngbar}. The 1- and 2-halo terms are 
\begin{equation}
    P_{\delta_\mathrm{g}\delta_\mathrm{g}}^\mathrm{1h}=\langle \hat{u}^\mathrm{gg}_k \rangle_ n\quad \mathrm{and}\quad P_{\delta_\mathrm{g}\delta_\mathrm{g}}^\mathrm{2h}=\langle b^{(1)}\hat{u}^\mathrm{g}_k \rangle_ n^2P_L(k,\chi), 
\end{equation}
respectively, where $\hat{u}^\mathrm{gg}_k$ and $\hat{u}^\mathrm{g}_k$ are defined in Eq.~\eqref{eq:uggk} and \eqref{eq:ugk}. In the low-$k$ limit, since $\hat{u}^{\mathrm{sat}}_k\approx1$, we have $\langle b^{(1)}\hat{u}^\mathrm{g}_k \rangle_ n\rightarrow b_\mathrm{g}(z)$ and thus $P_{\delta_\mathrm{g}\delta_\mathrm{g}}^\mathrm{hm}\approx b_\mathrm{g}^2P_L$. In the high-$k$ regime, the scale dependence is determined by the HOD and departs from that of the \verb|halofit| power spectrum, as can be seen on the right panel of Figure~\ref{fig:pkshm}.

In fact, the observed power spectrum of galaxies includes an extra contributions coming from lensing magnification. We have
\begin{equation}
    \delta_\mathrm{g}^\mathrm{obs}=\delta_\mathrm{g}+\mu_\mathrm{g}\quad \mathrm{with}\quad \mu_\mathrm{g} = (5s-2)\delta_\mathrm{m} \quad\mathrm{and}\quad s \equiv\frac{\mathrm{dlog_{10}N}}{\mathrm{d}\mathcal{M}}
\end{equation}
where $\mathcal{M}$ is the apparent magnitude of the galaxies, evaluated near the magnitude limit of the survey. The extra-contributions are  $P_\mathrm{\delta_g\mu_g}^\mathrm{hm}=(5s-2)P_\mathrm{\delta_g\delta_m}^\mathrm{hm}$ and $P_\mathrm{\mu_g\mu_g}^\mathrm{hm}=(5s-2)^2P_\mathrm{\delta_m\delta_m}^\mathrm{hm}$.

\subsubsection{Useful Angular Power Spectra}\label{sss:useful_cls}

\begin{figure}
\centering
    \includegraphics[width=0.7\columnwidth]{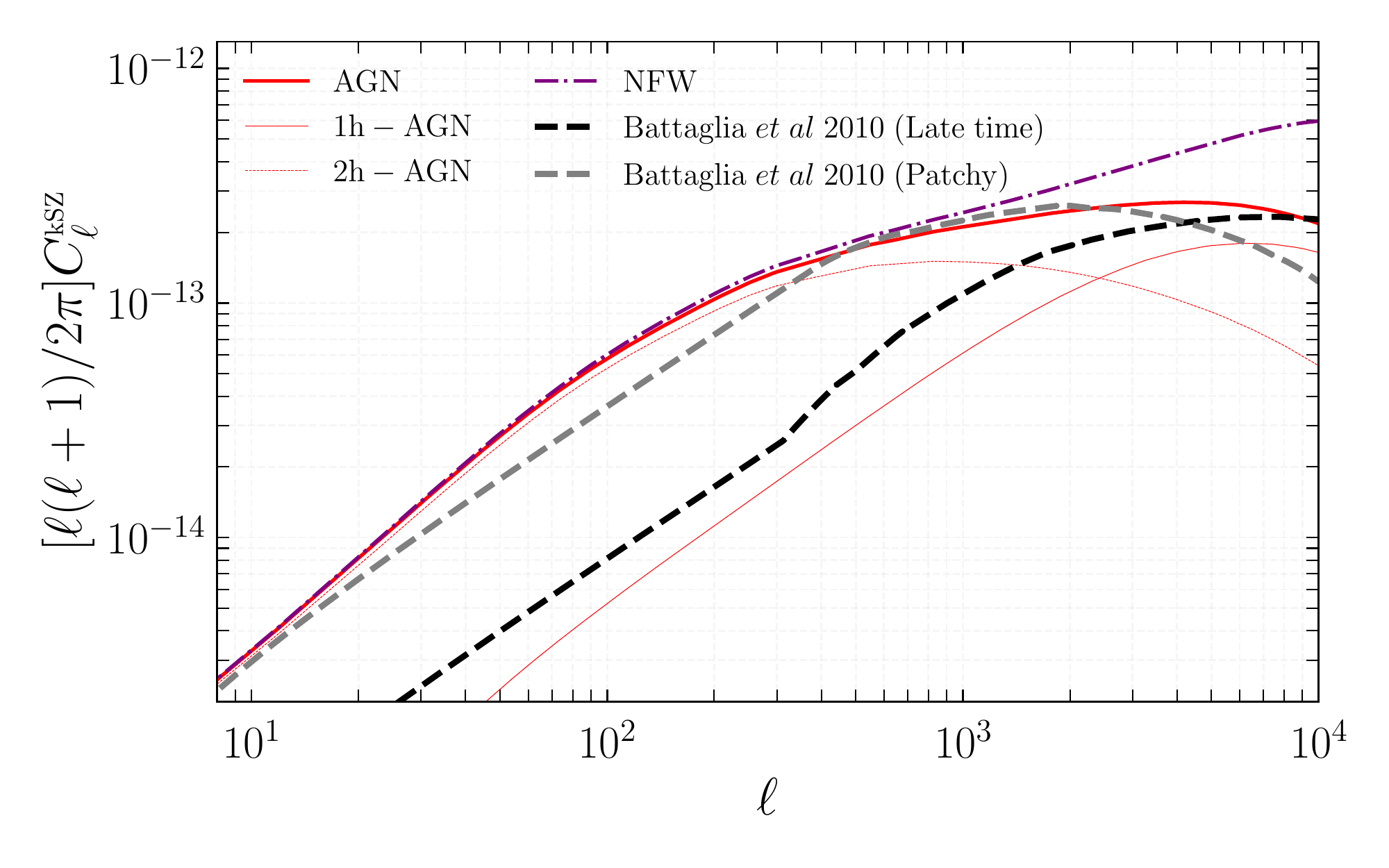}
    \vspace{-0.3cm}

    \caption{The dimensionless angular power spectrum of the kSZ effect from hydrodynamical simulations and using the halo model formula in the high-$k$ regime (see Eq.~\ref{eq:clksz2}). For the halo model results we assumed the gas distributed according the \textit{AGN Feedback} model (red solid lines) and NFW-like profile (purple line). The simulation results are the thick dashed lines and are from \cite{Battaglia_2010}. The black dashed line represents the contribution from halos while the grey dashed line is the contribution from reionization (it is not used in our analysis, but shown here for comparison).} 
    \label{fig:cls_ksz}
\end{figure}

Angular power spectra are obtained from the 3D power spectra using Eq.~\eqref{eq:clxy}. Each tracer has its own projection kernel. For the kSZ effect, $W^\mathrm{kSZ}$ is given in Eq.~\eqref{eq:clksz2} along with the angular power spectrum expression in the high-$k$ limit. We show the kSZ angular power spectrum in Figure~\ref{fig:cls_ksz} for different gas profile assumptions. The low-$\ell$ limit does not depend on the assumption. The halo model power spectrum computed with  the \textit{AGN Feedback} gas profile agrees reasonably well with results from simulations \citep{Battaglia_2010}.

For CMB lensing, the projection kernel is 
\begin{equation}
 W^\mathrm{\kappa_{_{CMB}}}(\chi)=\frac{3}{2}\frac{\Omega_\mathrm{m}(H_0/c)^2}{\chi^2}(1+z)\chi \frac{\chi_\star-\chi}{\chi_\star}.\label{eq:wkcmb}
\end{equation}
where $\chi_\star$ is the comoving distance to last scattering. 

For galaxy weak lensing, the projection kernel is 
\begin{equation}
 W^\mathrm{\kappa_{_{g}}}(\chi)=\frac{3}{2}\frac{\Omega_\mathrm{m}(H_0/c)^2}{\chi^2}(1+z) \chi I_s(\chi)\quad\mathrm{with}\quad I_s(\chi)=\int_z^{+\infty}\mathrm{d}z_\mathrm{s} \varphi^\prime_s(z_\mathrm{s})\frac{\chi_s-\chi}{\chi_s}\label{eq:wkg}
\end{equation}
where $\varphi^\prime_s$ is the normalized redshift distribution of source galaxies, and we used the notation $\chi_s=\chi(z_\mathrm{s})$ for the comoving distance to redshift $z_\mathrm{s}$. (Note that, formally, Eq.~\eqref{eq:wkg} reduces to  Eq.~\eqref{eq:wkcmb} if $\varphi^\prime(z_\mathrm{s})=\delta_D(z_\mathrm{s}-z_\star)$.).

\begin{figure}
    \includegraphics[width=1.\columnwidth]{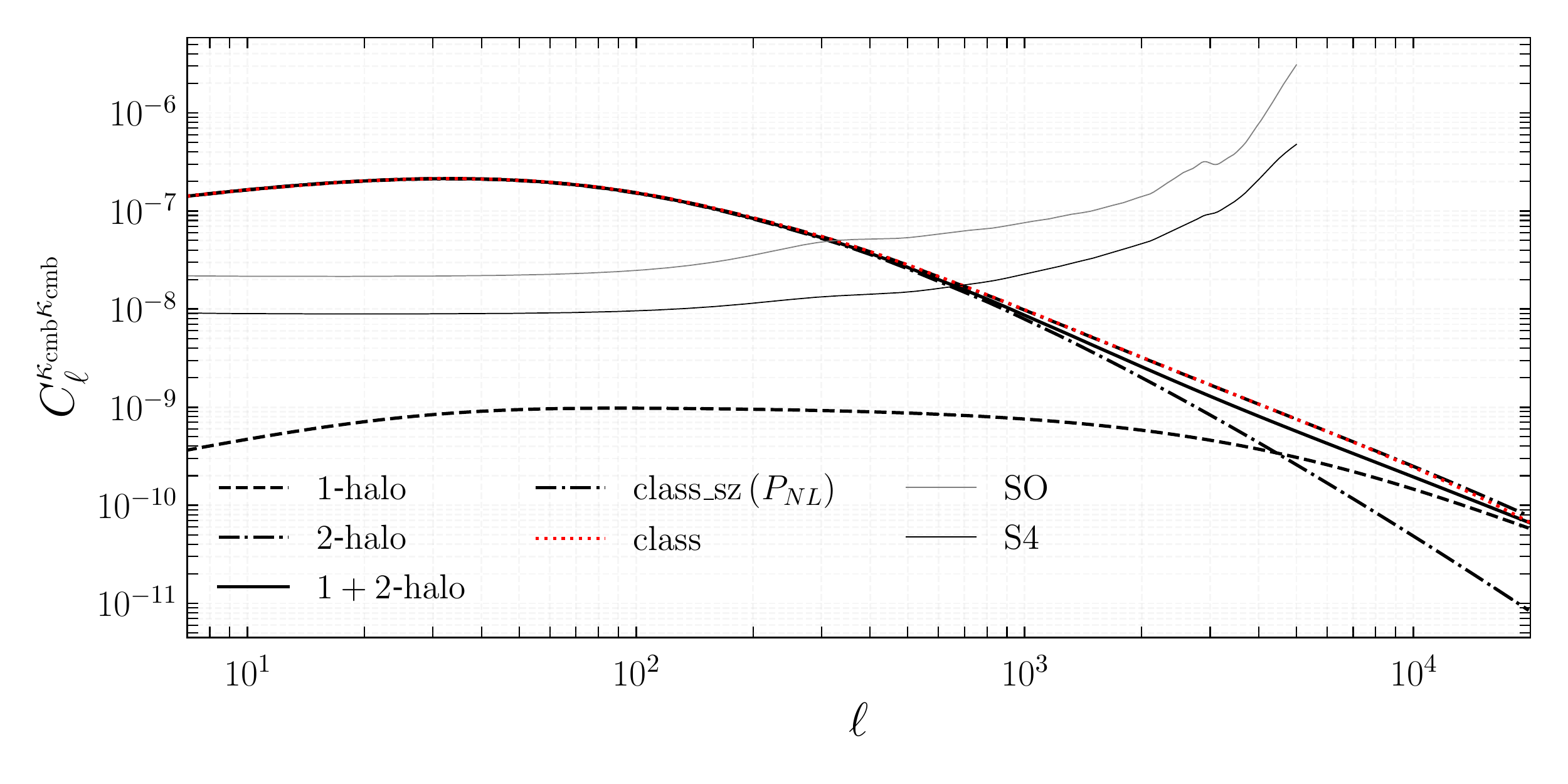}
    \caption{Dimensionless angular power spectrum of CMB lensing convergence with noise estimates for SO and S4. The red dotted line shows the {\texttt{class}} prediction which is based on {\texttt{halofit}}. The black lines are the {\texttt{class\_sz}} predictions using halo model (solid, dashed and dashed-dotted lines) and {\texttt{halofit}} (thin dashed line). Note that for this plot we used $z_\mathrm{max}=15$. 
    } 
    \label{fig:cls_kk}
\end{figure}

For galaxy number density, the projection kernel is 
\begin{equation}
    W^{\delta_\mathrm{g}}(\chi) = \frac{H}{\chi^2c }\varphi_\mathrm{g}^\prime(z)\quad\mathrm{with}\quad\varphi_\mathrm{g}^\prime(z) = \frac{1}{N_\mathrm{g}^\mathrm{tot}}\frac{\mathrm{d}N_\mathrm{g}}{\mathrm{d}z}\quad \mathrm{where}\quad N_\mathrm{g}^\mathrm{tot}=\int \mathrm{d}z\frac{\mathrm{d}N_\mathrm{g}}{\mathrm{d}z}.\label{eq:wg}
\end{equation}
Here, $\frac{\mathrm{d}N_\mathrm{g}}{\mathrm{d}z}$ is the galaxy redshift distribution of the survey which is from data. 
For lensing magnification, the projection kernel is
\begin{equation}
 W^\mathrm{\mu_{_{g}}}(\chi)=\frac{3}{2}\frac{\Omega_\mathrm{m}(H_0/c)^2}{\chi^2}(1+z) \chi I_\mathrm{g}(\chi)\quad\mathrm{with}\quad I_\mathrm{g}(\chi)=\int_z^{+\infty}\mathrm{d}z_g \varphi^\prime_\mathrm{g}(z_\mathrm{g})\frac{\chi_\mathrm{g}-\chi}{\chi_\mathrm{g}}.\label{eq:wmug}
\end{equation}
where $\varphi^\prime_\mathrm{g}$ is the normalized redshift distribution of lens galaxies, and we used the notation $\chi_\mathrm{g}=\chi(z_\mathrm{g})$ for the comoving distance to redshift $z_\mathrm{g}$.
in Figure~\ref{fig:cls_kk} we show a comparison of CMB angular power power spectra computed with the halo-model and the \verb|halofit| power spectrum. The noise curves for SO and S4 are also plotted. 

\section{Comparison with Previous Works}\label{app:comp}
\subsection{Choices of Wiener  Filter}\label{app:filters}
\begin{figure}
    \includegraphics[width=1.\columnwidth]{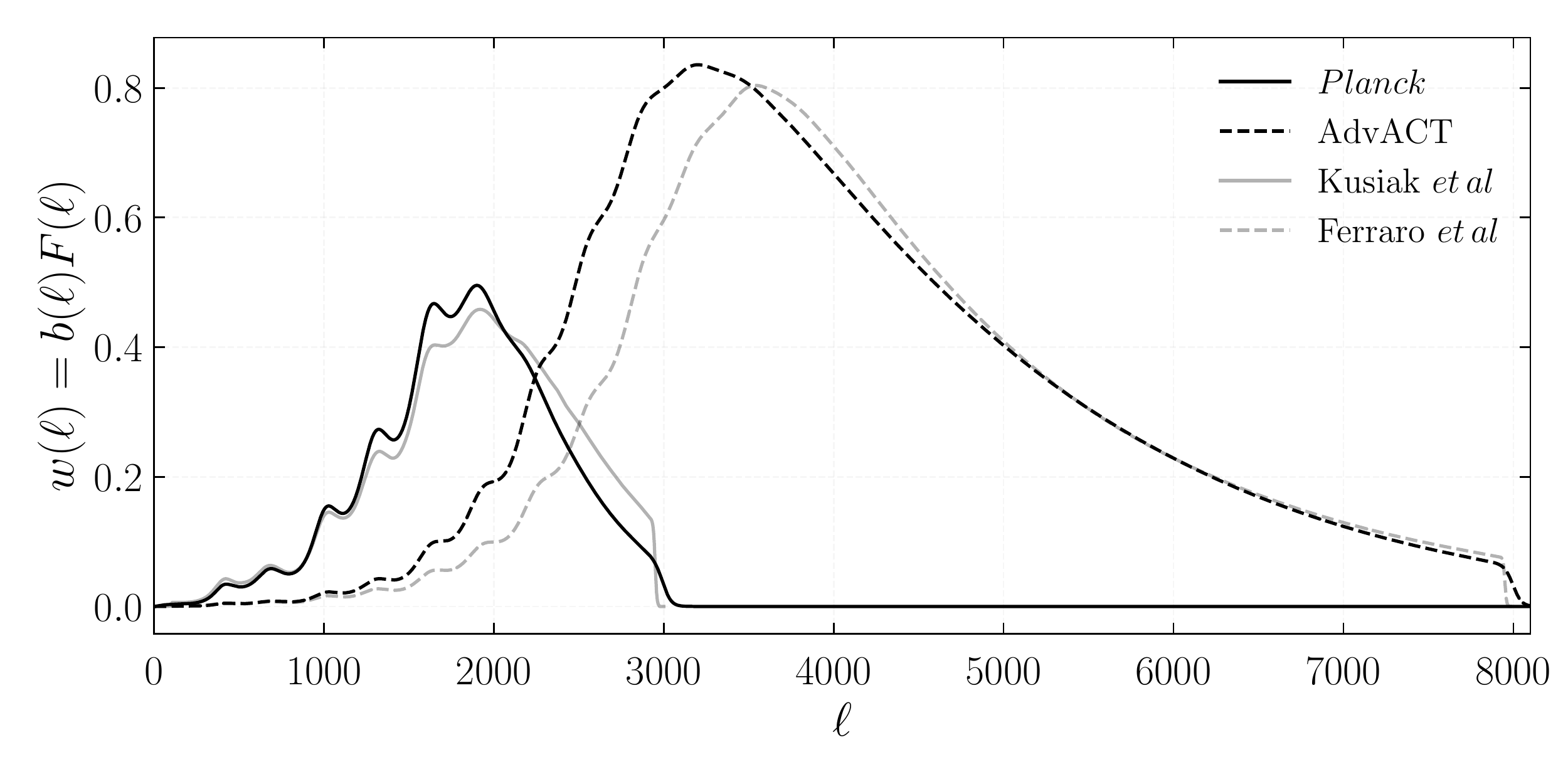}
    \vspace{-0.6cm}
     \caption{Comparison of Wiener filters (multiplied by the respective beams).}    
    \label{fig:filter_comp_figure}
\end{figure}

In all previous works using the projected-field kSZ estimator, the filter ansatz was 
\begin{equation}
    w^\mathrm{prev.}(\ell)=b(\ell)\frac{C_\ell^\mathrm{kSZ}}{C_\ell^\mathrm{tot}}.
\end{equation}
This differs from our filter definition (see Eq.~\ref{eq:filter}) as we use a square root in the numerator. These two choices yield similar SNR, but the version with the square root is always better, yielding larger SNRs by $\approx 3-5\%$.\footnote{The denominator of the filter actually used in previous works also differed slightly from $C_\ell^\mathrm{tot}$, due to a missing beam factor in the noise contribution to the total power spectrum.  The only effect due to this is a slight change in the high-$\ell$ turnover of the filter.}

To justify the filter definition of Eq.~\eqref{eq:filter} we provide a heuristic derivation based on Appendix B of \cite{Smith_2007}. Since the projected field kSZ power spectrum is an integral over the bispectrum, we expect the optimal estimator to be similar to the optimal bispectrum estimator.
In general, the optimal bispectrum estimator is given by
\begin{equation}
    \hat{\mathcal{C}}=\frac{1}{F}(\hat{\mathcal{C}}_3 -\hat{\mathcal{C}_1}),
\end{equation}
where $F$ is a normalization (whose exact expression is not important here) and where $\hat{\mathcal{C}}_3$ and $\hat{\mathcal{C}}_1$ are a three-point and one-point term. The one-point term can be significant when the sky coverage is partial and the noise is inhomogeneous, but we can omit it for the purpose of this discussion. The three-point term contains the scale-dependent information we are interested in. It is given by
\begin{equation}
    \hat{\mathcal{C}}_3 \approx \sum_{\ell_i,m_i}B_{\ell_1 \ell_2 \ell_3}\mathcal{G}_{m_1 m_2 m_3}^{\ell_1 \ell_2 \ell_3}\tilde{a}_{\ell_1 m_1}\tilde{a}_{\ell_2 m_2} \tilde{X}_{\ell_3 m_3}\label{eq:c3opt}
\end{equation}
where $B$ is the theoretical/predicted hybrid bispectrum, $\mathcal{G}$ is a Wigner-$3j$ symbol \citep[see Eq. 6 of][]{Smith_2007} and the tilde fields $\tilde{a}$ and $\tilde{X}$ are the harmonic coefficients of inverse-variance weighted CMB and LSS maps, i.e.,
\begin{equation}
    \tilde{a}_{\ell m}=(C_\ell^{\Theta\Theta}+C_\ell^{\mathrm{kSZ}}+N_\ell^{\Theta\Theta})^{-1} a_{\ell m},\quad\mathrm{and}\quad
    \tilde{X}_{\ell m}=(C_\ell^{XX}+N_\ell^{XX})^{-1} X_{\ell m}.
\end{equation}
The inverse variance weighting ensures that the poorly measured modes are filtered out and hence contribute less to the variance of the  estimator. From Eq.~\eqref{eq:c3opt} we see that the optimal filter $w_{\ell_1}$ to apply to the temperature map $a_{\ell_1 m_1}$ should have the scale dependence of $B_{\ell_1 \ell_2 \ell_3}/C_{\ell_1}^\mathrm{tot}$. Since the term that contributes most to the bispectrum in the range of scales of interest here is the 1-halo term, we can approximate the scale dependence as \begin{equation}
    w_{\ell_1}\sim \frac{1}{C_{\ell_1}^\mathrm{tot}}\int \mathrm{d}n_1 \hat{u}_{\ell_1}^\mathrm{e}(m_1)\hat{u}_{\ell_2}^\mathrm{e}(m_1)\hat{u}_{\ell_3}^X(m_1)\sim \frac{\sqrt{C_{\ell_1}^\mathrm{kSZ}}}{C_{\ell_1}^\mathrm{tot}}.
\end{equation}

\subsection{Comparison with forecats from Ferraro et al.}\label{ap:comp_ferraro}
\begin{figure}
    \includegraphics[width=0.5\columnwidth]{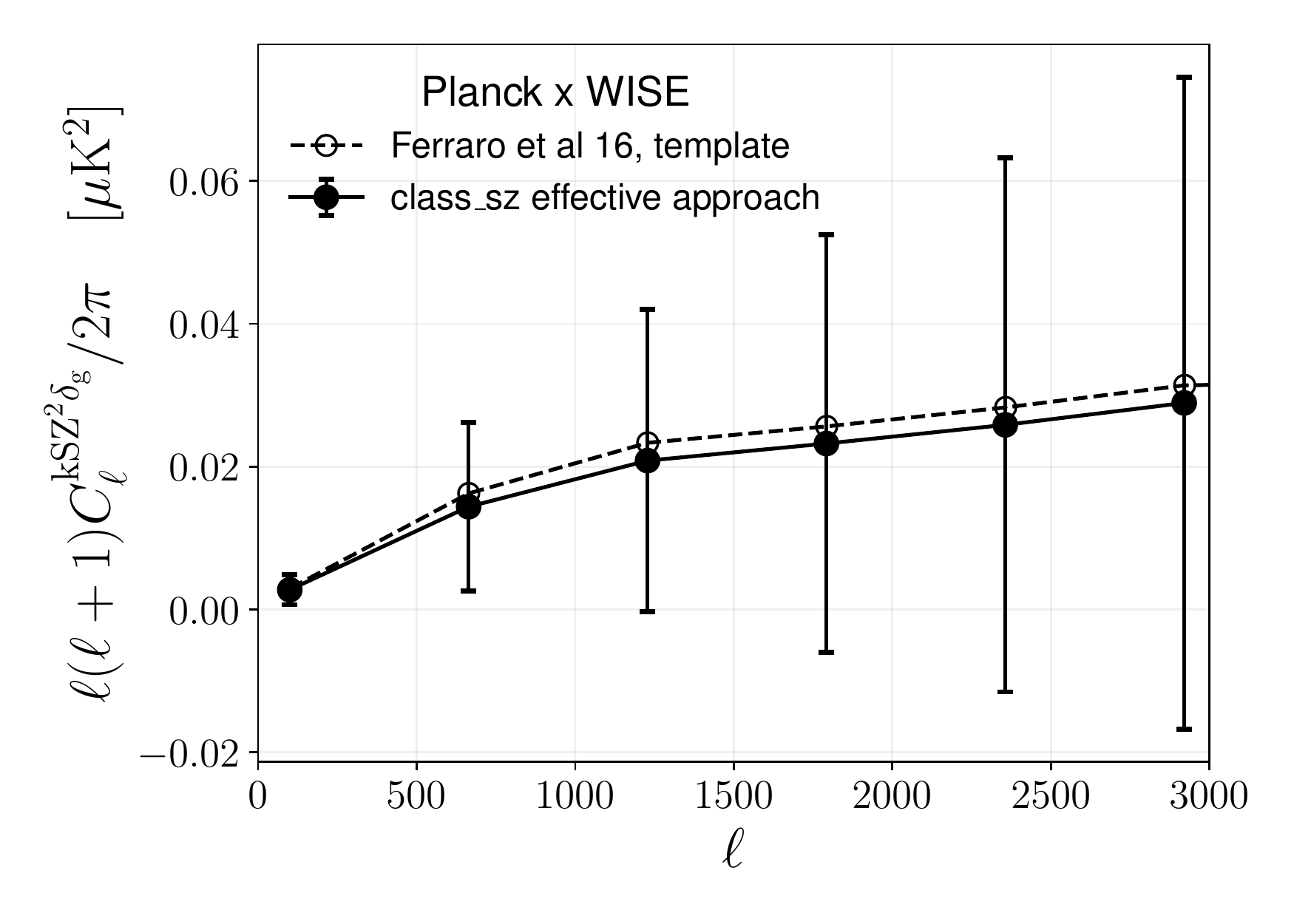}\includegraphics[width=0.5\columnwidth]{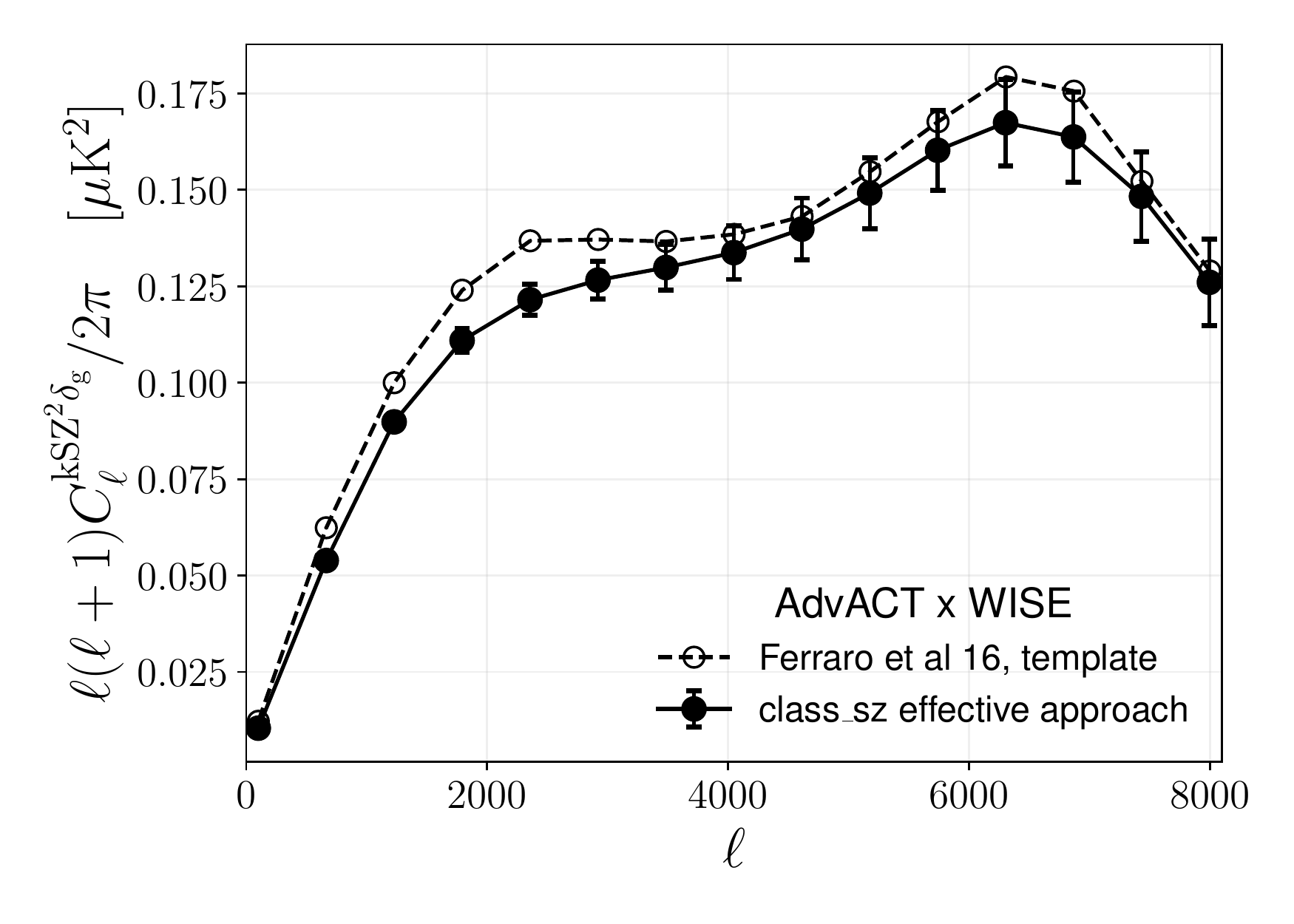} 
    \vspace{-0.6cm}
    \caption{Projected-field kSZ power spectrum in cross-correlation with galaxy density assuming \textit{Planck}$\times$\textit{WISE} (left) and \textit{AdvACT}$\times$\textit{WISE} (right). The solid lines with error bars are the {\texttt{class\_sz}} predictions for the power spectrum and covariance matrix and the dashed lines are computed with an independent code that was used in \citep{Ferraro2016,Hill2016,Kusiak:2021hai}. Here the computations are done within the \textit{effective approach} (see Subsection~\ref{ss:eff}), using the the filters of Figure~\ref{fig:filter_comp_figure} (grey lines).}  
    \label{fig:cls_comp_snr}
\end{figure}

We compared our \textit{effective approach} calculations with the code used in \cite{Ferraro2016,Hill2016,Kusiak:2021hai}. We found a good agreement over all scales for \textit{Planck}$\times$\textit{WISE} and AdvACT$\times$\textit{WISE} (relative difference of $\approx$10\%). For this comparison we used the same settings as in \cite{Ferraro2016}. In particular, we used the filters labeled `Kusiak et al' and `Ferraro et al' in Figure~\ref{fig:filter_comp_figure}, a sky fraction $f_\mathrm{sky}$, an effective galaxy bias $b_\mathrm{g}^\mathrm{eff}=1.13$, the linear matter power spectrum in the velocity dispersion, and the normalized galaxy redshift distribution of Figure~10 of \cite{Ferraro2016}. For the covariance matrix calculation we used a shot noise of $C_\ell^\mathrm{SN}=1.26\times 10^{-7}$ (which corresponds to 50 million galaxies over half the sky). The power spectra are plotted together on Figure~\ref{fig:cls_comp_snr}.

\cite{Ferraro2016} presented SNR forecasts that differ from the ones we obtained in Section~\ref{s:forecasts}. For AdvACT$\times$\textit{WISE} they obtained an SNR on $f_\mathrm{free}$ of 232. In the same settings as them (see paragraph above) we get nearly the same estimate, finding 223. This is roughly six times larger than our most optimistic forecast with AdvACT$\times$\textit{unWISE} (note that the SNR on $f_\mathrm{free}$ is two times $\mathrm{SNR_{tot}}$ if we do not take into account degeneracies with HOD parameters and assume the galaxy bias to be tightly constrained). This significant difference is mainly explained by the fact that the filter in \cite{Ferraro2016} uses a $10 \, \mu\mathrm{K}$-arcmin pixel noise level while we use $20 \, \mu\mathrm{K}$-arcmin (which is more consistent with current ILC noise estimates) and that our fiducial model for the projected-field kSZ power spectrum assumes the \textit{AGN Feedback} gas density profile whereas \cite{Ferraro2016} uses the \textit{effective approach } calculation. The difference of filter and the difference of fiducial model change the SNR by roughly the same amount.

\bibliographystyle{aasjournal}
\bibliography{main}{}

\end{document}